\documentclass{aa}
\usepackage{graphicx}
\usepackage{natbib}
\usepackage{float}

\usepackage{color}
\usepackage{multirow}

\usepackage[para,online,flushleft]{threeparttable}
\usepackage{txfonts}

\bibpunct{(}{)}{;}{a}{}{,} 

\usepackage{afterpage}
\usepackage{newtxtext,newtxmath}
\usepackage{booktabs, multirow}
\usepackage{comment}
\usepackage{amsmath,siunitx}
\usepackage{soul}
\usepackage[table]{xcolor} 
\usepackage{changepage,threeparttable} 
\usepackage[utf8]{inputenc}
\usepackage[T1]{fontenc}
\usepackage{tablefootnote}
\usepackage{footnote}
\usepackage{placeins}
\DeclareRobustCommand{\VAN}[3]{#2}
\let\VANthebibliography\thebibliography
\def\thebibliography{\DeclareRobustCommand{\VAN}[3]{##3}\VANthebibliography}

\usepackage{graphicx}   
\usepackage{amsmath}    
\usepackage{amssymb}    

\usepackage[]{hyperref}
\hypersetup{
  colorlinks,
  citecolor=blue,
  linkcolor=red,
  urlcolor=blue}
\usepackage{caption,subcaption,graphicx}

\renewcommand\thesubfigure{(\alph{subfigure})}
\newcommand{\PRESTO}{\texttt{PRESTO}}

\newcommand{\TEMPO}{\texttt{TEMPO}}

\newcommand{\PULSARMINER}{\texttt{PULSAR\_MINER}}
\newcommand{\SPIDERTWISTER}{\texttt{SPIDER\_TWISTER}}

\newcommand{\dmunit}{pc\,cm$^{-3}$}

\newcommand{\BWeff}{{\rm BW}_{\rm eff}}

\newcommand{\squeezeup}{\vspace{-2.5mm}}

\usepackage[normalem]{ulem}

\begin{document} 

   \title{Upgraded GMRT survey for pulsars in globular clusters.}
   \subtitle{I: Discovery of a millisecond binary pulsar in NGC~6652}
   \titlerunning{Search for pulsars with the uGMRT}
   \author
   {T.~Gautam \inst{\ref{1}} \thanks{Member of the International Max Planck Research School (IMPRS) for Astronomy and Astrophysics at the University of Bonn}
    \and A.~Ridolfi\inst{\ref{1},\ref{2}}
    \and P.~C.~C.~Freire\inst{\ref{1}}
    \and R.~S.~Wharton\inst{\ref{3}}
    \and Y.~Gupta\inst{\ref{4}}
    \and S.~M.~Ransom\inst{\ref{5}}
    \and L.~S.~Oswald\inst{\ref{6},\ref{7}}
    \and M.~Kramer\inst{\ref{1}}
    \and M.~E.~DeCesar\inst{\ref{8}}
    }

    \institute{
Max-Planck-Institut f\"{u}r Radioastronomie, Auf dem H\"{u}gel 69, D-53121 Bonn, Germany\label{1}
\and
INAF -- Osservatorio Astronomico di Cagliari, Via della Scienza 5, I-09047 Selargius (CA), Italy\label{2}
\and
NASA Postdoctoral Program Fellow, Jet Propulsion Laboratory, California Institute of Technology, Pasadena, CA 91109, USA\label{3}
\and
National Centre for Radio Astrophysics, Tata Institute of Fundamental Research, Pune 411007, Maharashtra, India\label{4}
\and
National Radio Astronomy Observatory, 520 Edgemont Rd., Charlottesville, VA 22903, USA\label{5}
\and
Department of Astrophysics, University of Oxford, Denys Wilkinson building, Keble road, Oxford OX1 3RH, UK\label{6}
\and
Magdalen College, University of Oxford, Oxford OX1 4AU, UK\label{7}
\and
George Mason University, Fairfax, VA 22030, resident at U.S. Naval Research Laboratory, Washington, D.C. 20375, USA\label{8}
\\
\email{tgautam@mpifr-bonn.mpg.de}
}

    \date{Received --; accepted --}
\abstract{}{}{}{}{} 
\abstract
{Globular clusters (GCs) contain a unique pulsar population, with many exotic systems that can form only in their dense stellar environments. 
Such systems are potentially very interesting for new tests of gravity theories and neutron-star mass measurements.}
{The leap in sensitivity of the upgraded Giant Metrewave Radio Telescope (uGMRT) in India, especially at low radio frequencies ($<$ 1 GHz),
motivated a new search for radio pulsars in a group of eight southern GCs. We aim to image these clusters in order to have independent measurements of the radio fluxes of known pulsars and
the identification of bright radio sources that could be pulsars missed by pulsation search pipelines due to their inherent limitations.}
{The observations were conducted at 650 MHz (Band 4 receivers) on Terzan~5, NGC~6441, NGC~6440, and NGC~6544, and at 400 MHz (Band 3 receivers) on NGC~6652, NGC~6539, NGC~1851, and M~30. Segmented acceleration and jerk searches were performed on the data. Simultaneously, we obtained
interferometric data on these clusters, which were later converted into radio images.}
{We discovered PSR J1835$-$3259B, a 1.83-ms pulsar in NGC~6652; this is in a near-circular wide orbit of 28.7-hr with an unidentified low-mass ($ \sim 0.2 \, M_{\rm \odot}$) companion, likely a helium white dwarf. We derived a ten-year timing solution for this system.
We also present measurements of scattering, flux densities, and spectral indices for some of the previously known pulsars in these GCs.}
{ A significant fraction of the pulsars in these clusters have steep spectral indices. Additionally, we detected eight radio point sources not associated with any
known pulsar positions in the radio images. There are four newly identified
sources, three in NGC~6652 and one in NGC~6539, as well as one previously
identified source in NGC~1851, NGC~6440, NGC~6544, and Terzan~5. 
Surprisingly, our images show that our newly discovered pulsar, PSR J1835$-$3259B, is the brightest pulsar in all GCs we have imaged; like
other pulsars with broad profiles (Terzan 5 C and O), its flux density in the radio images is much larger than in its pulsations. This indicates that their pulsed emission is only a fraction of their total emission. {The detection of radio sources outside the core radii but well within the tidal radii of these clusters show that future GC surveys should complement the search analysis by using the imaging capability of interferometers, and preferentially synthesise large number of search beams in order to obtain a larger field of view.}
}

\maketitle


\section{Introduction}

With immense stellar densities of up to $10^6$ stars per cubic parsec, the cores of globular clusters (GCs) are unique environments for tidal captures and the exchange encounters of stellar systems \citep{Sigurdsson_Phinney1995,Pooley+2003}. Interactions of old neutron stars (NSs) with low-mass stars form a large population of low-mass X-ray binaries (LMXBs) in GCs, about three orders of magnitude more numerous per unit stellar mass than in the Galactic field \citep{Clark+1975}.

If the LMXB is not disturbed by close stellar encounters (which are exceedingly rare in the Galactic disc), then the NS in the binary will continue accreting mass from its evolving companion \citep{Alpar+1982, Radhakrishnan_Srinivasan1982, Bhattacharya_vandenHeuvel1991}. When accretion stops, we have a
binary system with a low-eccentricity orbit consisting of a pulsar with a spin period of a few milliseconds and generally very small spin-down rates; these are known as `millisecond pulsars' (MSPs). Their companions are either light white dwarf (WD) stars or a low-mass non-degenerate star (in which case, we likely have an eclipsing system). In some cases, the MSP has no companion at all, possibly because it was completely ablated by winds from the pulsar. Almost all pulsars in the Galactic disc with spin periods smaller than 8-ms fit into these groups. 

The large number of LMXBs per unit stellar mass in GCs should result in a similarly enhanced MSP population.
Since the discovery of the first GC pulsar, PSR~B1821$-$24 \citep{1987Natur.328..399L}, a total of 246 pulsars in 36 GCs have been discovered to date\footnote{As of 2022 April 6, see \url{https://www3.mpifr-bonn.mpg.de/staff/pfreire/GCpsr.html}}; this population includes about a third of all known MSPs. As expected, most of these are found in GCs with a large stellar encounter rate, $\Gamma$ \citep{1987IAUS..125..187V}.

The pulsar populations in the different GCs are strikingly different.
In 47~Tucanae, the spin periods of the 29 known pulsars range between 2 and 8-ms \citep{Ridolfi+2016,Freire+2017,Ridolfi+2021}. The same is true for other GCs, particularly those with low-density cores. The cluster 47~Tucanae is important because, although it has a large $\Gamma$ (hence the large number of MSPs), it has a low interaction rate `per binary', $\gamma$ \citep{Verbunt_Freire_14}. This means that once an LMXB forms, there is little chance of it being disturbed again; this results in a set of MSPs similar to the Galactic MSP population described above.

However, in GCs with very high core densities, such as core-collapse clusters, $\gamma$ is much higher. This means that once an LMXB forms, there is a much higher probability that it will be disturbed, either as an LMXB or later as an MSP - WD system. This results in a large number of binary disruptions, with many isolated and partially recycled pulsars present (instead of the fully recycled pulsars seen in low-density GCs).

Such perturbations can create `exotic' binary systems, which have unconventional properties. In all clusters, relatively close interactions increase the orbital eccentricities of many MSP - WD binaries by orders of magnitude compared to those observed in the Galactic disc \citep{1992RSPTA.341...39P}. In high-$\gamma$ clusters, secondary exchange encounters -- where a previously recycled pulsar acquires a new companion -- can create eccentric pulsar binaries with high-mass degenerate companions such as another NS or a heavy WD (NGC~1851A, \citealt{Freire+2004,Ridolfi+2019}; NGC~6544B, \citealt{Lynch+2012}; NGC~6652A, \citealt{DeCesar+2015} and NGC~6624G, \citealt{Ridolfi+2021}), which are unlike any observed in the Galactic disc. This mechanism could potentially create an MSP-black hole (BH) system, which would offer a unique laboratory for tests of the fundamental properties of gravitational physics \citep{1999ApJ...514..388W,2014MNRAS.445.3115L}. For this reason, searching for pulsars in high-$\gamma$ GCs is a high priority.

One important feature of the MSP population in GCs is that we only detect its brightest members. The reason for this is the large distances to GCs, typically of the order of 10 kpc. This means that a large pulsar population remains undetected, yet there are likely several thousands in the Galactic GC systems (e.g. \citealt{Bagchi+2011,2015aska.confE..47H}). 
The main barrier to detecting this large population is primarily a lack of sensitivity. Therefore, whenever new observing systems with higher sensitivities become available, there is a large increase in the number of known pulsars in GCs.

The previous wave of discoveries happened in the mid-2000s. This was brought about by the use of sensitive, broadband L (1-2 GHz) and S-band (2-4 GHz) receivers, leading to quite a few new discoveries at Parkes \citep{2000ApJ...535..975C,Possenti+2005}, Arecibo \citep{2007ApJ...670..363H}, the Green Bank Telescope (GBT; \citealt{Ransom+2004,2008ApJ...675..670F}), and the Giant Metrewave Radio telescope (GMRT; \citealt{Freire+2004}). The inauguration of a new generation of extremely sensitive radio telescopes such as FAST \citep{Nan+2013} and MeerKAT \citep{Booth_Jonas2012,Jonas2016,Camilo2018} and major upgrades to existing facilities, such as the installation of technologically advanced digital back-ends and new broad-band receivers with lower system temperatures, is now resulting in another wave of discoveries, with close to 100 new pulsars found as of April 2022 (e.g. \citealt{Ridolfi+2021,2021arXiv210608559P}).

One of the facilities that have recently become available is the upgraded GMRT \citep{Gupta+2017}, henceforth uGMRT. The GMRT has long been one of the most sensitive telescopes at low radio frequencies, and for that reason, it has played an important role in discovering steep-spectrum pulsars (e.g. \citealt{Freire+2004,Bhattacharyya+2016,Joshi+2009,2021ApJ...910..160B}). With the recently installed sensitive receivers operating at lower system temperatures covering a wide bandwidth of $\sim$ 200 MHz (more than ten times that of the GMRT), along with the installation of a real-time coherent de-dispersion facility, the uGMRT represents a significant improvement in sensitivity compared to the GMRT. For that reason, it will have a critical role in discovering steep-spectrum pulsars.
As described in detail below, the new coherent de-dispersion facility is especially helpful for clusters with higher dispersion measures (DMs); in earlier GMRT surveys, radio pulsations from high-DM pulsars were smeared out at low frequencies due to intra-channel dispersion.
In addition, the interferometric nature of the GMRT allows the simultaneous recording of visibility data from the full array;
this results in radio images of entire clusters and the detection of radio sources that may be missed by the limited size of phase array beams - some of which could be undiscovered pulsars.

These instrumental improvements and the low-frequency bands, which complement those of the other ongoing GC pulsar surveys, motivated us to start a GC pulsar survey with this instrument.
Indeed, all recent sensitive surveys for pulsars in GCs were made in the L or S band. Such high-frequency surveys can be biased towards detecting relatively flat-spectrum sources compared to the low-frequency surveys with similar observational setups. Thus, low-frequency surveys are crucial in the detection of previously undetected steep-spectrum pulsars if they exist, and they help characterise the known pulsars at these frequencies.  

In this paper, we report the first results from this survey, from observations of eight GCs at 400 MHz (Band 3) and 650 MHz (Band 4) with the new GMRT Wideband Backend (GWB; \citealt{Reddy+2017}) system of the uGMRT. In Section \ref{sec:survey_details}, we discuss the survey, including the GC selection criteria, the observations performed to date, estimates of the survey sensitivity, and the data analysis. In the following sections, we describe some of the survey results, which illustrate the variety of topics that can be addressed with the type of data we obtained. The discovery and timing analysis of PSR J1835$-$3259B, the first pulsar found in this survey, are discussed in Section \ref{sec:results}. In Section \ref{subsec:characterization}, we discuss some of the characteristics of the previously known pulsars in these clusters that were detected in our observations, namely their flux densities, spectral indices and scattering timescales. Section \ref{sec:imaging_gc} presents the imaging analysis and discusses the point radio sources found in the radio images of these clusters.  Finally, Section \ref{sec:conclusions} provides the summary and conclusions of this paper.

\section{Survey details}
\label{sec:survey_details}
\subsection{Target selection}

The considerations made above regarding the interaction rate per binary were the primary guides of our target selection. We chose GCs known to contain 
eccentric and, in most cases, binary pulsars with high mass ($> 0.5 \, M_{\odot}$) companions (NGC~1851, Terzan~5, NGC~6441, NGC~6440, NGC~6539, NGC~6544), or suspected of having them, such as M~30 (see \citealt{2004ApJ...604..328R}; see Table~\ref{tab:cluster_details}).
Searching such clusters for more exotic binaries is especially valuable for scientific follow-up as it is in such binaries that we have been able to precisely measure the NS masses \citep{Lynch+2012,Ridolfi+2019} - and in one case (M15C) even perform a test of GR \citep{2006ApJ...644L.113J}. Furthermore, as mentioned before, the secondary exchange encounters that formed these binaries could in principle form MSP - MSP or MSP - BH systems.

Another advantage of this group of GCs is that the known DMs make the search process easier. Furthermore, the timing of some pulsars known in these clusters has already benefited from our additional
GMRT observations \citep{Ridolfi+2019}.
Finally, they are located south of the Arecibo and FAST survey areas, where the sensitivity of the uGMRT is competitive, particularly at lower radio frequencies.

Previous low-frequency GMRT GC surveys \citep{Freire+2004} were restricted to low-DM clusters because of intra-channel dispersive smearing. If we use a filterbank as a back-end, this is given by
\begin{equation}
    \tau_{\rm DM} = 8.3 \times \, \Delta \nu \times \rm \, DM \, \nu^{-3} \mu\rm s,
\end{equation}
where $\Delta \nu$ represents bandwidth of an individual channel in MHz and
$\nu$ represents the centre frequency of the band (GHz), with $\nu \gg \Delta \nu$.
Clusters with large DM values exhibit larger pulse smearing within a frequency channel,
which increases further for the lower frequencies used in previous GMRT surveys ($\sim 320\, \rm MHz$, \citealt{Freire+2004}). For instance, a cluster with a DM >100 $\rm pc \, cm^{-3}$ would lead to a smearing timescale of more than 1.27-ms at 400 MHz (considering a channel bandwidth of 97.6 KHz). Because of this, some of the high-DM clusters have not been searched effectively for pulsars at these lower frequencies: instead, all pulsars in the high-DM GCs Terzan 5, NGC 6440 and NGC 6441 have been found with the S-band system of the GBT or, more recently, in
MeerKAT L-band surveys \citep{Ridolfi+2021}.

The use of the uGMRT real-time coherent de-dispersion pipeline coherently removes most of this intra-channel dispersive smearing for clusters with a known DM. This is done by de-convolving the voltages detected by the receiver with the inverse of the transfer function of the interstellar medium. As shown by \cite{Ridolfi+2019}, this greatly increases the effective time resolution of the data and the signal-to-noise ratio (S/N) of the folded pulse profiles. This means that we can substantially increase the DMs of the clusters we survey at lower frequencies, despite the much larger bandwidths of the uGMRT receivers. Five of the eight GC targets mentioned above have high DMs (>100 $\rm pc \, cm^{-3}$); the three high-DM clusters we mentioned above have a DM $> \, 200 \, \rm pc \, cm^{-3}$.

For observations at low frequencies, the pulse profile is also smeared by interstellar scattering, 
which we discuss in Section \ref{sec:survey_sensitivity}. The coherent de-dispersion technique does not allow the removal of this effect; the pulsars will only remain detectable if scattering time, $\tau_{\rm sc,}$ is generally shorter (preferably much shorter) than the pulse period. In order to determine the effect of scattering on the pulse profiles, we observed four of these clusters at 650 MHz. Pulsars whose profiles are negligibly scattered at this frequency will be observed at even lower frequencies in future observations. This process is also important in determining the best timing band for individual pulsars.

\begin{table*}
\caption[]{Properties of clusters observed, from \citealt{Harris_1996} (2010 revision).}
\label{tab:cluster_details}
\footnotesize
\centering
\renewcommand{\arraystretch}{1.0}
\vskip 0.1cm
\begin{tabular}{lllcccccc}
\hline
\hline
{Cluster} & {Coordinates of cluster} &{Core radius ($\rm R_c$)}& {Tidal radius ($\rm R_{\rm t}$)}&{Distance (kpc)} &{Previously Known} &{Binaries} &{Core-} \\ & {center (RA, Dec)} & {(arcmin)}& {(arcmin)} & & {Pulsars} & & {collapsed}\\\midrule
{NGC~1851} & 05:14:06.69, $-$40:02:48.89 & 0.06 & 6.52 & 12.1 & 1 & 1 & no\\
{Terzan~5} & 17:48:04.80 , $-$24:46:45.00 & 0.18 & 6.66&10.3 & 38 & 19 & no \\
{NGC~6440} & 17:48:52.68, $-$20:21:39.70 & 0.13 & 5.83& 8.4 & 6 & 3 & no\\
{NGC~6441} & 17:50:13.06 , $-$37:03:05.20 & 0.11 & 7.14& 11.7 & 4 & 2 & no\\
{NGC~6539} & 18:04:49.89, $-$07:35:24.69 & 0.54 & 20.88 & 8.4 & 1 & 1 & no\\
{NGC~6544} & 18:07:20.58 , $-$24:59:50.40 & 0.05 & 2.13 & 2.7 & 2 & 2 & yes\\
{NGC~6652} & 18:35:44.86, $-$32:59:25.10 & 0.10 & 6.31 & 10.0 & 1 & 1 & yes\\
{M30} & 21:40:22.40, $-$23:10:48.79 & 0.06 & 18.97 & 8.0 & 2 & 2 & yes \\

\hline
\hline
\end{tabular}
\end{table*}

\subsection{Observations}
We observed the eight target GCs with the GMRT from April 2017 to September 2018.
The low-DM clusters NGC~6652 and M30 were observed at 400 MHz, and the others at 650 MHz (see Table~\ref{tab:observations}).
The observations were performed with a bandwidth of 200 MHz and utilised the digital GWB system. We recorded a `phased-array' (PA) voltage stream in which the voltages from the antennae used in each observation are added coherently with time delays appropriate for the region of the sky targeted. From these voltages, a first processing pipeline at GMRT calculates the total intensities for each 0.097 MHz channel and integrates them for 81.92 $\mu$s; this is known as the PA data. 
 To maintain the correct phasing of the array, we observed a phase calibrator (chosen closest to the target) for three mins after every 1-hr in Band-4 (650 MHz) observations and after every 30 mins in Band-3 (400 MHz) observations. This is shorter than our intended integration times (see Table~\ref{tab:observations}); for this reason, when we do the calibration we keep taking data continuously, in order to maintain time coherence for the whole observation. The samples taken during the time the telescope is not pointed at the target were later replaced with the median bandpass values using \texttt{PRESTO}'s \texttt{rfifind} routine.
Since all target GCs have known associated DMs, we could also use a second processing pipeline (available at GMRT since September 2017) that divided the band into 0.195 MHz channels and coherently de-dispersed each channel's voltage stream in real-time at the DM of the cluster \citep{2016ExA....41...67D}. After de-dispersion, the total power for each channel was integrated for 20.48 $\mu$s, resulting in the `coherently de-dispersed' (CD) data.

We used 12 of the central square antennas (closely distributed set of antennas at the central region of the GMRT configuration) to create the PA beam, which gave us a maximum baseline of 1.1 km. This resulted in PA and CD beams of sizes $\ang{;2.3;}$ and $\ang{;1.5;}$ at 400 MHz and 650 MHz, respectively, and allowed us to cover the cores (where most of the compact objects are expected to be found) of all GCs (the core radius of the largest GC, NGC~6539, is $\sim$ $\ang{;0.5;}$).

In parallel, u-v visibilities are also calculated from the voltages of all functioning antennae in the array (up to 30 of them), this gave us baselines with lengths up to $\sim$25 km. These were recorded every 16-s and allowed the imaging of a good fraction of the primary beam of the antennae, with a spatial resolution that is significantly better than that of the central square.
To enable high sensitivity for faint pulsars, each cluster was observed for more than 30 mins.

\begin{table*}
\caption[]{Observation details. $^{\ast}$Only the longest epoch was searched in this survey; we plan to search the whole dataset thoroughly in the future.}
\label{tab:observations}
\footnotesize
\centering
\renewcommand{\arraystretch}{1.0}
\begin{tabular}{lcccccc}

\hline
\hline
{Cluster} &{Observation} &{Frequency} &{Integration} &{Target position} &{Coherent DM} \\
          &{(Epoch)}     &{(MHz)}     &{Time (s)}    &                & {(\dmunit)} \\
\midrule
{NGC~1851}$^{\ast}$ & 58051 &400 &2400 &J0514$-$4002A &52.15  \\
{Terzan~5}          & 58332 &650 &6660 & Cluster center &238.73 \\
{NGC~6440}          & 58363 &650 &7498 &J1748$-$2021A &223.00  \\
{NGC~6441}          & 58332 &650 &6660 &J1750$-$3703C &232.00  \\
{NGC~6539}$^{\ast}$ & 57829 &400 &9744 &B1802$-$07 &186.32  \\
{NGC~6544}          & 58363 &650 &7500 &J1807$-$2459A &135.50  \\
{NGC~6652}          & 58084, 58165 &400 &2400 &J1835$-$3259A &63.35  \\
{M30}               & 58102 &400 &1938 &J2140$-$2310B &25.09 \\

\hline
\hline
\end{tabular}
\end{table*}

\subsection{Survey sensitivity}
\label{sec:survey_sensitivity}
The minimum detectable flux density in this survey for a pulsar can be calculated with the radiometer equation \citep{Dewey+1985}:

\begin{equation}
\label{eq:radiometer_equation}
S = \frac{ {\rm S/N} \,\,  \beta  \,\, T_{\rm sys}} { G \sqrt{n_{\rm pol}  \  \BWeff ~ \Delta t_{\rm obs}}}  \,\, \sqrt{\frac{\zeta}{1 - \zeta}},
\end{equation}
where the total system temperature, $ T_{\rm sys}$ is 130 K for 400 MHz and 102.5 K for 650 MHz.
The total gain of the 12 antennae of the central array, $G$,  is 4.2 K $\rm Jy^{-1}$;
the number of polarisations summed $n_{\rm pol}$ is 2; the minimum S/N is chosen to be 10; the length of a typical observation, $\Delta \rm t_{\rm obs}$, is 2 hrs; the usable bandwidth $\rm BW_{\rm eff}$ is taken as 180 MHz (considering $\sim$ 20 MHz loss due to MUOS  satellites); and the signal loss due to digitisation, $\beta$, is close to (and assumed to be) 1, because our samples had 16 bits. $\zeta$ is the observed duty cycle of the pulse defined as the ratio of observed width and pulse period.

The observed pulse width can be affected by the interstellar medium in the form of scattering, dispersion smearing across individual frequency channels, dispersion smearing due to the finite DM step size of the time series, and finite time sampling. Since the CD data is coherently de-dispersed at the nominal DM of the cluster, there was very little contribution to pulse broadening from dispersive smearing across each channel. Assuming a negligible contribution from scattering, only contributions from finite sampling time and DM step size are added in the estimation of $\zeta$; this means that these sensitivity curves represent a best-case scenario.
The intrinsic pulse width is assumed to be 8$\%$ of the pulse period, which is a sensible number between the typical values of MSPs and slow pulsars (a similar value is also used for the sensitivity estimate of a GC survey with MeerKAT, \citealt{Ridolfi+2021}).

\begin{figure}
\centering
        \includegraphics[width=\columnwidth]{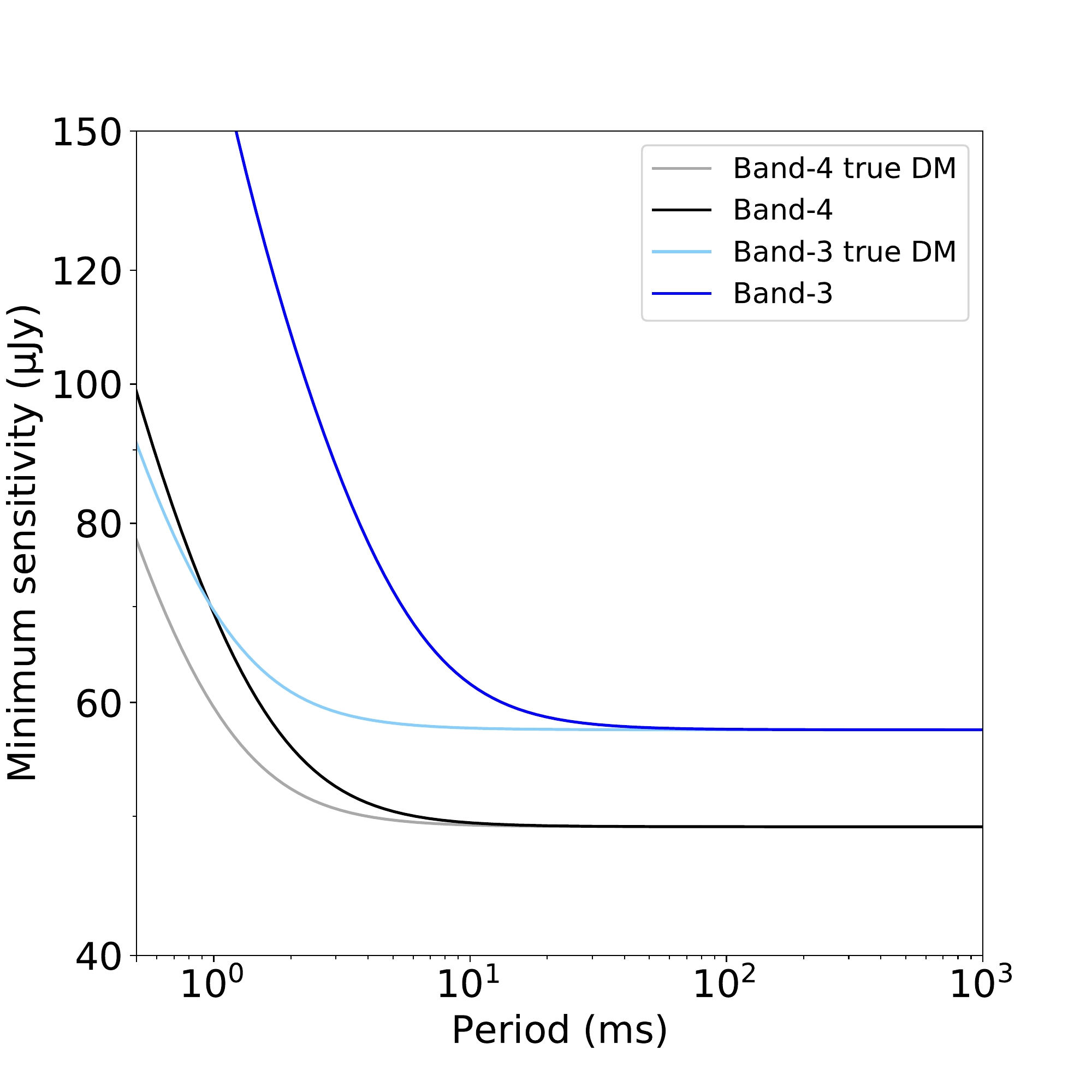}
        \caption{Minimum detectable flux density as a function of spin periods. Black and blue lines indicate sensitivity curves for 650 MHz and 400 MHz receivers. Lighter shades represent respective sensitivity curves if the DM trial value is the same as the true DM of a pulsar candidate.}
        \squeezeup
        \label{fig:sensitivity_curve}
\end{figure}

Figure \ref{fig:sensitivity_curve} shows the minimum sensitivity of 400 and 650 MHz surveys. For slow pulsars ($P$ > 50 ms), sensitivity is up to 50 $\mu$Jy for Band-4 and 60 $\mu$Jy for Band-3 observations, while for fast pulsars ($P$ $\sim$ 5 ms) it is up to a minimum flux density of $\sim$ 100 $\mu$Jy. 
Bold sensitivity curves represent the case where the pulsar candidate’s actual DM is considered the farthest from the trial DM value at which the time series are created, that is, the true DM value is assumed at the edge of the DM step size. Lighter shaded curves represent the sensitivity if a pulsar candidate’s actual DM happens to be the same as the trial DM value; in which case, there is a lower sensitivity threshold.

\subsection{Data Reduction}
\label{sec:data_red}
For the 650 MHz observations, the receivers' response had large variations, leading to significant power changes within the bandpass. Therefore, the initial data reduction involved flattening the band shape. This was done using the data reduction software \texttt{GPTool}\footnote{\url{https://github.com/chowdhuryaditya/gptool}} by re-normalising the data of each frequency channel to the bandpass average of that channel.

The next step involved down-sampling the dataset from its native digitisation of 16-bit to the more commonly used 8-bit format to make it compatible with the commonly used pulsar software. The bitshift for this conversion was chosen such that the test pulsar in each epoch had the highest S/N. We used \texttt{ugmrt2fil}\footnote{\url{https://github.com/alex88ridolfi/ugmrt2fil.git}} to down-sample and then convert the data into \texttt{SIGPROC}'s \citep{Lorimer+2011} filterbank format, this allowed us to use the latest version of PRESTO and perform more advanced acceleration and jerk searches.

\subsection{Search technique}
\label{sec:search_tech}
The search procedure used in this survey is based upon the pulsar searching package \PRESTO\footnote{\url{https://www.cv.nrao.edu/~sransom/presto/}} \citep{Ransom+2002}; the routines mentioned below are from this package. We used a modified version of an automated pipeline \PULSARMINER\footnote{\url{https://github.com/alex88ridolfi/PULSAR\_MINER}}~v1.1. This pipeline consists of the following steps.

The first step consist of removing the effect of radio interference due to terrestrial signals. This was done by using the routine \texttt{rfifind}\footnote{\url{https://www.cv.nrao.edu/~sransom/PRESTO_search_tutorial.pdf}}, which creates a mask by identifying abnormalities of mean and variances in user-defined, averaged, time, sub-integration, and frequency channel ranges in the data. This step is crucial to identifying narrow-band RFI signals. The masks generated from this routine were used throughout the search analysis to zap bad channels and time integrations from the de-dispersed time series and the original data. In our case, typically around 10-15 \% of the data were removed at this step.
In order to zap the broadband RFI signals, which may be periodic but do not have a dispersion peak at a finite DM due to their terrestrial nature, we created a 0-DM time series using the routine \texttt{prepdata} and searched for strong periodic signals (and their harmonics) in the data. The frequencies were then noted in a `.zaplist' file, which was used later in the analysis to remove these signals from the finite DM time series. In addition, the frequencies and harmonics of already known pulsars in these clusters were also added to this list to reduce the resulting number of candidates from these searches.

The second step involves creating a de-dispersed time series of the masked data for a range of DM values. Since the DMs of all the clusters searched were known beforehand, we searched these clusters with a DM range from $\rm DM_{\rm min}-5 \, \rm pc \rm \, cm^{-3}$ to $\rm DM_{\rm max}+5 \, \rm pc \rm \, cm^{-3}$ , where $\rm DM_{\rm min}$ and $\rm DM_{\rm max}$ are the minimum and maximum DMs of known pulsars in each cluster. We used  \texttt{PRESTO}'s \texttt{DDplan.py} routine to calculate the optimum DM step size and down-sampling factor (to reduce the effective time resolution). This step accounted for the dispersion smearing across each channel and helped reduce the total computational time. This scheme was then used to create the time series with the \texttt{prepsubband} routine.

The third step performs a Fourier transform on each time series to extract the periodic signals. This was done with the \texttt{realfft} routine. The pipeline then removed the red noise in the Fourier power from each of these power spectra to flatten fluctuations due to receiver or acquisition systems.

In the next step, these power spectra were searched for periodic components using \texttt{PRESTO}'s \texttt{accelsearch} routine. It search for frequencies and their harmonics (with power summed up to the 8th harmonic in this search) and shortlist strong periodic signals.
In binary pulsars, the orbital acceleration along the line of sight ($a$) cause a change in the Doppler shift that smears the pulsed signal in the Fourier domain by a number of Fourier bins: $z\, = \, \frac{T^2a}{Pc}$ , where $P$ and $T$ are the observed spin period and the observation time. In our survey, the maximum values of $z$, \texttt{zmax} used are $\pm \,$  30, 300, and 600. To efficiently use the computing power, we first prioritised the searches with lower values of \texttt{zmax}, ($\pm \,$30, 300), and then as computing power became available with higher values ($\pm \,$ 600). This resulted in sets of searches with maximum accelerations of $\sim$ 3.5, 34.7, and 70 $\rm m\,s^{-2}$ for a 5-ms pulsar in a 1-hr-long observation.
To be sensitive to highly relativistic binaries where the effect of orbital variations can smear the signal in an epoch, each observation was segmented into three to four chunks of $\sim$ 15-20 mins; this extended the acceleration range up to 560 $\rm m\,s^{-2}$ and 1100 $\rm m\,s^{-2}$, respectively. The routine \texttt{accelsearch} can only search for constant acceleration across the integration length, which is an assumption valid only for about one-tenth of the orbit \citep{1991ApJ...368..504J}. Thus, to search for compact binaries with improved detection sensitivity in longer time segments, we used the `Jerk search' method \citep{Andersen&Ransom2018}. This algorithm search for a linear change of the acceleration with time ($\dot{a}$) within the observation span and thus help regain the S/N of the folded profile. For this search, we used the parameter \texttt{wmax} ($\frac{\dot{a}T^3}{Pc}$) of 600, this made our search sensitive for a maximum acceleration change of 1.23 $\rm m\,s^{-2}$ across a 15-min integration time. Jerk searches were also performed on both 15-20 min segments and the full data span. The candidates from these searches were limited to a period range of 1-ms to 15-s.

The resulting candidates were then shortlisted using \texttt{ACCEL\_sift.py}. It removes duplicates and harmonics from the candidate list and sorts the candidates based on their Fourier significance. We used the default 3$\sigma$ threshold that is sufficient to find real pulsar candidates given the large number of trials) and DM peak.

Finally, these candidates were folded both in their corresponding time series and the original filterbank files using the \texttt{prepfold} routine. The respective detection parameters for each of the candidates were taken from the files produced by \texttt{accelsearch}. Nearly 500-1000 candidates were shortlisted, depending upon the observation length and RFI severity in each observation. Based on the typical total number of trials in our Fourier domain searches (following Equation 6 of \citealt{Andersen&Ransom2018}), we obtain a false-alarm threshold S/N \citep{2012hpa..book.....L} of 8.5. The candidate viewing was then performed by visually inspecting folded pulse profiles of each pulsar candidate. This was followed by several checks for harmonics to confirm their nature. 

Since the data were coherently de-dispersed, there was negligible intra-channel dispersion smearing; thus, the time resolution was not limited by these smearing timescales.
Because of the high time resolution, we needed a smaller DM step size in order not to degrade that resolution via DM smearing when adding the channels. This step is especially crucial at low frequencies as the dispersion smearing become more pronounced. Hence, a large number of time series had to be searched. Thus, to efficiently search the clusters, we used the \texttt{Hercules}\footnote {\url{https://docs.mpcdf.mpg.de/doc/computing/clusters/systems/Radioastronomy.html}} computing cluster located at the MPG's computer centre in Garching. The modified version of \texttt{PULSAR\_MINER}\footnote{\url{https://github.com/tgautam16/jerksearch.git}} was able to perform parallel processing with 184 available nodes, each with 24 CPU cores, where each time series was allotted an individual node (using \texttt{SLURM}\footnote{\url{https://slurm.schedmd.com}}) to perform acceleration and jerk searches. Likewise, the folding process took place on a single node, making use of 48 threads that folded 24 candidates concurrently.

\section{A new pulsar: PSR J1835-3259B in NGC~6652}
\label{sec:results}
\begin{figure*}
\centering
    \includegraphics[angle=270,origin=c,width=1.8\columnwidth,trim={0cm 0cm 0.5cm 0cm},clip]{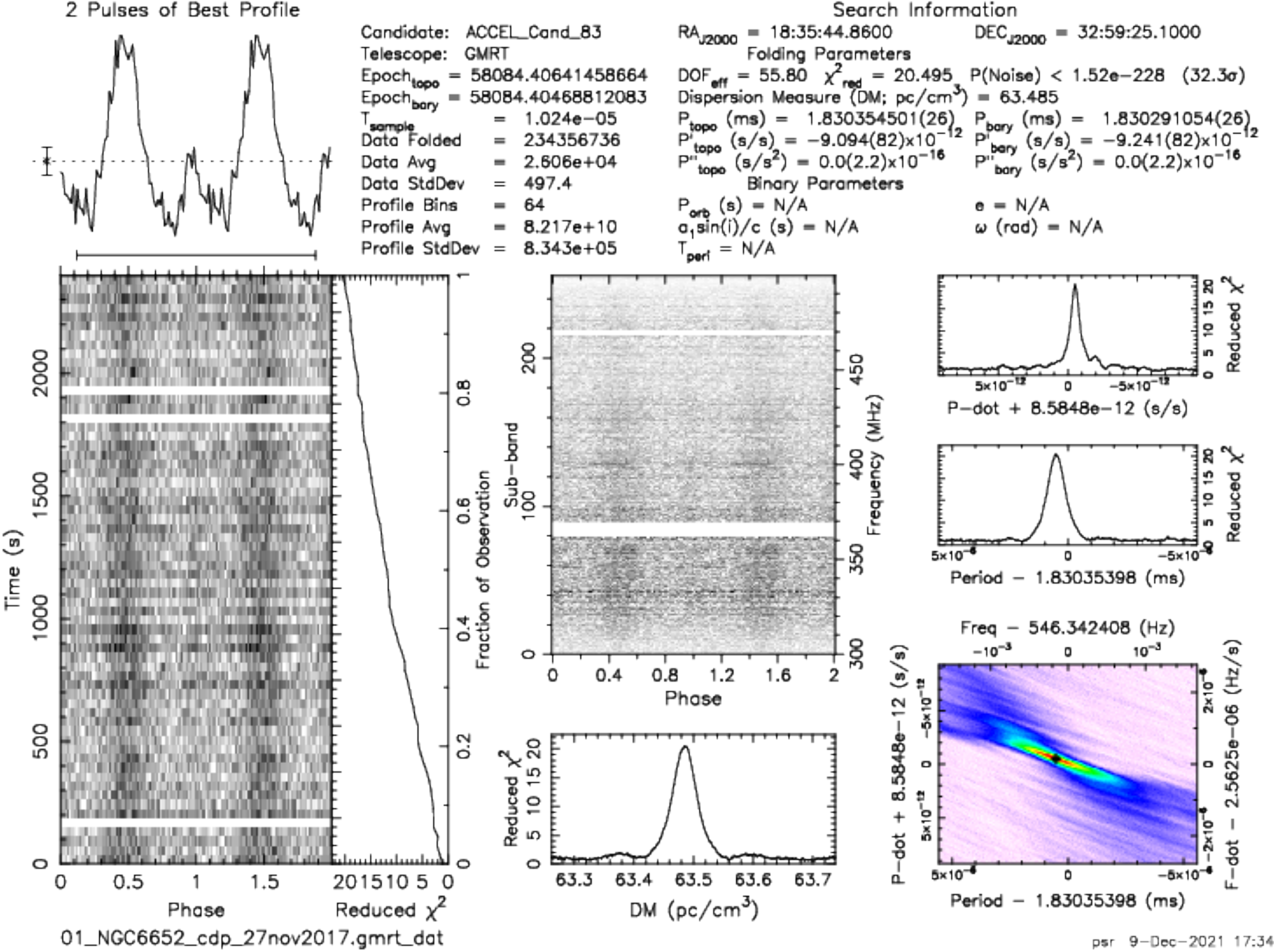}
    \vspace{-2cm}
        \caption{Discovery plot of NGC~6652B folded with \texttt{prepfold} showing time vs. phase and frequency vs. phase intensity in greyscale.}
        \label{fig:discovery-plot}
\end{figure*}

\subsection{Discovery, follow-up, and timing}
\label{subsec:discovery}

The search analysis of these clusters resulted in one confirmed discovery, PSR J1835$-$3259B in NGC~6652. The letter `B' indicates that this is the second pulsar discovered in this cluster, the first being PSR J1835$-$3259A \citep{DeCesar+2015}; we henceforth refer to both as NGC~6652A and NGC~6652B. Interestingly, did not detect NGC~6652A, neither in imaging (see Section~\ref{sec:imaging_gc}) nor in our PA and CD datasets, even though its position is well within the positions of those beams.
This means that it likely has a flat radio spectrum.

\begin{figure}
\centering
\begin{subfigure}[b]{0.237\textwidth}
   \includegraphics[width=\columnwidth]{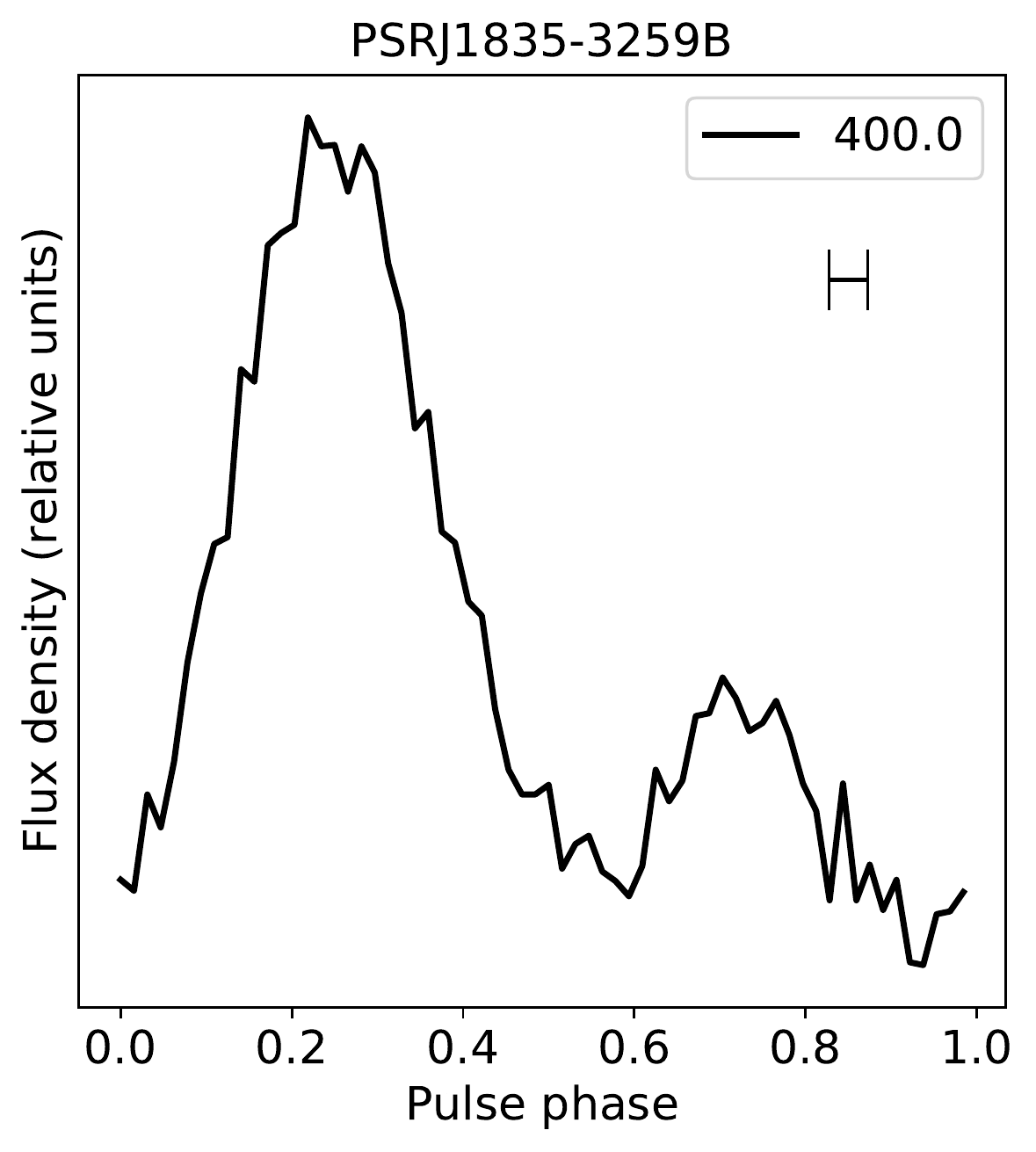}
\end{subfigure}
\begin{subfigure}[b]{0.237\textwidth}
    \includegraphics[width=\columnwidth]{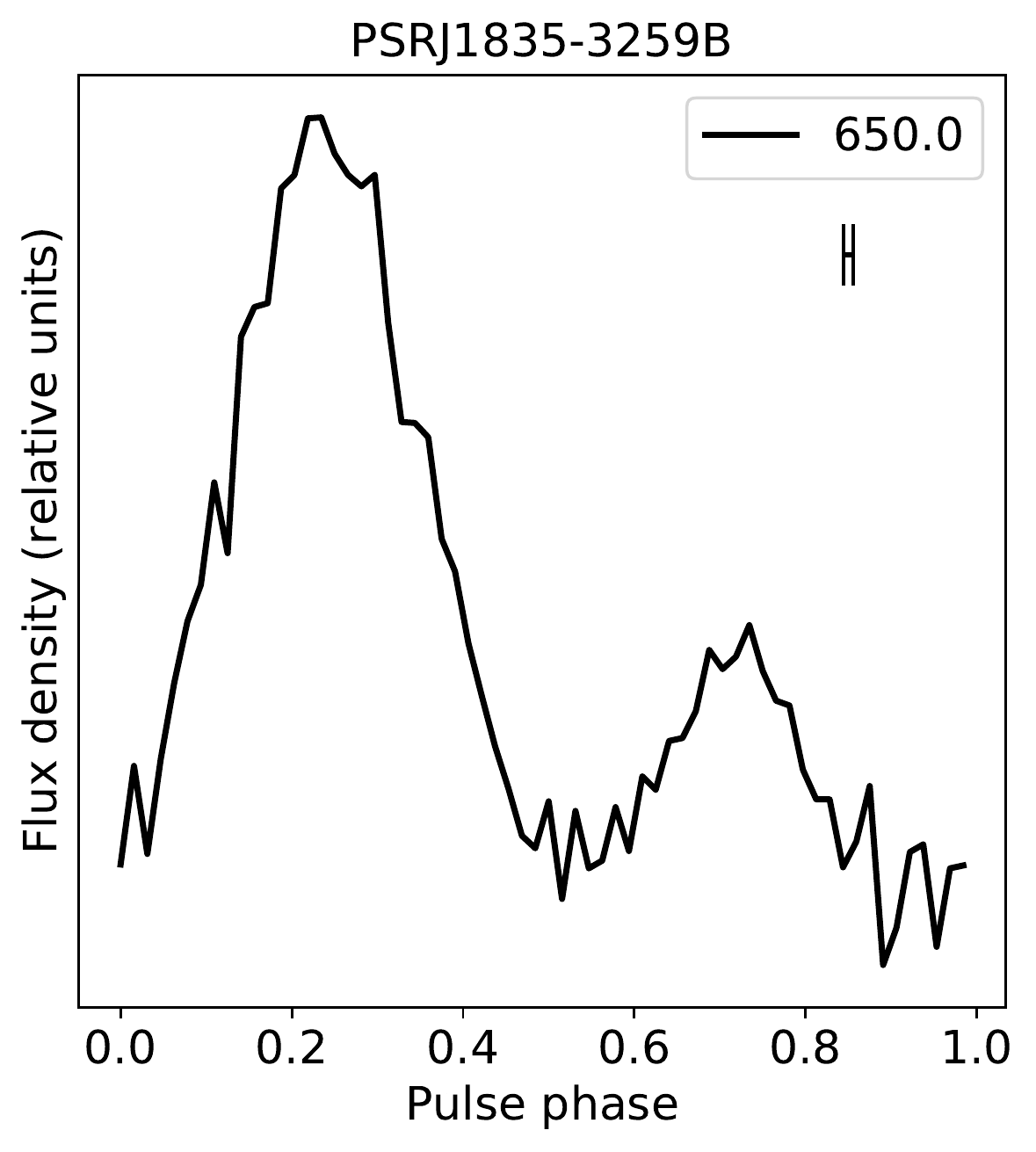}
\end{subfigure}
\begin{subfigure}[b]{0.237\textwidth}
   \includegraphics[width=\columnwidth]{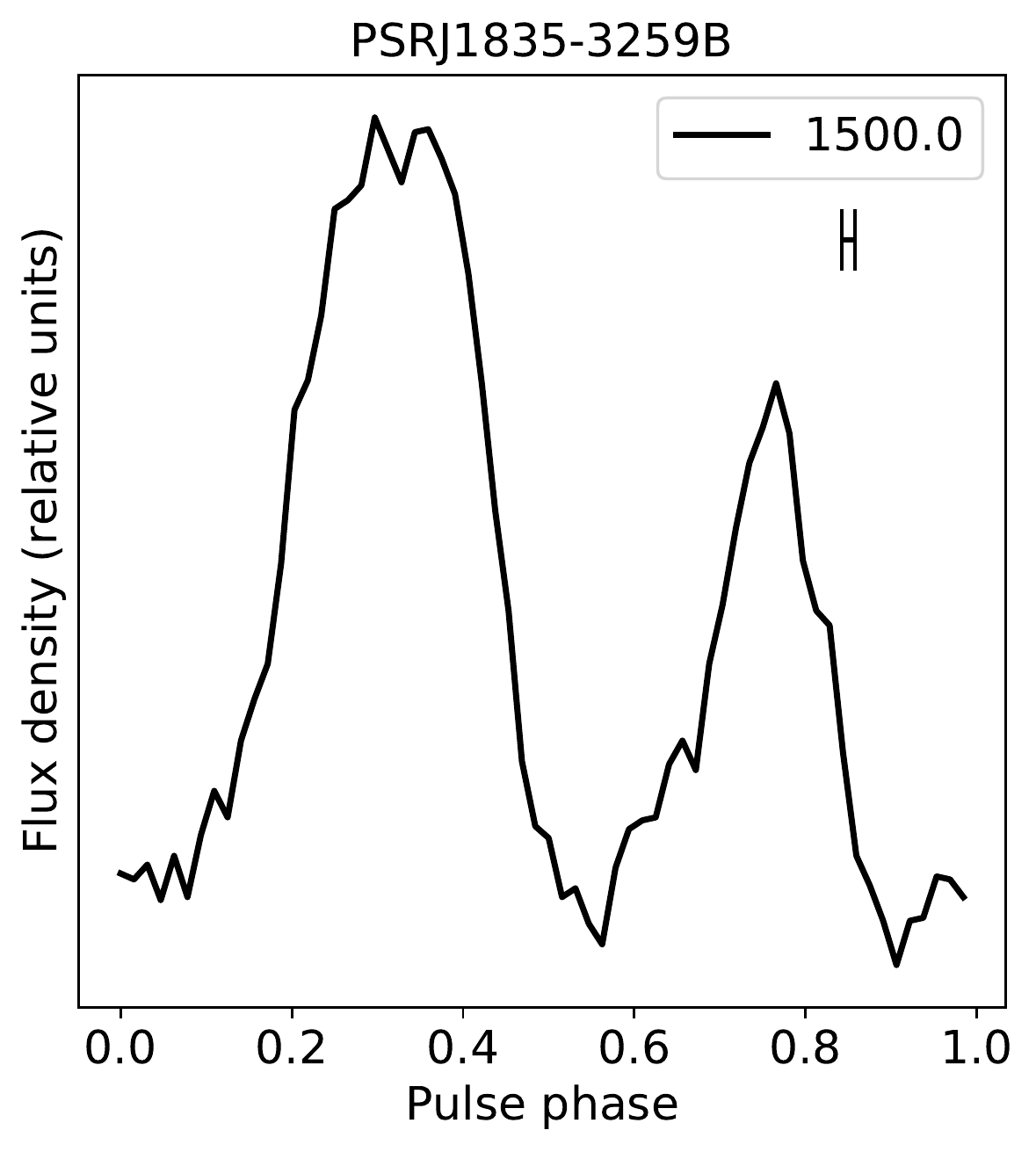}
\end{subfigure}
\caption{Intensity profiles as function of rotational phase of NGC~6652B obtained with Band-3 receiver of GMRT at 400 MHz, Band-4 receiver of GMRT at 650 MHz, and L-band receiver of GBT at 1500 MHz. Flux number on y-axis is relative. The horizontal bars show the time resolution of the system for the DM of the pulsar. For the GBT profile, this is merely the adopted bin size of the profile. The profiles were visually aligned.}
\label{fig:timing_profiles}
\end{figure}

NGC~6652B is a 1.83-ms pulsar found in a 40-min GMRT observation of its parent cluster taken on 27 November, 2017 with the 400 MHz (Band 3) receivers, at a DM of 63.48 pc cm$^{-3}$. It was found by the acceleration search with a \texttt{zmax} of 30; the initial detection (see \texttt{prepfold} plot in Figure \ref{fig:discovery-plot}) had an acceleration of 1.5 $\rm m\,s^{-2}$, clearly indicating that it is a member of a binary system. The pulsar was visible in another 400 MHz observation taken in February 2018, which confirmed its discovery and was also subsequently detected in GMRT 650 MHz observations. Later, it was detected in archival GBT L- and S-band observations taken from 2011 to 2020 with the Green Bank Ultimate Pulsar Processing Instrument (GUPPI) \citep{Ransom+2009} back-end that were originally made to time NGC~6652A.
Figure \ref{fig:timing_profiles} shows the variation of radio intensity with spin phase of the pulsar (its integrated `pulse profile') at three different frequencies. In all of them, it shows a clear interpulse, which becomes more pronounced at higher radio frequencies.

\begin{figure}
\centering
    \includegraphics[width=\columnwidth]{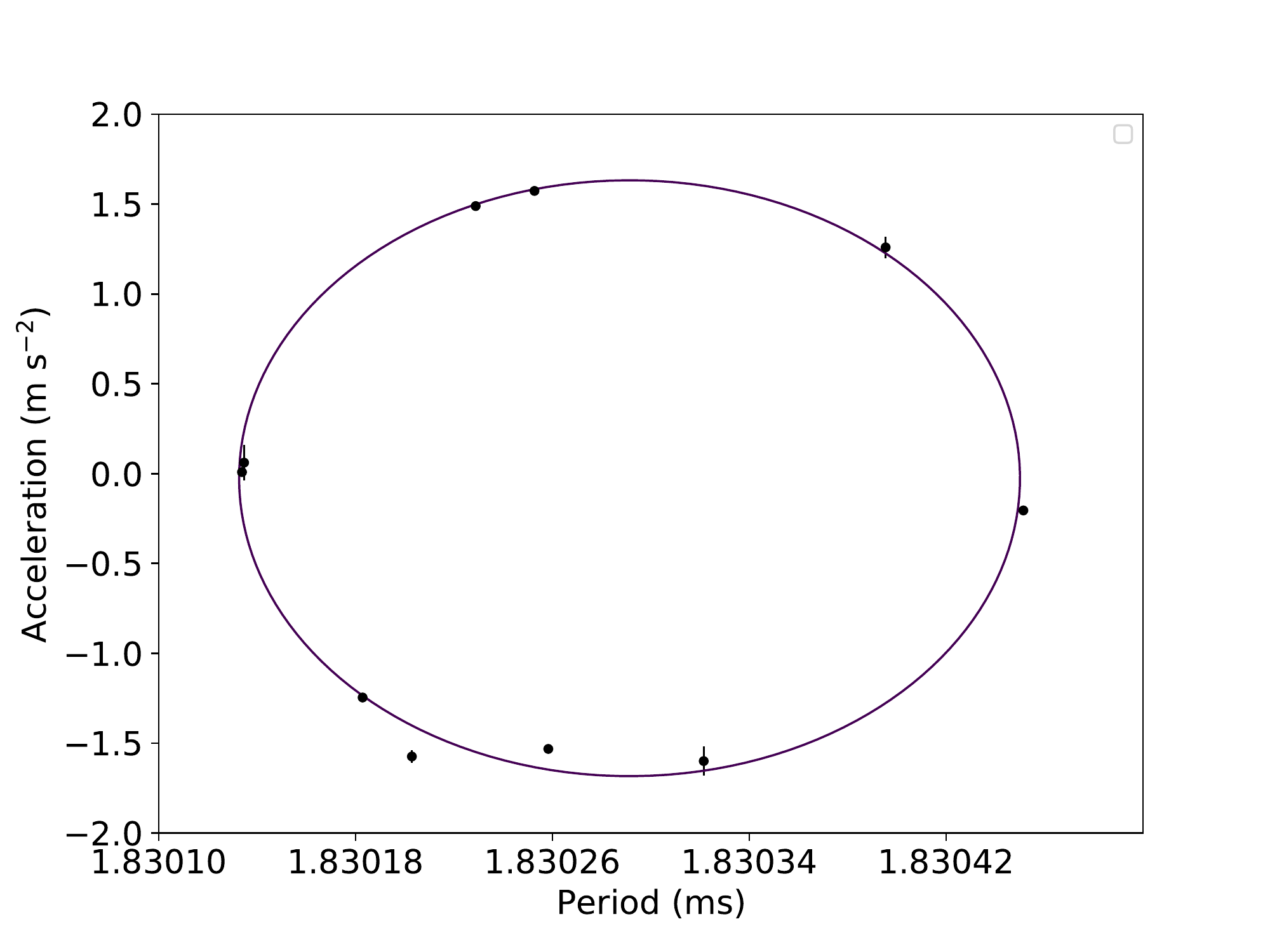}
        \caption{Plot of period vs. acceleration for timing observations taken on 27 November 2017; 18, 21, and 28 January, 2020; and 5, 14, and 15 February 2020. Line-of-sight acceleration is calculated using observed period derivatives, $a_l = \frac{\dot{P} c}{P}$ where $P$ and $\dot{P}$ are the observed spin period and spin period derivative in each detection. The pulsar turns out to be in an orbit with $P_{\rm b}  \sim \rm \, 1.197 \rm \, d$ and $x\, \sim$ 1.43 lt-s.}
        \squeezeup
        \label{fig:p-a-plot}
\end{figure}

For new binary pulsars, it is important to determine the nature of the binary companion, along with the orbital, spin, and kinematic properties of the system to find the potential of the pulsar for future scientific studies. We followed up this system with regular uGMRT observations using Band 3 (400 MHz) from January-March 2020.

Each 1-hr-long observation was made with exactly the same settings as the discovery observation, also recording CD data from the central array. This allowed us to maintain consistency among these and discovery observations. The pulsar was detected in 12 out of 13 observations. The flux of this pulsar in PA data was nearly 3-5 times smaller than the discovery observation, and hence we had to use the jerk search method to re-detect it. To identify the cause of these variations, we compared the flux density of this source in the radio images to that in the timing observations. Within measurement uncertainties, the imaging flux densities were consistent with a constant value, confirming that the flux density of this pulsar did not vary between these observations. We have not been able to determine the origin of this discrepancy, and, in particular, whether it was a problem in the recording of the PA data.

For initial estimates of the orbital parameters, we used the velocity-acceleration method discussed by \cite{Freire+2001}: we plotted the barycentric period and acceleration values from each epoch where the pulsar was visibly bright (Figure~\ref{fig:p-a-plot}). We fit an ellipse to these detections and find the best fit that is consistent with a nearly circular orbit.
This fit yields approximate estimates of the intrinsic spin period of the pulsar, $P_0$ (1.83029 ms), orbital period $P_{\rm b}$ (1.1605 days), projected semi-major axis, $x$ (1.3984 lt-s), and the time of passage at the ascending node, $T_{\rm asc}$ (58084.5910).
We then refined the estimates of $T_{\rm asc}$ for all detections using \SPIDERTWISTER\footnote{\url{https://github.com/alex88ridolfi/SPIDER_TWISTER}} (see Section 2 of \citealt{Ridolfi+2016}).
With these more precise values of $T_{\rm asc}$ for each epoch, we refined the orbital period $P_{\rm b}$ based on the periodogram method discussed in \cite{Freire+2001} and also used by \cite{Ridolfi+2016}, which looks for common sub-multiples of the differences of all $T_{\rm asc}$ that are consistent with our initial estimate of $P_{\rm b}$.
With these first-order parameters in place, we were then able to look at the precise variation of spin period as a function of time; this was done with \texttt{PRESTO}'s routine \texttt{fit\_circular\_orbit.py}. This resulted in much improved estimates for $P_{\rm b}$, $x$, and $T_{\rm asc}$, with which we created a starting ephemeris to fold all pulsar observations.  

We used the \texttt{PSRCHIVE}\footnote{\url{http://psrchive.sourceforge.net}} package \citep{Hotan+2004,vanStraten+2012} tools for the timing analysis after this point. The folded profiles were converted to `archive' files using \texttt{psrconv}, and these archives were then cleaned of narrow-band RFI using \texttt{pazi}. 
Standard templates of the pulse profile were then made separately for the GMRT 400 and 650 MHz observations, and together for the GBT L-band and S-band observations. This was done by adding the folded archive files from bright observations together using \texttt{psradd} to develop integrated profiles; these are shown in Figure \ref{fig:timing_profiles}. These profiles were then interactively fitted with multiple von Mises functions to create analytic noiseless profiles using \texttt{paas}.
These standard templates hence created were used to calculate the topocentric times of arrival (ToAs) using \texttt{pat}.
We made ToAs for four different sub-bands at Band 3 of the GMRT, two sub-bands for Band 4 of the GMRT, and two sub-bands for the L- and S-bands of the GBT; this allowed precise tracking of DM variations. We made ToAs every few mins; the time scale varied due to S/N considerations.

These ToAs were then analysed with the \TEMPO\footnote{\url{http://tempo.sourceforge.net}} software, where we first used the orbital and spin parameters determined above.
The program calculates the differences between the ToAs and the arrival time predicted by the timing model (the timing residuals) and then minimises the residuals by adjusting the parameters in the timing model. Binary pulsars with low eccentricities have high correlations between the longitude of periastron, $\omega$ and epoch of periastron $T_0$; for this reason, we used the ELL1 model \citep{Lange+2001} to describe the time delays of the pulses caused by the orbital motion as a function of time. In this model, the Keplerian parameters $\rm \omega$, $T_0$, and $e$ are replaced by the time of ascending node, $T_{\rm asc}$, which can be measured precisely even for circular orbits, and the two Laplace-Lagrange parameters, $\epsilon_1 \, = \, e  \sin{\omega}$ and
$\epsilon_2 \, =\, e \, \cos{\omega}$.  

\begin{figure}
\centering
        \includegraphics[width=1.05\columnwidth]{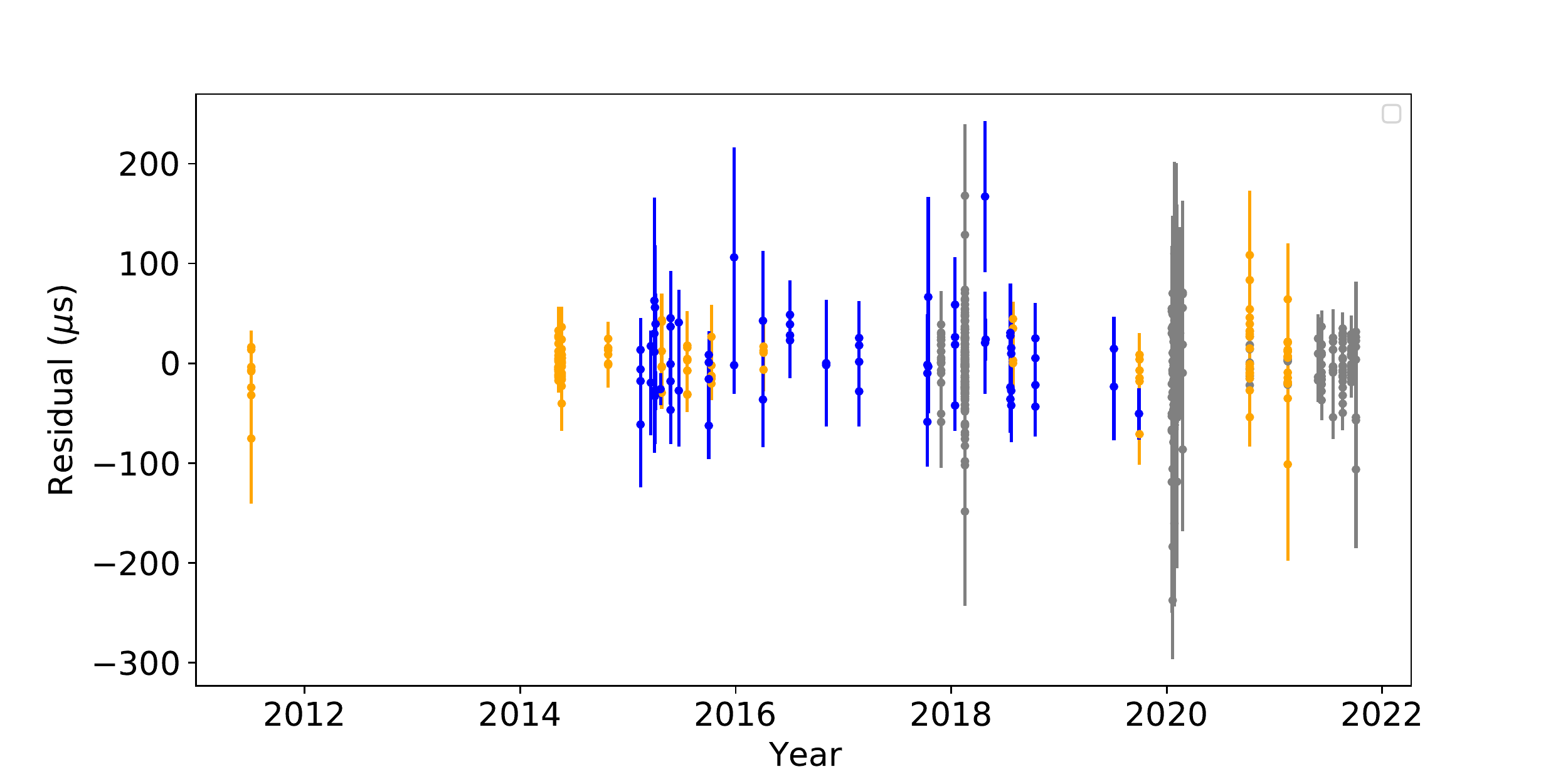}
        \caption{Residuals obtained with phase-connected solution of NGC~6652B in Table~\ref{tab:timing_solution_J1835-3259B}. Grey points are ToAs derived from data taken at frequencies below 1000 MHz, while orange points represent GBT ToAs derived from L-band data within 1000-1700 MHz, and blue points show S-band data above 1700 MHz.}
        \squeezeup
        \label{fig:residual-plot}
\end{figure}

In the first stage of timing a newly discovered pulsar, the estimated spin, orbital, and astrometric parameters are not precise enough to calculate the exact number of rotations between the ToAs of different observations, but they can still fit the orbital parameters to obtain residuals that are flat within each observation, that is, we can determine the number of rotations between ToAs within each observation. This was done using \texttt{TEMPO} by including time offsets between observations. With the resulting ephemeris, we could confidently re-fold all our observations and proceed to phase-connect all of them; that is, we could determine the rotation counts between observations (for a detailed description of the process, see \citealt{2018MNRAS.476.4794F}). We then move onto connecting timing observations spaced further and further apart until we were able to connect all the GMRT observations for this pulsar, plus all GBT observations. The resulting timing solution, which spans a total of ten years, is presented in Table~\ref{tab:timing_solution_J1835-3259B}. The residuals obtained with this solution are shown in Figure~\ref{fig:residual-plot}.

The timing solution includes a refined DM estimate of 63.5083 $\rm pc \rm \, cm^{-3}$ plus a variation modelled by five significant DM derivatives. The fit for proper motion of the system is consistent with the proper motion estimate of GC NGC~6652 mentioned in \citet{Vasiliev_Baumgardt+2021}.

\begin{table}
\caption{Timing solution of NGC~6652B.}
\label{tab:timing_solution_J1835-3259B}
\begin{center}{\footnotesize
\setlength{\tabcolsep}{-4.0pt}
\renewcommand{\arraystretch}{1.2}
\begin{tabular}{l c}
\hline
\hline
Right Ascension, $\alpha$ (J2000)                                     \dotfill &   18:35:45.4704(1)                                                       \\
Declination, $\delta$ (J2000)                                         \dotfill &   $-$32:59:25.48(1)                                                      \\
Proper Motion in $\alpha$, $\mu_\alpha$ (mas yr$^{-1}$)               \dotfill &   $-$7.7(5)                                                              \\
Proper Motion in $\delta$, $\mu_\delta$ (mas yr$^{-1}$)               \dotfill &   $-$4(4)                                                                \\
Spin Frequency, $f$ (s$^{-1}$)                                        \dotfill &   546.36058616286(10)                                                    \\
1st Spin Frequency derivative, $\dot{f}$ (Hz s$^{-2}$)                \dotfill &   $-$1.2962(1)$\times 10^{-14}$                                          \\
Reference Epoch (MJD)                                                 \dotfill &   58736.720                                                           \\
Start of Timing Data (MJD)                                            \dotfill &   55750                                                              \\
End of Timing Data (MJD)                                              \dotfill &   59490                                                              \\
Dispersion Measure, DM (pc cm$^{-3}$)                                 \dotfill &   63.5083(6)                                                             \\
1st derivative of DM, DM1 (pc cm$^{-3}$ yr $^{-1}$)
     \dotfill &   0.0237(6)                                                             \\
2nd derivative of DM, DM2 (pc cm$^{-3}$ yr $^{-2}$)
     \dotfill &   0.0055(9)                                                             \\
3rd derivative of DM, DM3 (pc cm$^{-3}$ yr $^{-3}$)
     \dotfill &   $-$0.0126(9)                                                             \\
4th derivative of DM, DM4 (pc cm$^{-3}$ yr $^{-4}$)
     \dotfill &   $-$0.015(2)                                                             \\
5th derivative of DM, DM5 (pc cm$^{-3}$ yr $^{-5}$)
     \dotfill &  $-$0.0050(6)                                                             \\
Solar System Ephemeris                                                \dotfill &   DE436                                                                  \\
Terrestrial Time Standard                                             \dotfill &   TT(BIPM)                                                               \\
Time Units                                                            \dotfill &   TDB                                                                    \\
Number of ToAs                                                        \dotfill &   526                                                            \\
Residuals RMS ($\mu$s)                                                \dotfill &   22.40                                                                  \\
\hline
\multicolumn{2}{c}{Binary Parameters}  \\
\hline
Binary Model                                                          \dotfill &   ELL1                                                                   \\
Projected Semi-major Axis, $x$ (lt-s)                         \dotfill &   1.430841(2)                                                            \\
First Laplace-Lagrange parameter, $\eta$                              \dotfill &   $-$2.7(4)$\times 10^{-5}$                                              \\
First Laplace-Lagrange parameter, $\kappa$                            \dotfill &   $-$2.2(3)$\times 10^{-5}$                                              \\
Epoch of passage at Ascending Node, $T_\textrm{asc}$ (MJD)            \dotfill &   58866.2655109(6)                                                       \\
Orbital Period, $P_{\rm b}$ (days)                                          \dotfill &   1.197863230(1)                                                         \\
Orbital Period derivative, $\dot{P}_{\rm b}$ (10$^{-12}$ s s$^{-1}$)  \dotfill &    0.86(1.08)                                                                  \\
\hline
\multicolumn{2}{c}{Derived Parameters}  \\
\hline

Spin Period, $P$ (s)                                                  \dotfill &   0.0018302930799293(3)                                                  \\
Maximum Spin Period derivative, $\dot{P}$ (s s$^{-1}$)                    \dotfill &   $ 6.65 \times 10^{-20} $                                           \\
Mass function, $f$(${M}_{\rm p}$) (${\rm M}_\odot$)                       \dotfill &    0.0021920077(95)                                                                \\
Minimum companion mass, $M_{\rm c, min}$ (${\rm M}_\odot$)            \dotfill &   0.18                                                                   \\
Median companion mass, $M_{\rm c, med}$ (${\rm M}_\odot$)             \dotfill &   0.21  \\
Eccentricity, $e$   
    \dotfill &   3.49(35) $\times 10^{-5}$ 
                    \\
Longitude of periastron, $\omega$ ($^{\circ}$)
    \dotfill & 230(5)  
                    \\
Time of periastron, $\rm T0$ (MJD)
    \dotfill & 58867.03(1)
                    \\
Offset from GC center in $\alpha$, $\theta_\alpha$                    \dotfill &   0.033                                                                  \\
Offset from GC center in $\delta$, $\theta_\delta$                    \dotfill &   0.019                                                                  \\
Total offset from GC center, $\theta_\perp$ (arcmin)                  \dotfill &   0.038                                                                  \\
Proj. distance from GC center, $r_\perp$ (pc)                         \dotfill &   0.111                                                                  \\
Proj. distance from GC center, $r_\perp$ (core radii)                 \dotfill &   0.383                                                                  \\

\hline
\hline
\end{tabular} }
\end{center} 
\end{table}

\subsection{Characteristics of the new binary pulsar}

As we can see in Figure~\ref{fig:timing_profiles}, the profile of NGC~6652B has two broad components approximately 180 degrees from each other, with emission occurring during most of the spin phase. The secondary component becomes more prominent at higher radio frequencies. A comparison with the time resolution of the system (horizontal bars in the plots in that figure) shows that these pulse components, being much wider than the time resolution, are intrinsically broad.
The role of scattering in this is limited since the profile has basically the same
characteristics at 1.5 GHz. 

The pulsar and its companion complete an orbit every 1.197 days.
The projected semi-major axis of its orbit is 1.43 lt-s. Assuming a pulsar mass of $1.4 \, M_{\odot}$, the companion has a minimum mass of $0.176\, M_{\odot}$, consistent with the 
$P_{\rm b}$--$M_{\rm WD}$ relation given by \cite{Tauris_Savonije+1999} for MSPs with helium WD companions.
The orbital eccentricity is small but significant, $(3.49 \, \pm \, 0.35 )\, \times \, 10^{-5}$, which is also
consistent with a helium WD companion. This indicates that the system has not interacted strongly with other stars
in the cluster since recycling.
We have not detected any radio eclipses for this system, even at 400 MHz in observations taken near superior conjunction ($\phi \sim 0.25 $), so nothing indicates that this is a `redback' binary.

\subsection{Position of the system}

\begin{figure}
\centering
        \includegraphics[width=\columnwidth]{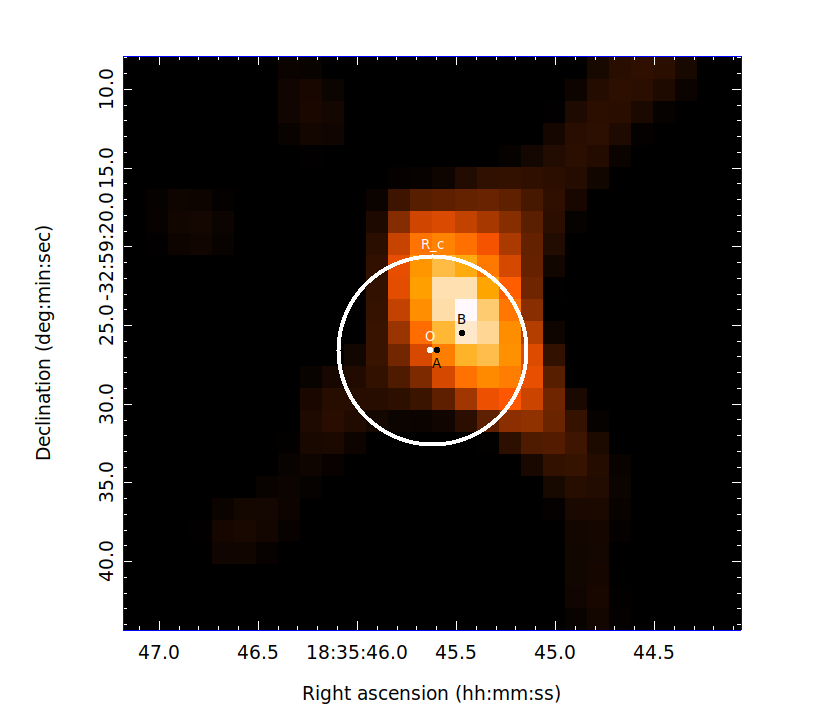}
        \caption{Pulsar positions in NGC~6652. Derived timing position of NGC~6652B is shown with black label B. The cluster centre is marked with the letter O; it is assumed to be at RA= 18h35m45.63s, and Dec=$-$32$\degr$59$\arcmin$26.6$\arcsec$. The pulsar is offset from centre of the cluster by 0.033 $\arcmin$ in RA and 0.018 $\arcmin$ in Dec. The position of NGC~6652A is also shown for comparison. The white circle represents the core radius ($0.1\arcmin$) of the cluster. The pixels show radio intensity in our 400 MHz (Band 3) interferometric image of the cluster, a radio source is clearly
          coincident with the position of NGC~6652B. NGC~6652A is not detected in the image, nor are its pulsations detectable in our PA and CD data.}
        \squeezeup
        \label{fig:cluster-position-6652}
\end{figure}

Its position with respect to NGC~6652A and the cluster centre is shown in Figure~\ref{fig:cluster-position-6652}; the total offset from the GC centre is 0.038$\arcmin$. The timing position coincides with that of a very bright radio source, which we identify preliminarily as the pulsar (see Section~\ref{sec:imaging_gc}).

NGC 6652 has a total of 12 known X-ray sources \citep{2012ApJ...751...62S}. The timing position of NGC~6652B coincides with one of them, a faint source named `H', which has the coordinates $\alpha \, = \, $18:35:45.473 and $\delta \, = $ $-$32:59:25.48; the positions are consistent within the astrometric precision of the X-ray source. In that study, the authors conclude that, although the faint sources are likely to be cataclysmic variables, some of them could be MSPs.

\subsection{Additional candidates}

In addition, we found a total of 24 candidates with a significance higher than 6$\sigma$ in clusters NGC~6440, NGC~6441, NGC~6544, and Terzan~5. These were followed up with additional observations with uGMRT from October to November 2019. However, none of these were re-detected in the follow-up observations. Possible reasons for the non-detection of these candidates could be unfavourable orbital phases (in case of eclipsing binaries) or the flux density subsiding below the detection threshold due to scintillation. Another cause could be the decrease in observed flux density due to varying beam shape when observed at a different elevation to that of the original observation. Nevertheless, a few candidates with high enough significance (> 7$\sigma$) and multiple re-detections in different segments of the observations are presented in Appendix \ref{appendix:a}. We encourage future GC pulsar search surveys in these clusters to investigate these candidates for re-detection.

\section{Characterising known pulsars}
\label{subsec:characterization}
From the available dataset, we re-detected 22 pulsars (out of 55 known) in the eight clusters we observed; their intensity profiles
are shown in Figure \ref{fig:known_profiles}. These detections include four pulsars that are in eclipsing binary systems, namely, Terzan~5A, Terzan~5O, NGC~6440D, and possibly the `black widow' system NGC~6544A, for which eclipses have not been observed at higher radio frequencies \citep{Lynch+2012}. We were only able to detect the ingress and egress of the eclipse in Terzan~5A in our data. No eclipses for the other pulsars occurred during our observations, because none of them coincided with a superior conjunction of those binaries.
Table \ref{tab:flux_estimates} presents the re-detections along with the minimum detection threshold for each of the epochs. These sensitivity thresholds yield an upper limit on the flux density of the pulsars that are not detectable in these observations. 

We estimated flux densities for all the re-detected pulsars (of which only Terzan~5A, NGC~6539A, and NGC~1851A had previously reported flux densities at such low frequencies; our estimates are comparable to those values), except for four of these (NGC~6441D, Terzan~5E, I, K) whose S/N were below 6$\sigma$; hence, the flux estimates are not reliable. For these sources as well, we placed the survey detection threshold of their host clusters (see Table~\ref{tab:flux_estimates}) as the upper limit on their flux densities. To estimate the flux density, each folded archive of these pulsars was first cleaned with \texttt{pazi} to remove any narrow-band RFI and maximise the S/N. Optimised estimates of the spin period and DM were then determined using \texttt{PSRCHIVE}'s \texttt{pdmp} for each pulse profile. The radiometer equation given by \cite{Dewey+1985} is used to determine the flux densities for all the pulsars at 400 MHz and 650 MHz correspondingly. The calculated flux densities were fit with a power law along with previously known flux values at higher frequencies (taken from the ATNF catalogue) to estimate the spectral indices for each of these pulsars. Expected flux densities at corresponding frequencies are determined (see Table \ref{tab:flux_estimates}) from the previously known spectral index estimate from ATNF and assuming $-$1.4 where the spectral index isn't known. 

Using the wide bandwidth of 200 MHz of our observations, we have also estimated the temporal broadening due to the scattering effect of ISM in the pulse profiles of bright pulsars. To estimate this, we modelled the pulse profiles of several sub-bands using the software \texttt{SCAMP-I}\footnote{\url{https://github.com/pulsarise/SCAMP-I}}. Full details of the method employed by this software can be found in \cite{Oswald+2021}. For each pulsar, we split the observing band into either two, four, or eight pulse profiles, depending on the signal-to-noise constraints, and modelled the profile in each sub-band independently. The modelling technique uses a Markov-chain Monte Carlo (MCMC) method to fit a single Gaussian, convolved with an isotropic scattering transfer function, to the pulse profile. Five parameters are constrained by the modelling process: the amplitude, mean, and standard deviation of the Gaussian representing the intrinsic pulse profile; any direct current (DC) offset of the profile baseline; and the scattering timescale, $\tau_{\rm scat}$. The MCMC method was run for 10,000 steps for each profile, and the chains were inspected by eye to determine whether they had converged. For the pulsars where the band could be split into a sufficiently large number of sub-bands to obtain multiple scattering timescales at different frequencies, we also obtained a scattering index. This was done by performing a least-squares power-law fit to the variation of scattering timescale with frequency. We report the estimates for scattering timescales, and, where obtainable, scattering indices, in Table~\ref{tab:flux_estimates}. We note that for the eclipsing binary pulsar Terzan~5A, we removed time integrations in the folded profile near the eclipsed portion to avoid any phase delays present due to wind emanating from the companion. Table~\ref{tab:flux_estimates} also presents the theoretical prediction of scattering timescales based on the empirical relation by \citealt{Bhat+2004}:
\begin{equation}
 \log \tau_{\rm sc} = -6.46 + 0.154 (\log {\rm DM}) + 1.07 (\log {\rm DM})^{2} -3.86 \log \left(\frac{\nu_c}{\rm GHz}\right)
,\end{equation}
where $\nu_c$ is the centre frequency of the observing band. We should note that in \cite{Bhat+2004} the scatter of fitted data has a variation of around an order of
magnitude with regard to this prediction, and hence our scattering estimates lie well within this variation. 
We note that more recent work, notably that of \cite{2015MNRAS.449.1570L}, has used an increased number of measurements of scattering timescales to undertake deeper investigations of the $\tau$-DM scaling relationship and update its characteristic shape. However, since all of the empirical scaling relationships in the literature are compatible within the spread of scattering timescale measurements collected thus far, the accuracy of the predictions given in Table~\ref{tab:flux_estimates} is sufficient for the purposes for which we are using them. Furthermore, the $\tau$-DM relationships given by \cite{2015MNRAS.449.1570L} are fixed at 1~GHz, whereas the form of the scaling relationship given in \cite{Bhat+2004} makes use of a global frequency-scaling index of $\alpha = 3.86$. This makes it convenient to use it to calculate estimates at our pulsar reference frequencies, even for cases where we do not have a measurement of $\alpha$.

\begin{figure*}
\begin{subfigure}[b]{0.24\textwidth}
    \includegraphics[width=\textwidth]{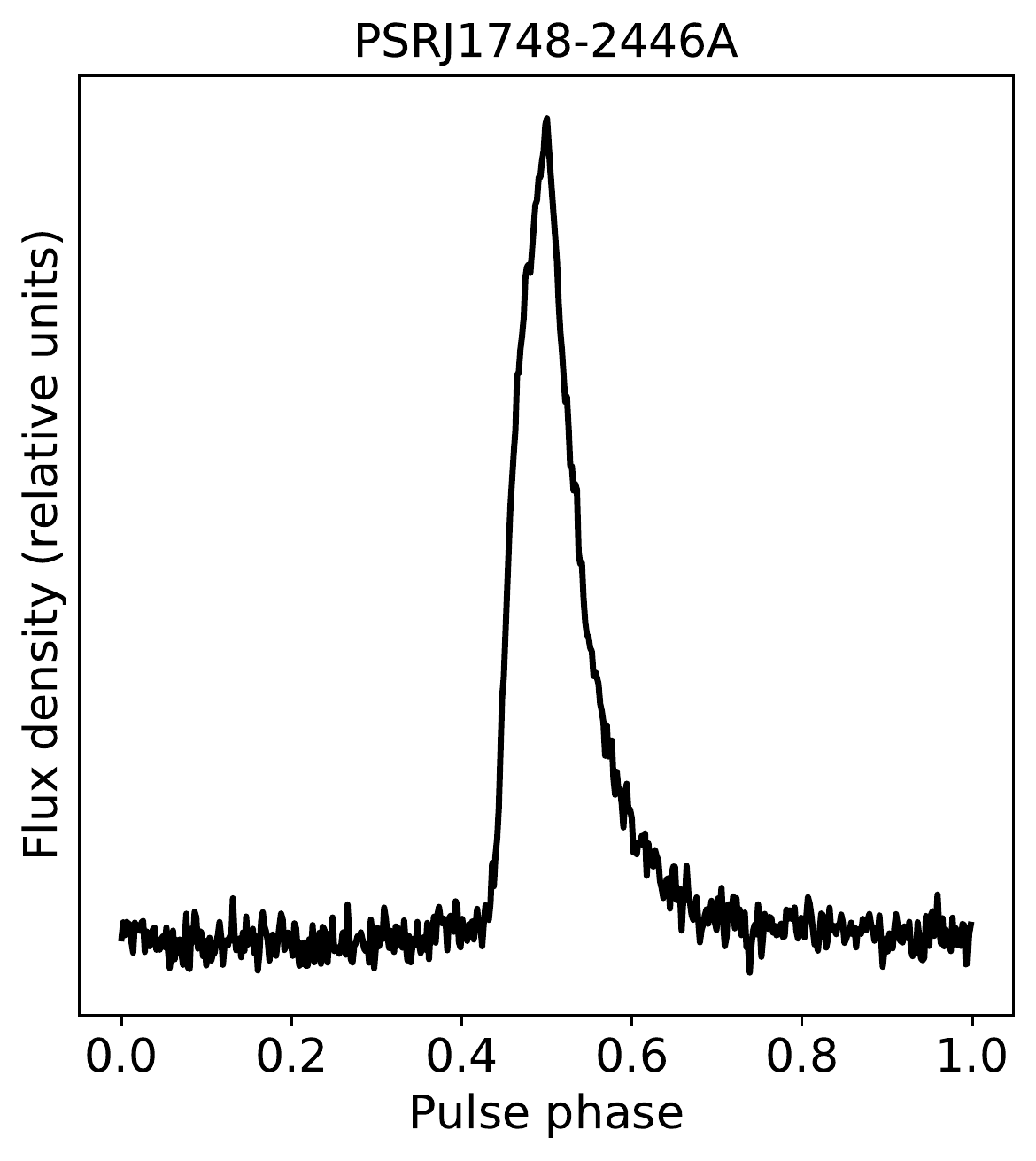}
    \qquad
    \label{fig:ter5a}
\end{subfigure}
\begin{subfigure}[b]{0.24\textwidth}
    \includegraphics[ width=\textwidth]{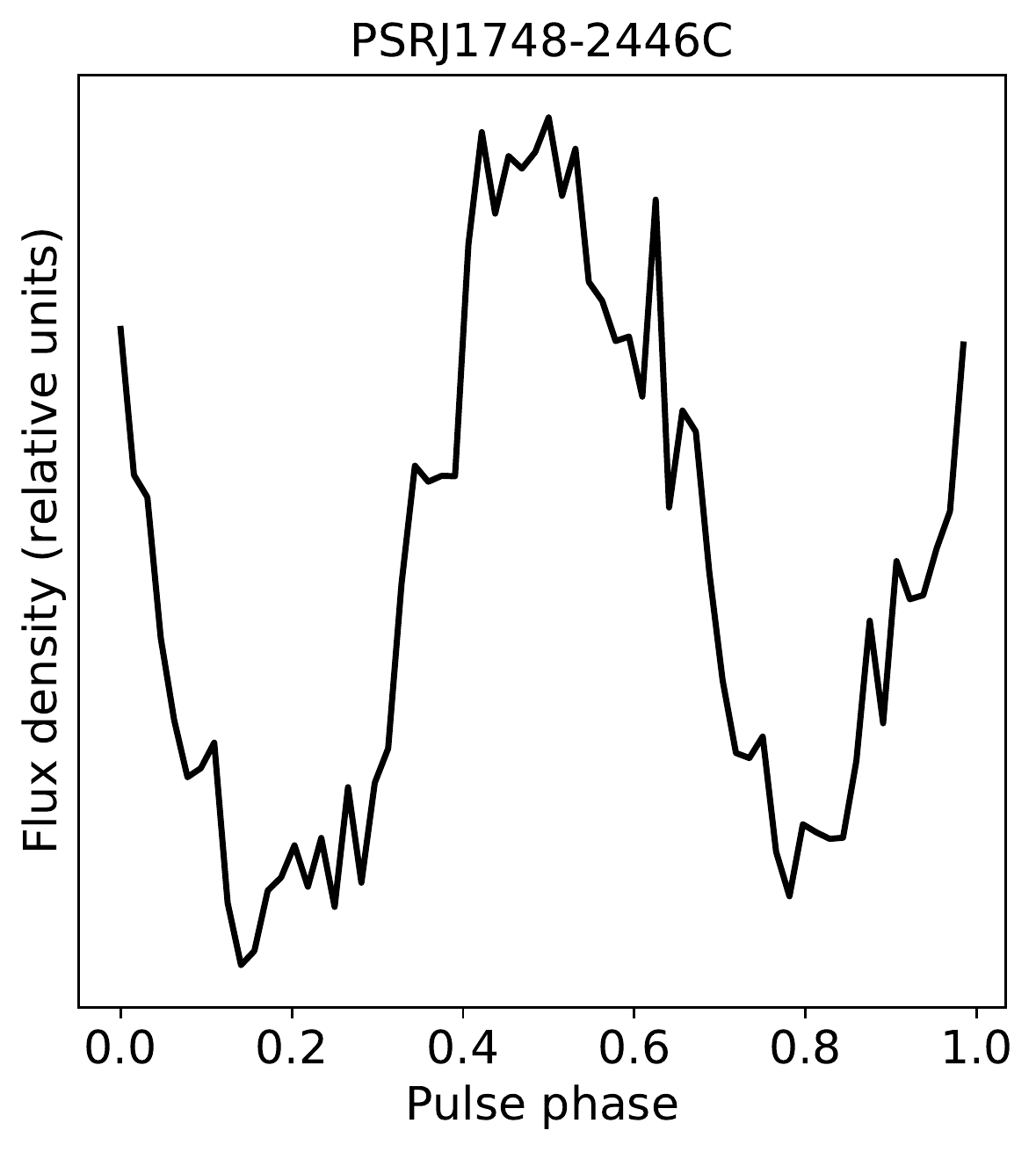}
    \qquad
    \label{fig:ter5c}
\end{subfigure}
\begin{subfigure}[b]{0.24\textwidth}
    \includegraphics[width=\textwidth]{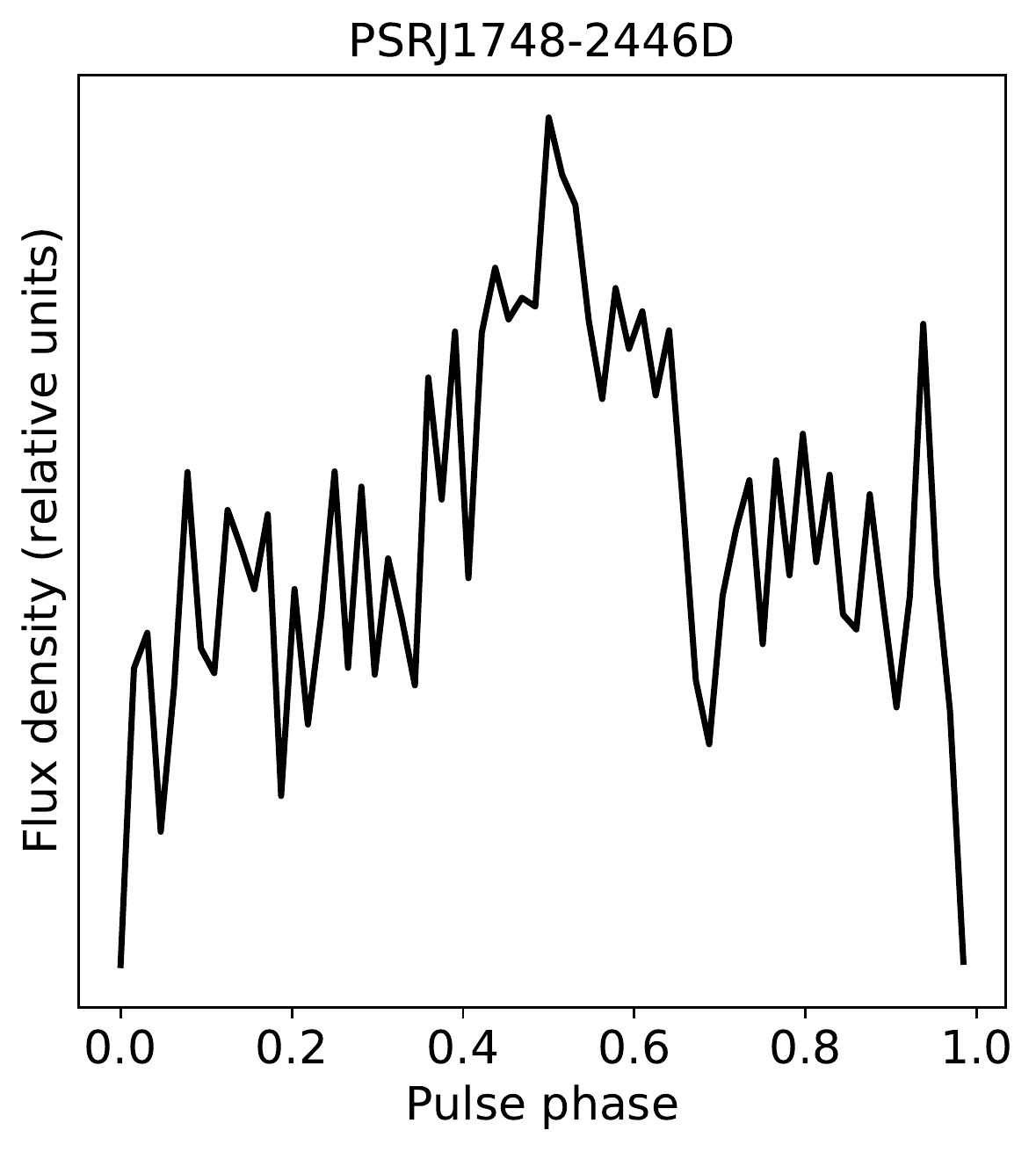}
    \qquad
    \label{fig:ter5d}
\end{subfigure}
\begin{subfigure}[b]{0.24\textwidth}
    \includegraphics[width=\textwidth]{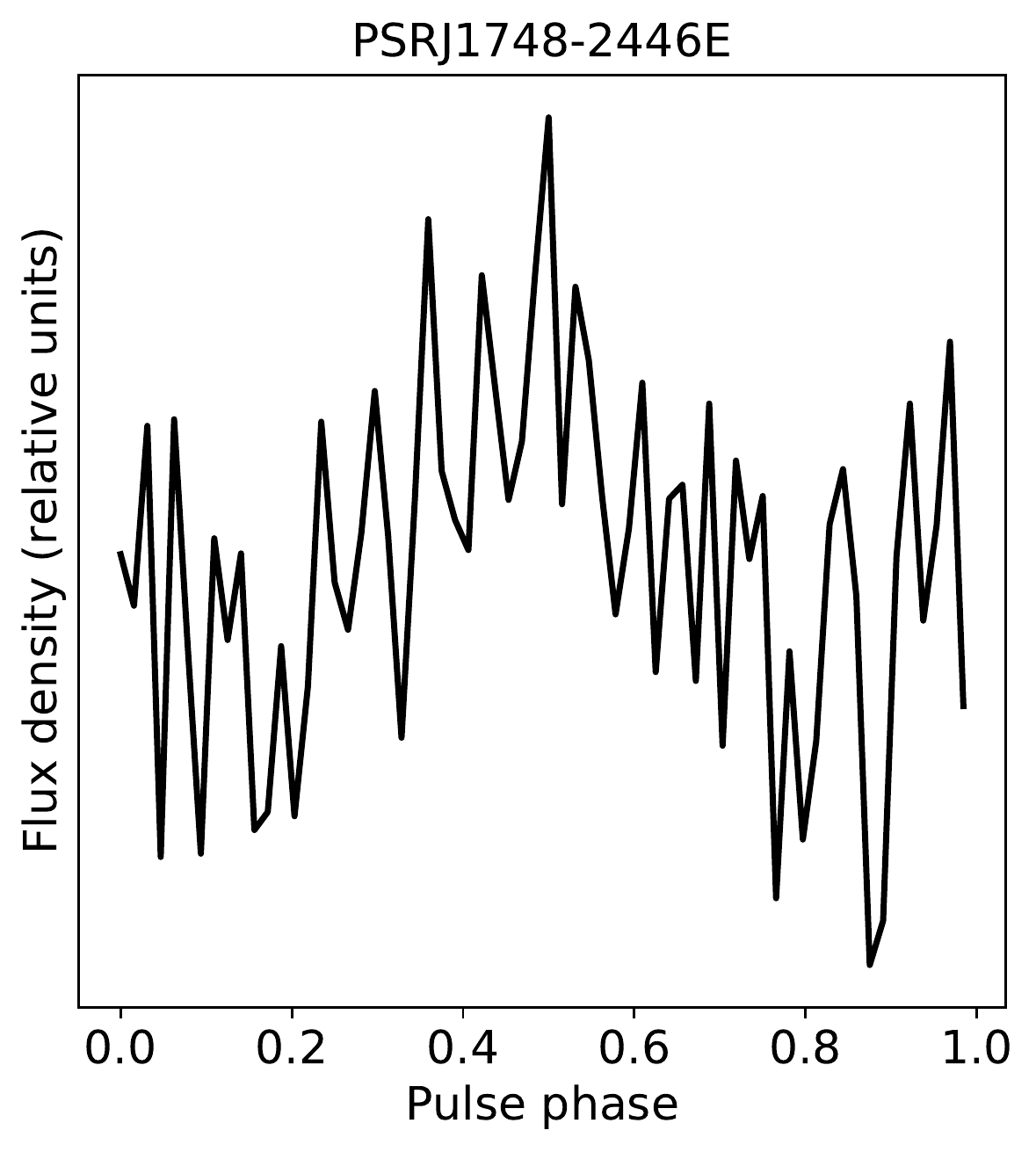}
    \qquad
    \label{fig:ter5e}
\end{subfigure}
\\ \vskip 0.5 cm
\begin{subfigure}[b]{0.24\textwidth}
    \includegraphics[width=\textwidth]{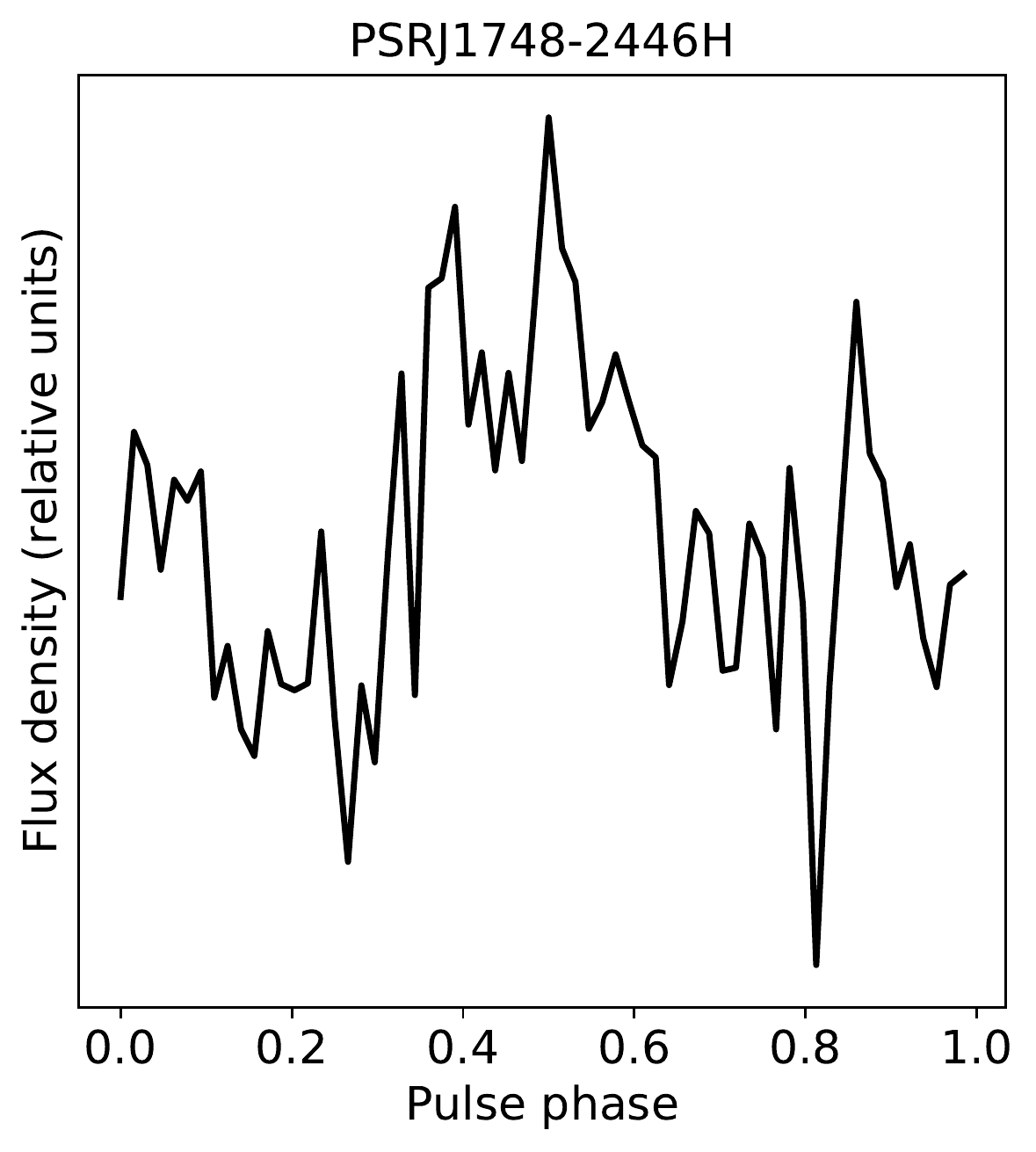}
    \label{fig:ter5h}
\end{subfigure}
\begin{subfigure}[b]{0.24\textwidth}
    \includegraphics[width=\textwidth]{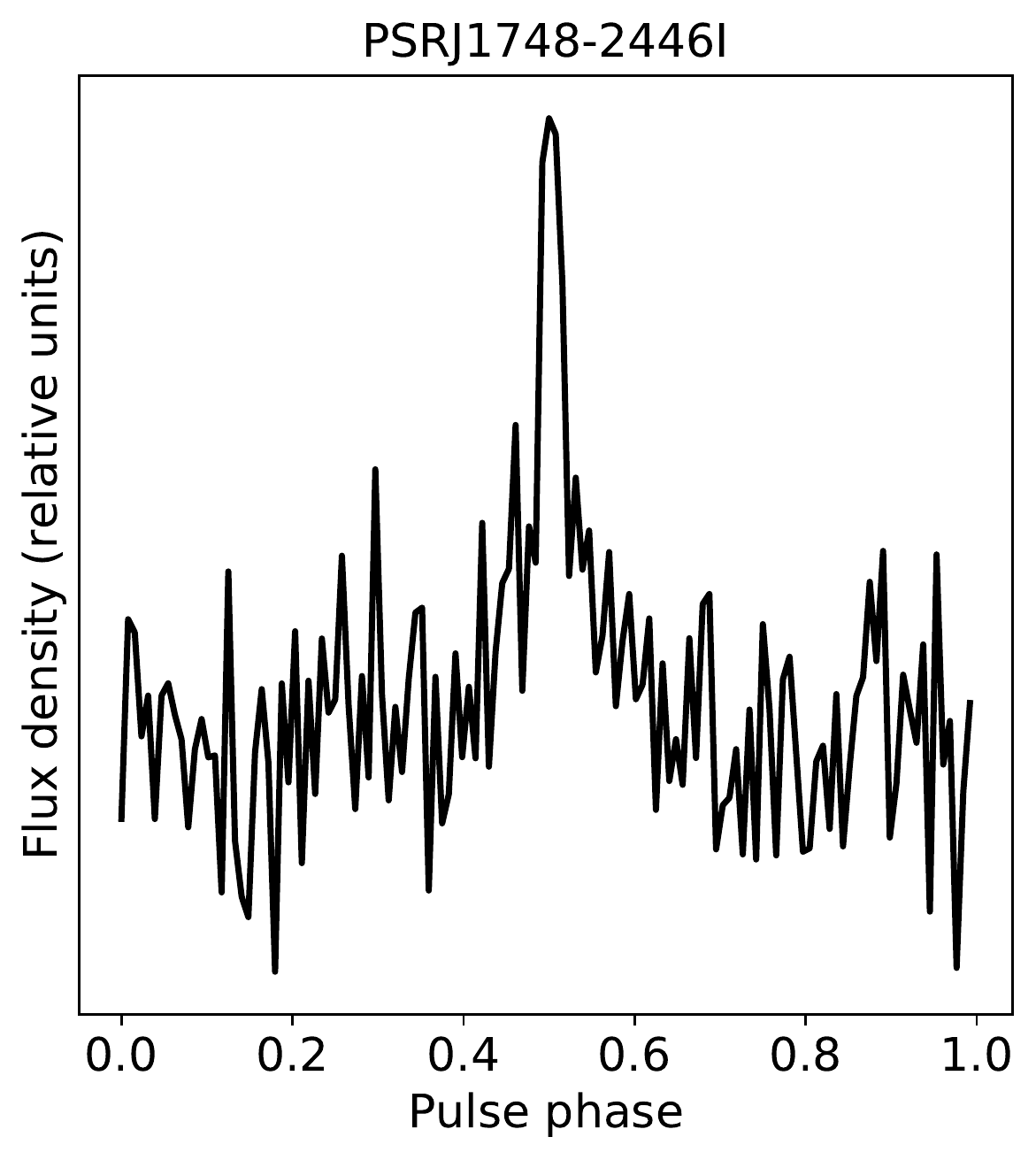}
    \label{fig:ter5i}
\end{subfigure}
\begin{subfigure}[b]{0.24\textwidth}
    \includegraphics[width=\textwidth]{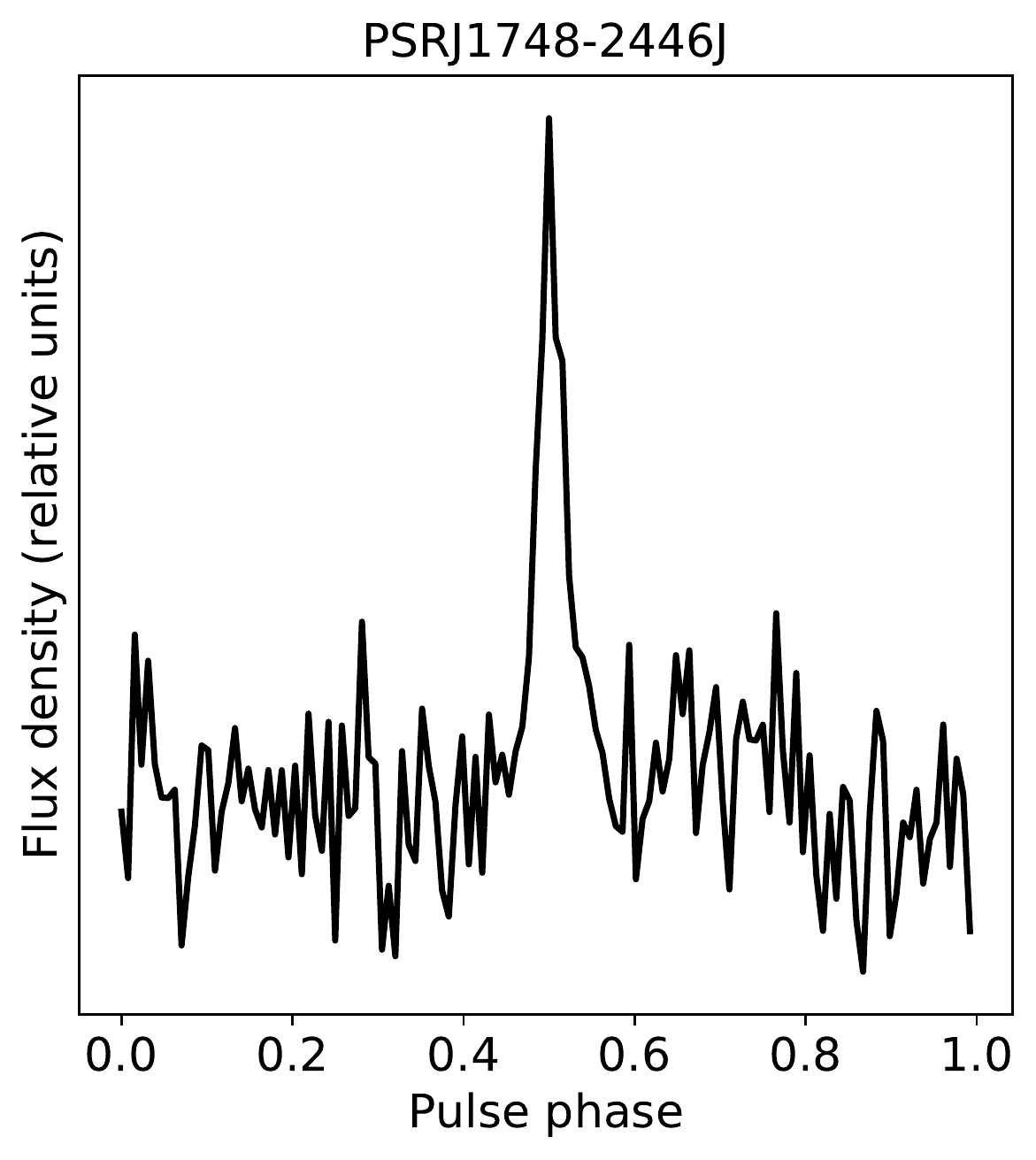}
    \label{fig:ter5j}
\end{subfigure}
\begin{subfigure}[b]{0.24\textwidth}
    \includegraphics[width=\textwidth]{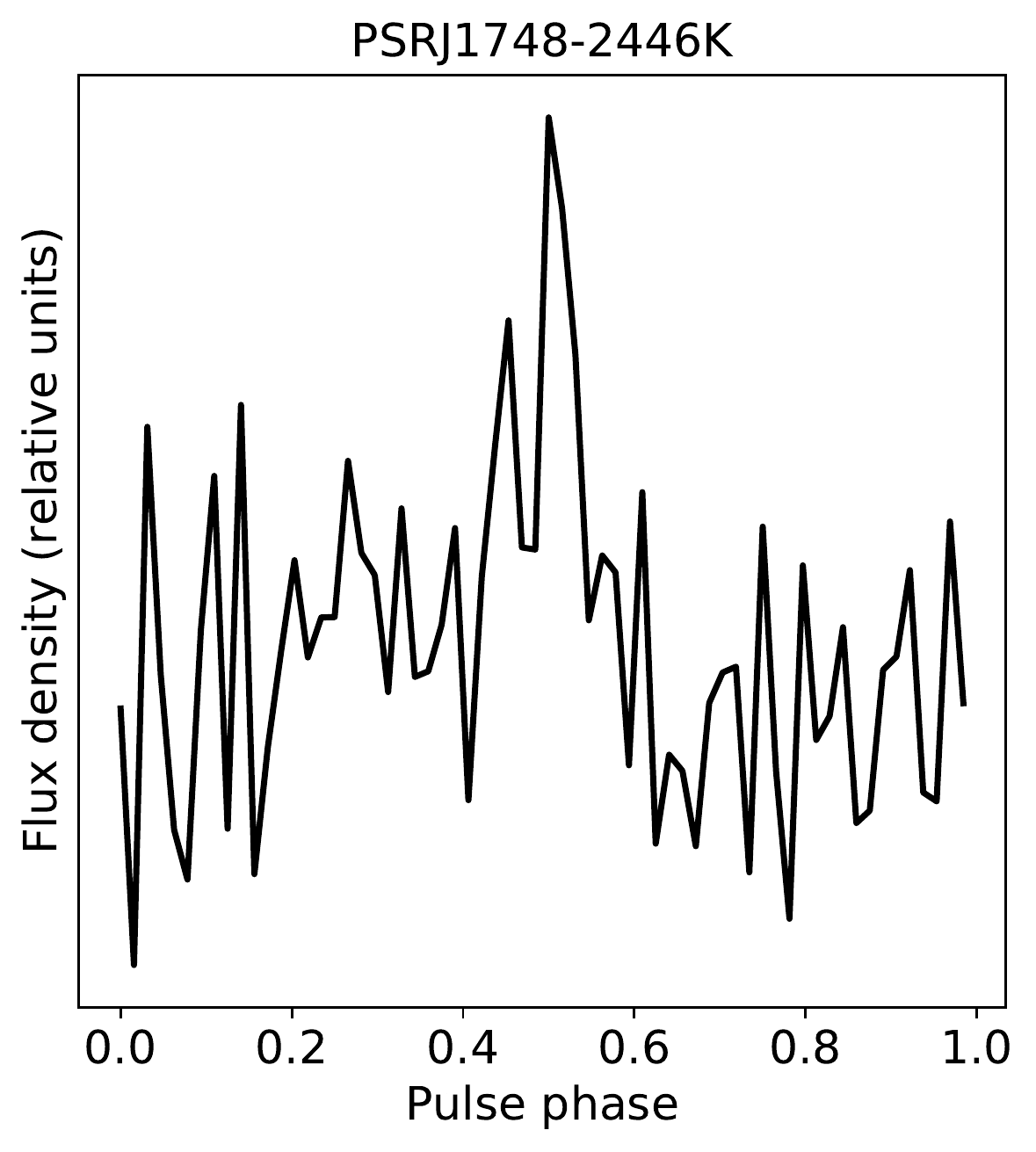}
    \label{fig:ter5k}
\end{subfigure}
\\ \vskip 0.5 cm
\begin{subfigure}[b]{0.24\textwidth}

    \includegraphics[width=\textwidth]{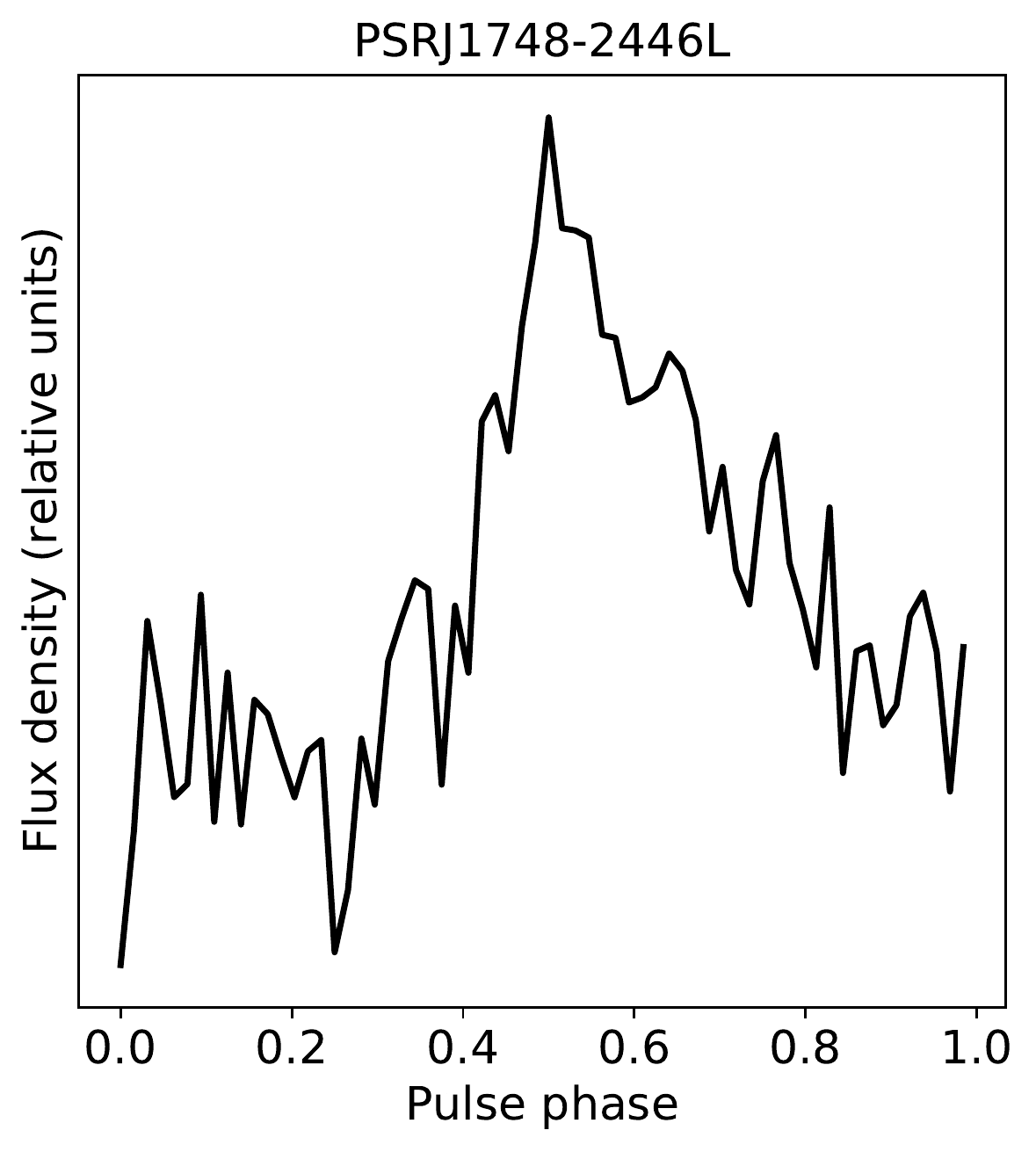}

    \label{fig:ter5l}
\end{subfigure}
\begin{subfigure}[b]{0.24\textwidth}

    \includegraphics[width=\textwidth]{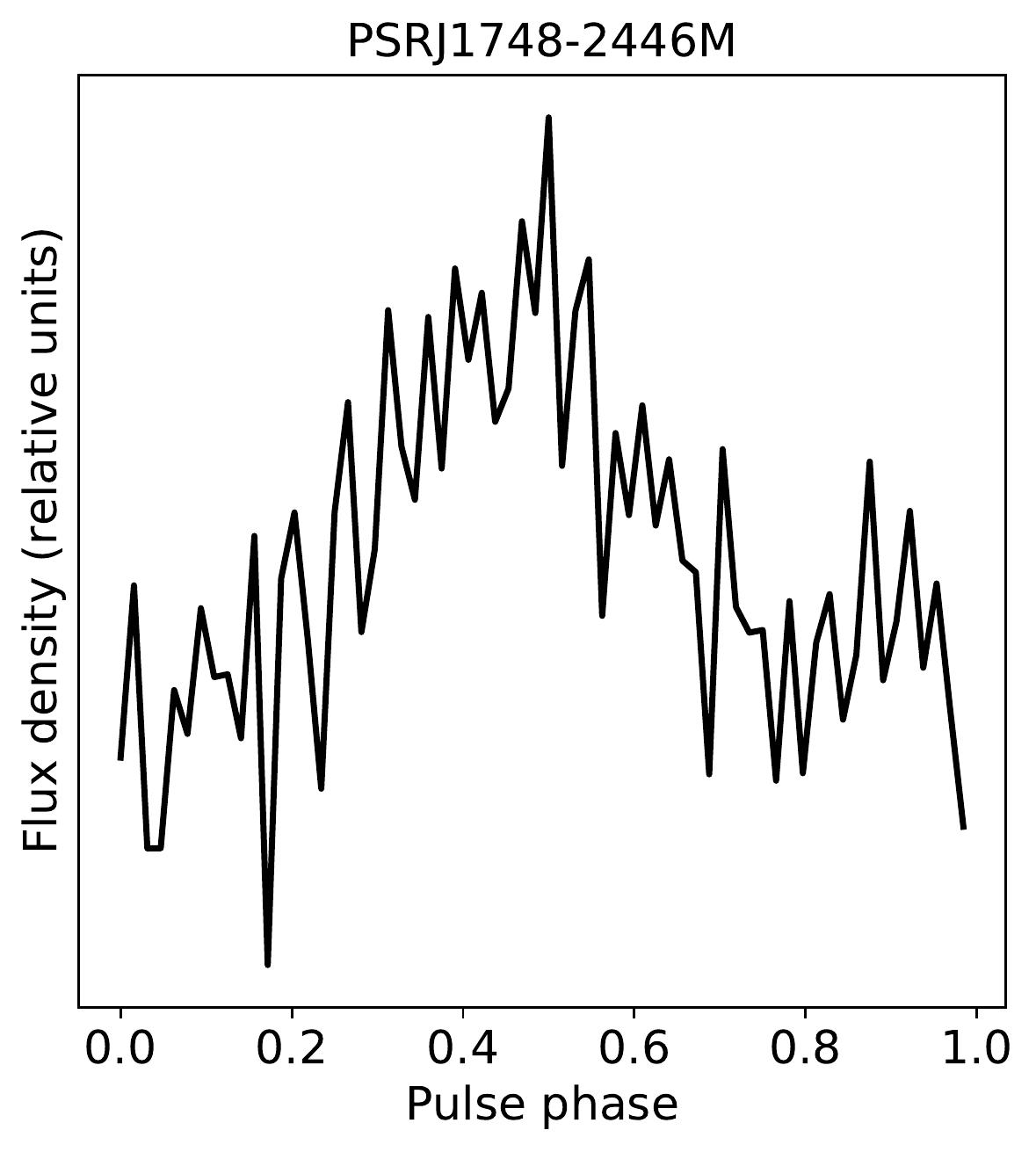}

    \label{fig:ter5m}
\end{subfigure}
\begin{subfigure}[b]{0.24\textwidth}

    \includegraphics[width=\textwidth]{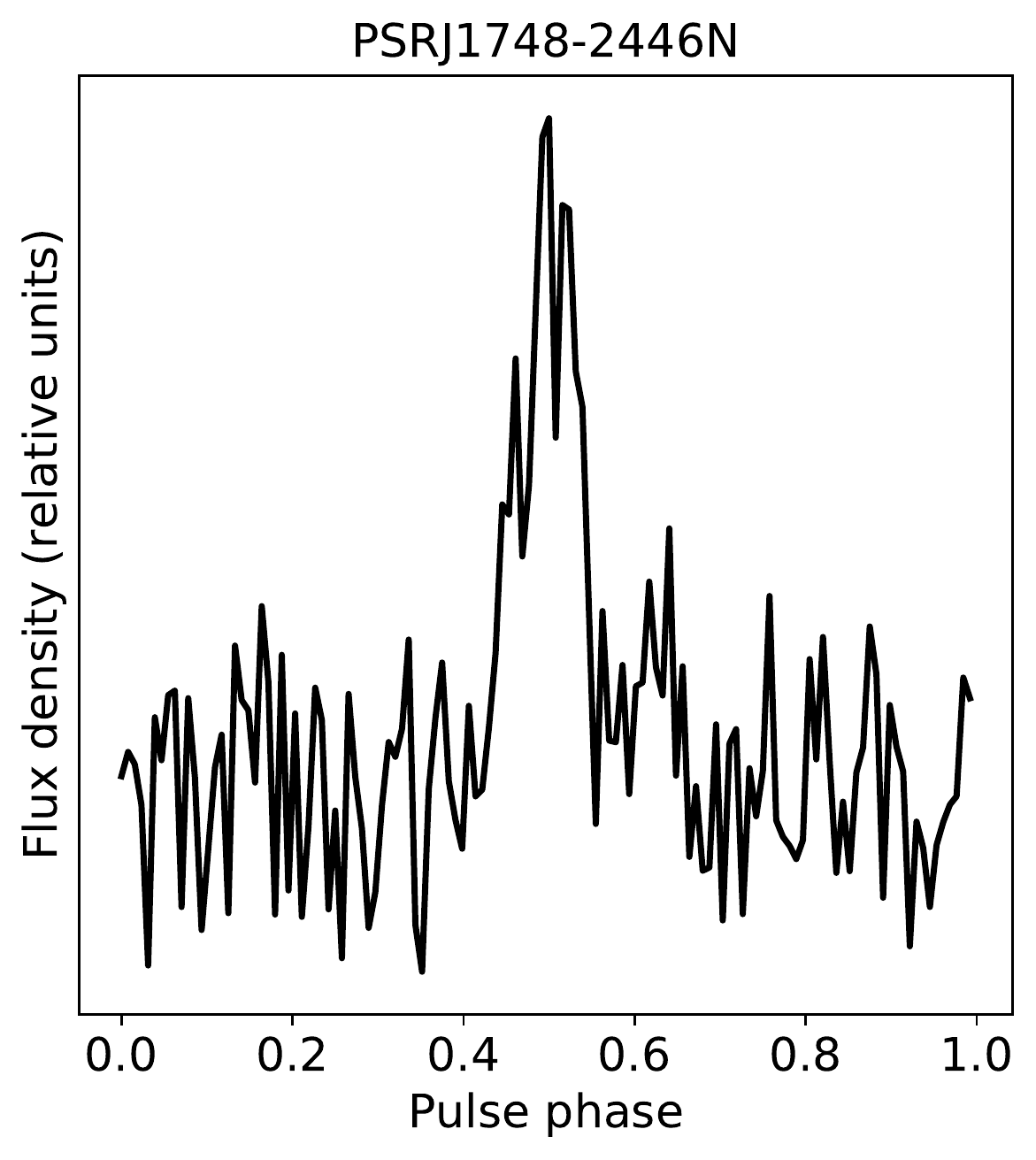}

    \label{fig:ter5n}
\end{subfigure}
\begin{subfigure}[b]{0.24\textwidth}

    \includegraphics[width=\textwidth]{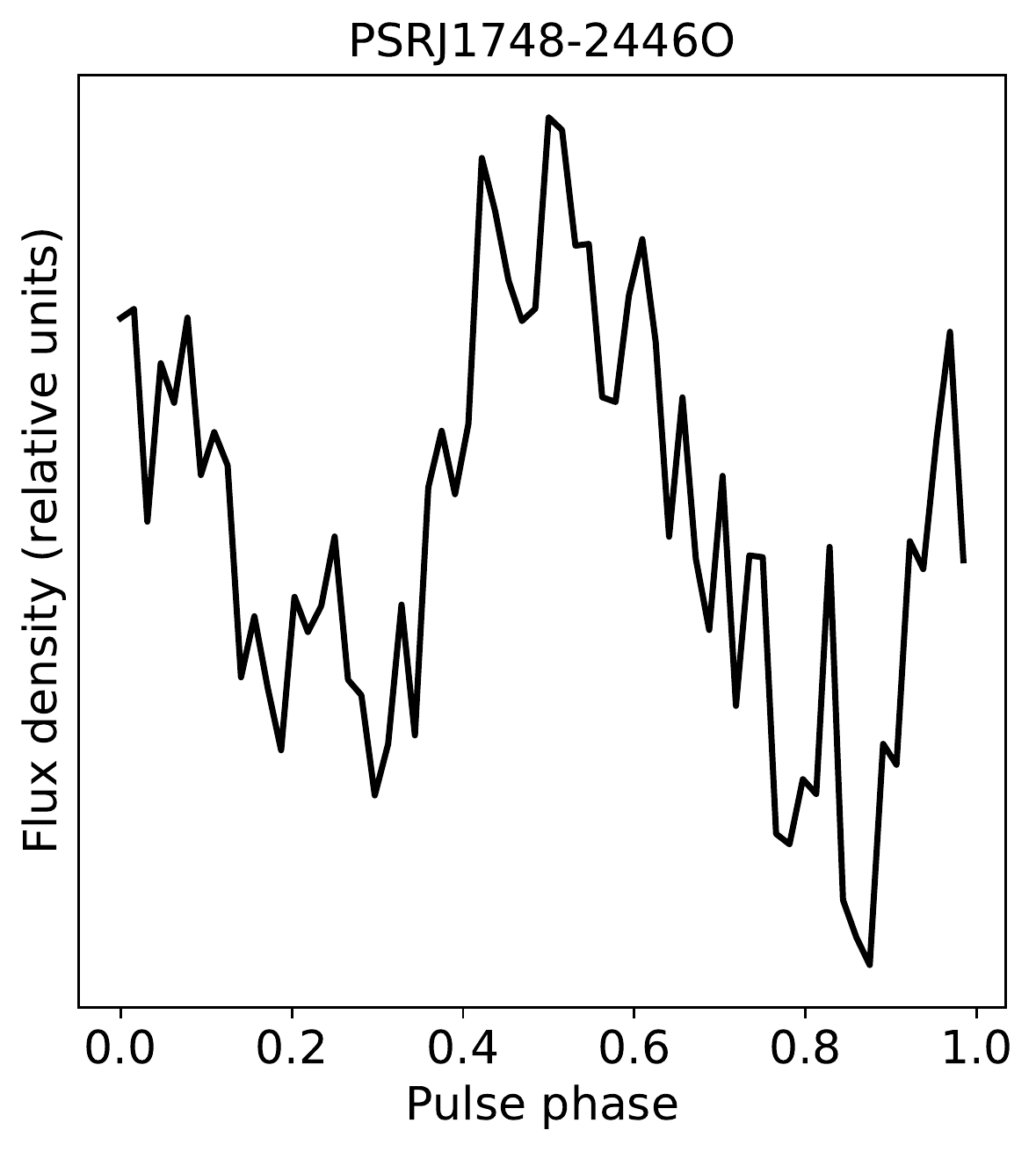}

    \label{fig:ter5o}
\end{subfigure}
\\ \vskip 0.5 cm
\begin{subfigure}[b]{0.24\textwidth}

    \includegraphics[width=\textwidth]{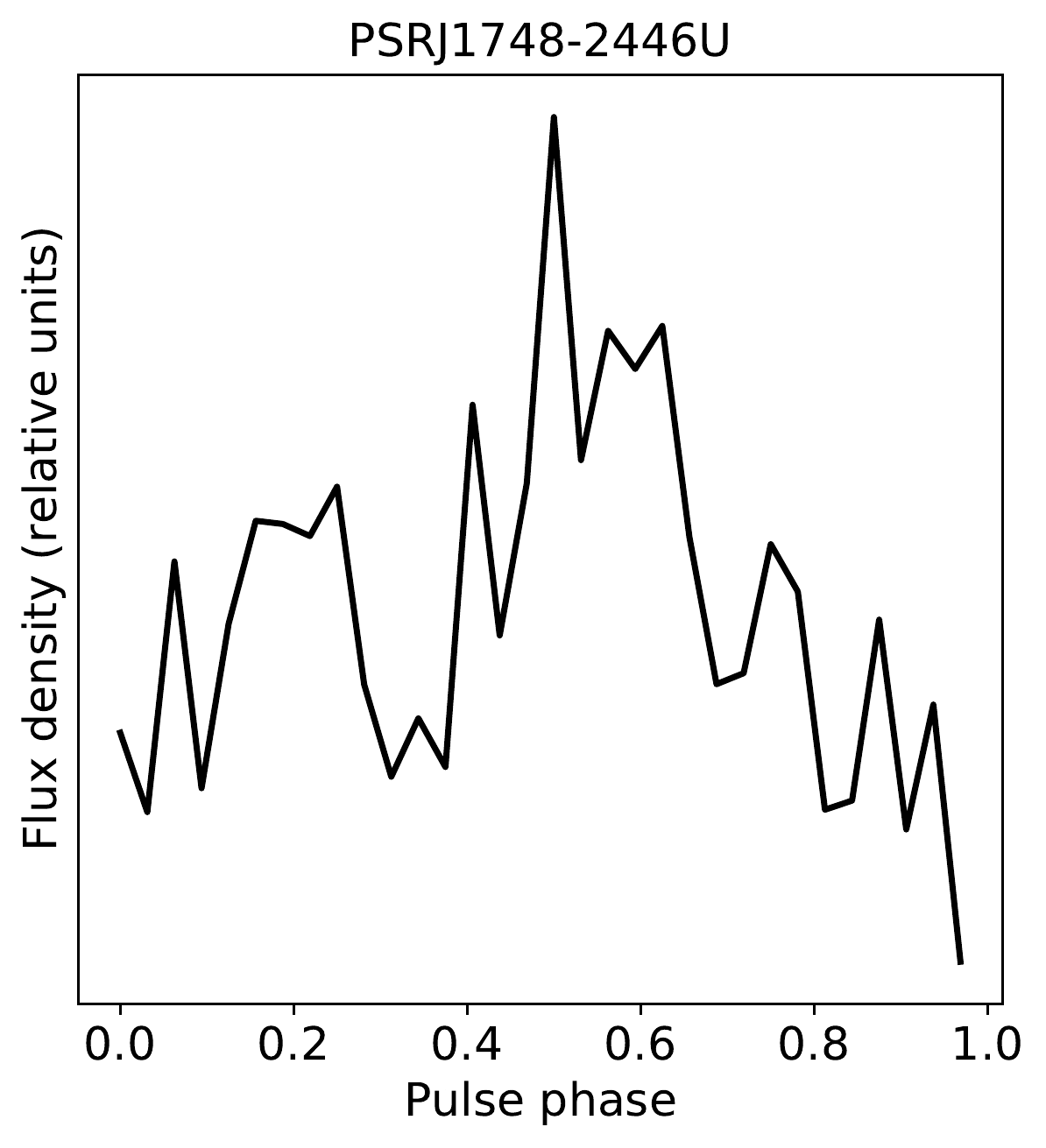}

    \label{fig:ter5u}
\end{subfigure}
\begin{subfigure}[b]{0.24\textwidth}

    \includegraphics[width=\textwidth]{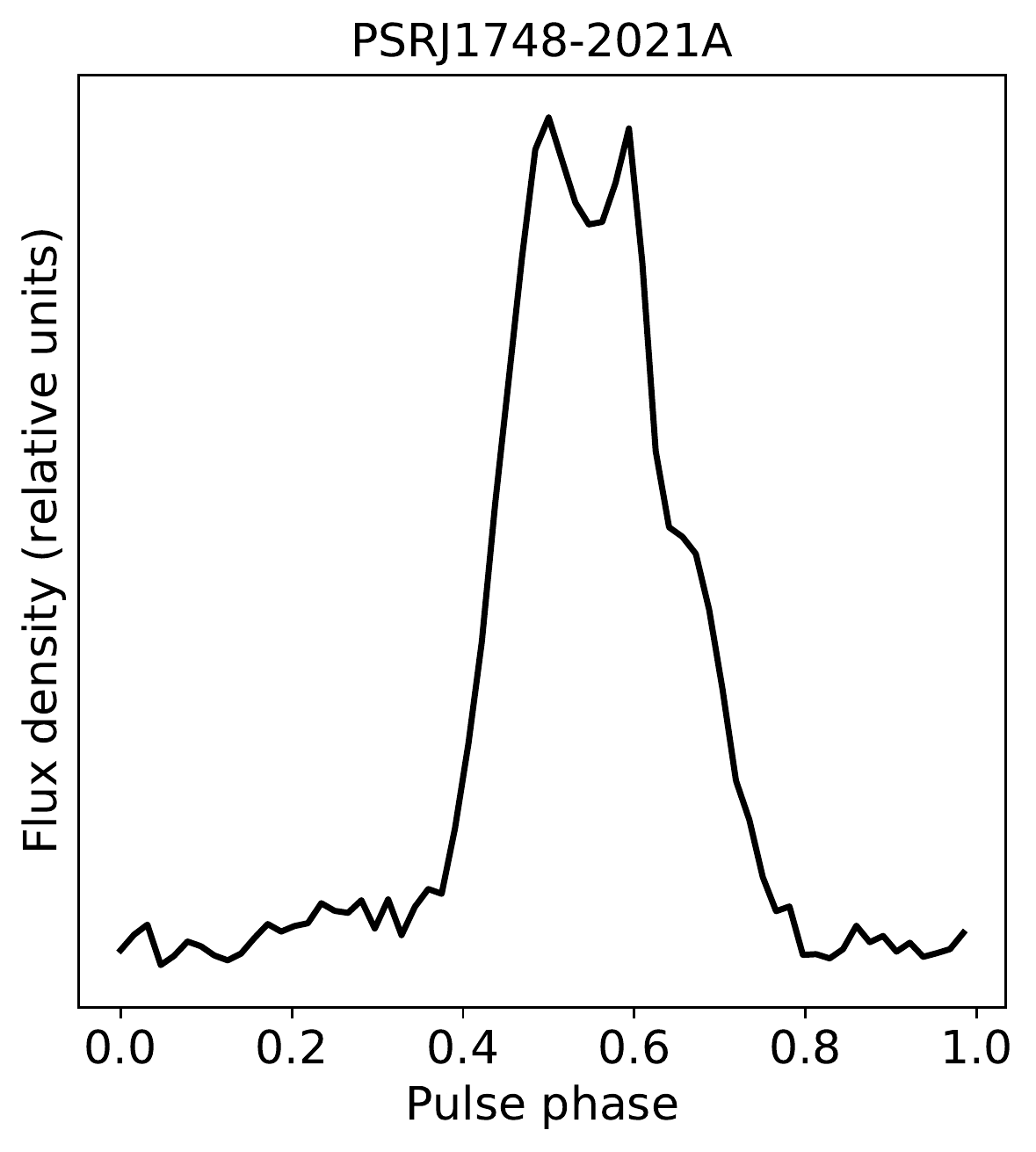}

    \label{fig:6440a}
\end{subfigure}
\begin{subfigure}[b]{0.24\textwidth}

    \includegraphics[width=\textwidth]{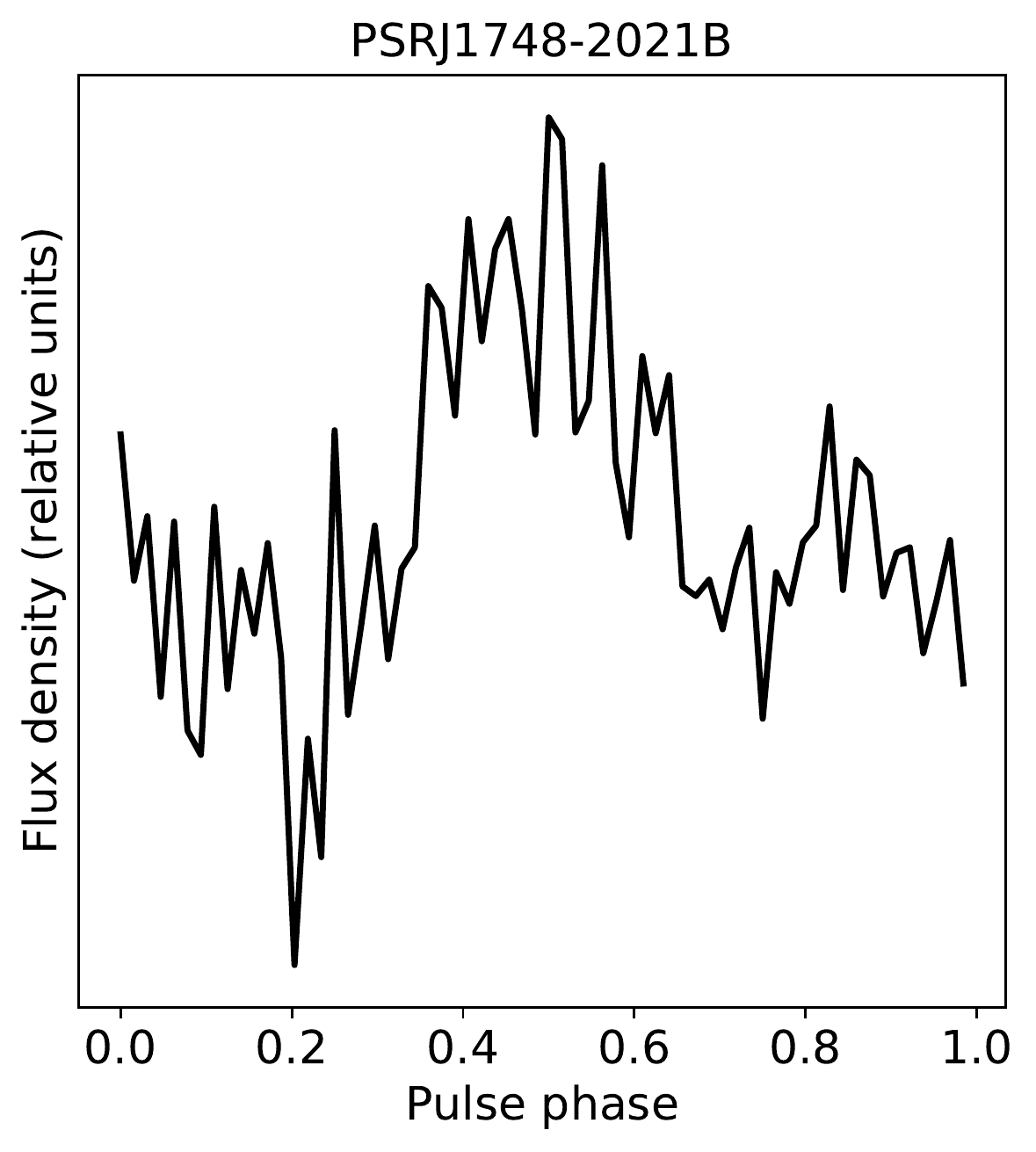}

    \label{fig:6440b}
\end{subfigure}
\begin{subfigure}[b]{0.24\textwidth}

    \includegraphics[width=\textwidth]{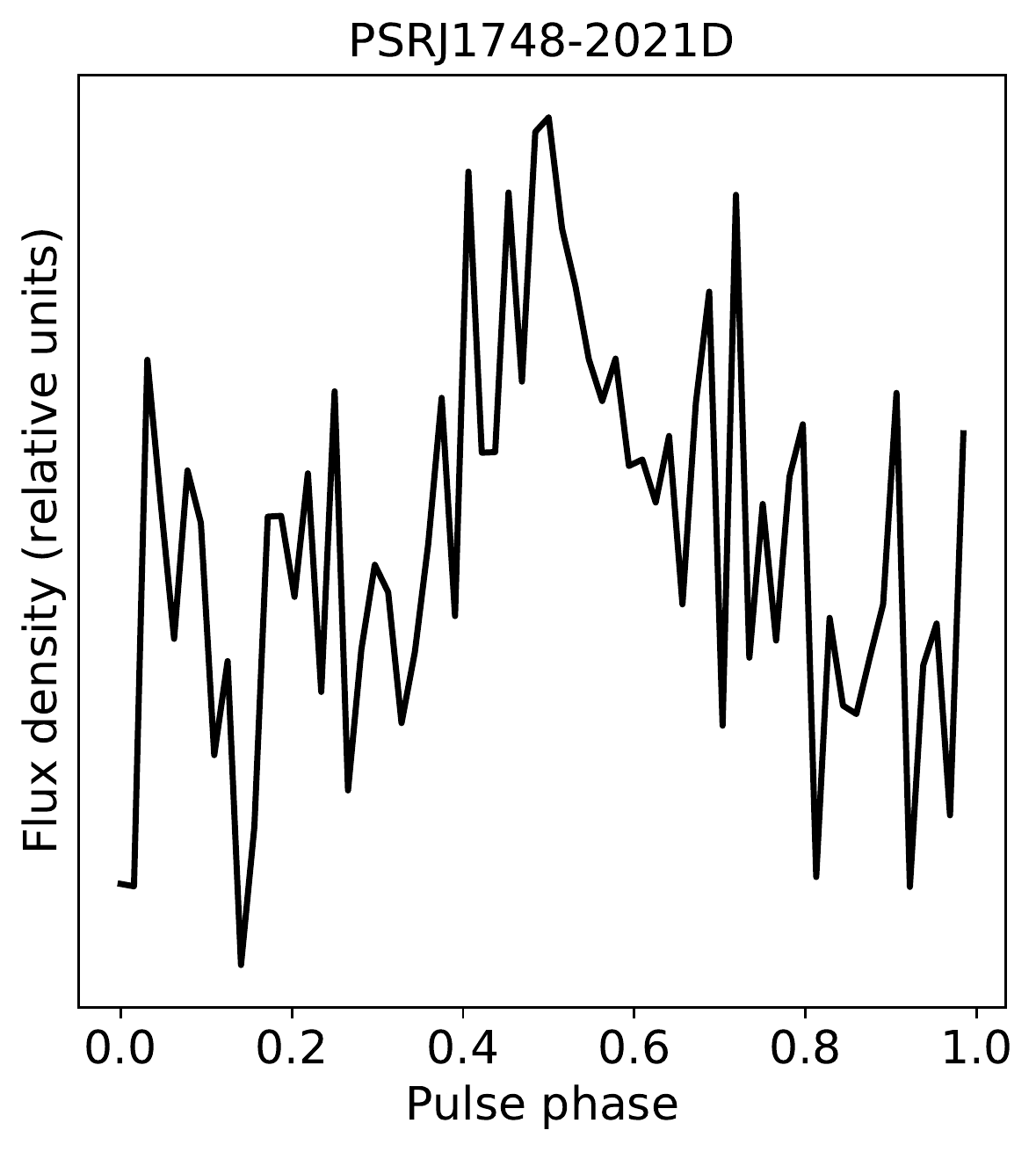}

    \label{fig:6440d}
\end{subfigure}

\caption {Intensity profile vs. rotational phase of previously known pulsars visible with uGMRT at sub-GHz frequencies.}
\end{figure*}
\begin{figure*}
\ContinuedFloat
\begin{subfigure}[b]{0.24\textwidth}

    \includegraphics[width=\textwidth]{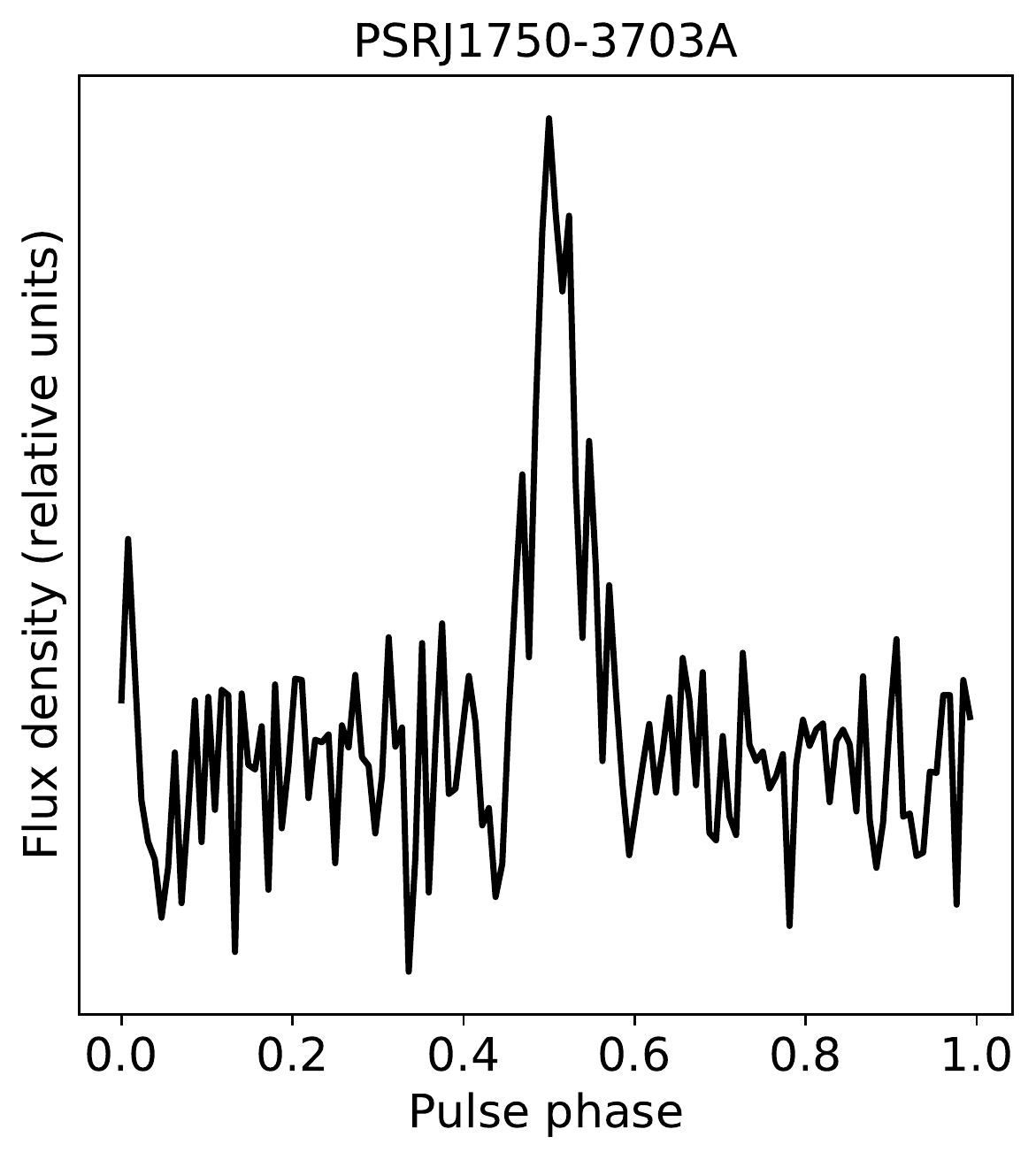}

    \label{fig:6441a}
\end{subfigure}
\begin{subfigure}[b]{0.24\textwidth}

    \includegraphics[width=\textwidth]{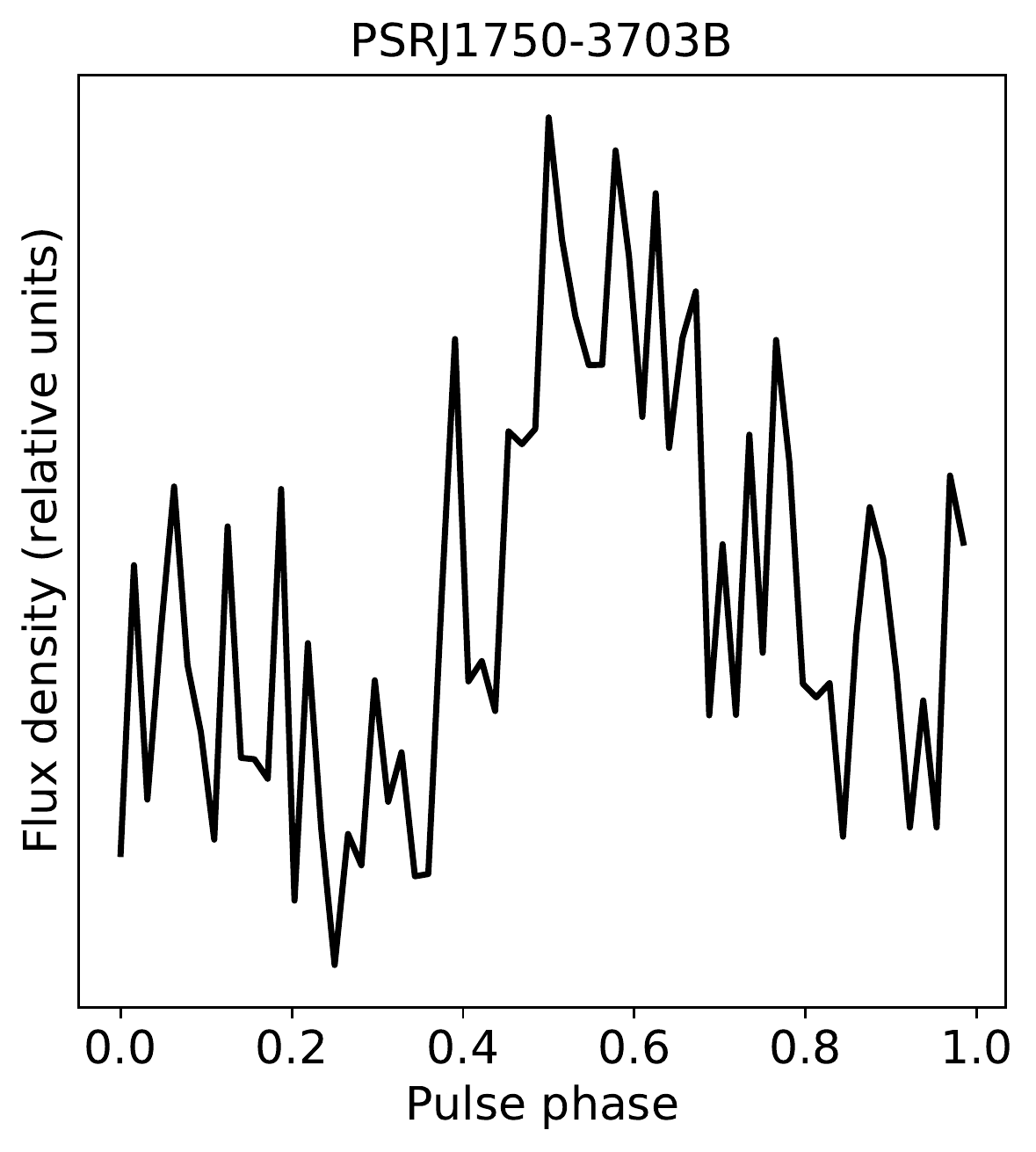}

    \label{fig:6441b}
\end{subfigure}
\begin{subfigure}[b]{0.24\textwidth}

    \includegraphics[width=\textwidth]{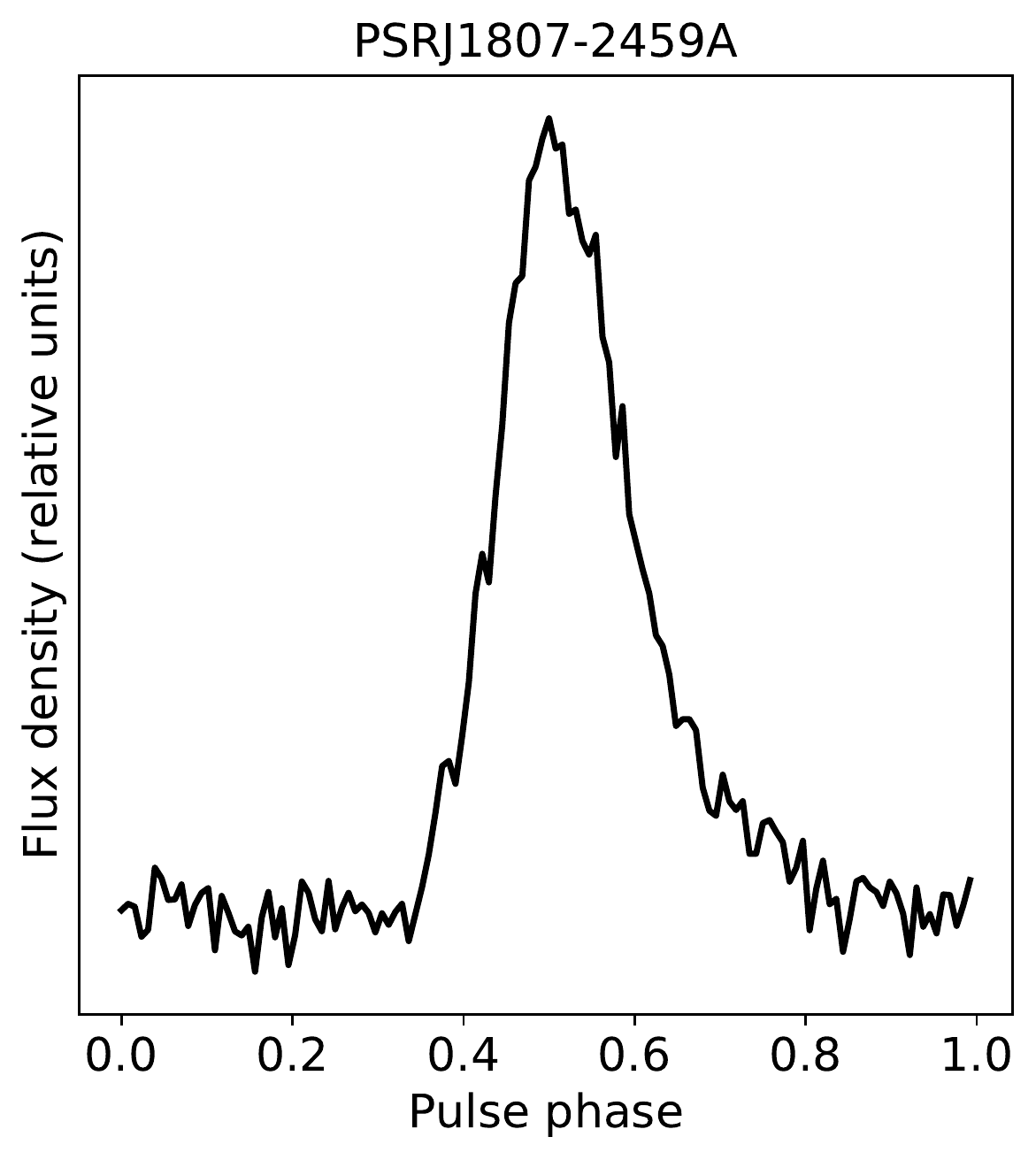}

    \label{fig:6544a}
\end{subfigure}
\begin{subfigure}[b]{0.24\textwidth}
 
    \includegraphics[width=\textwidth]{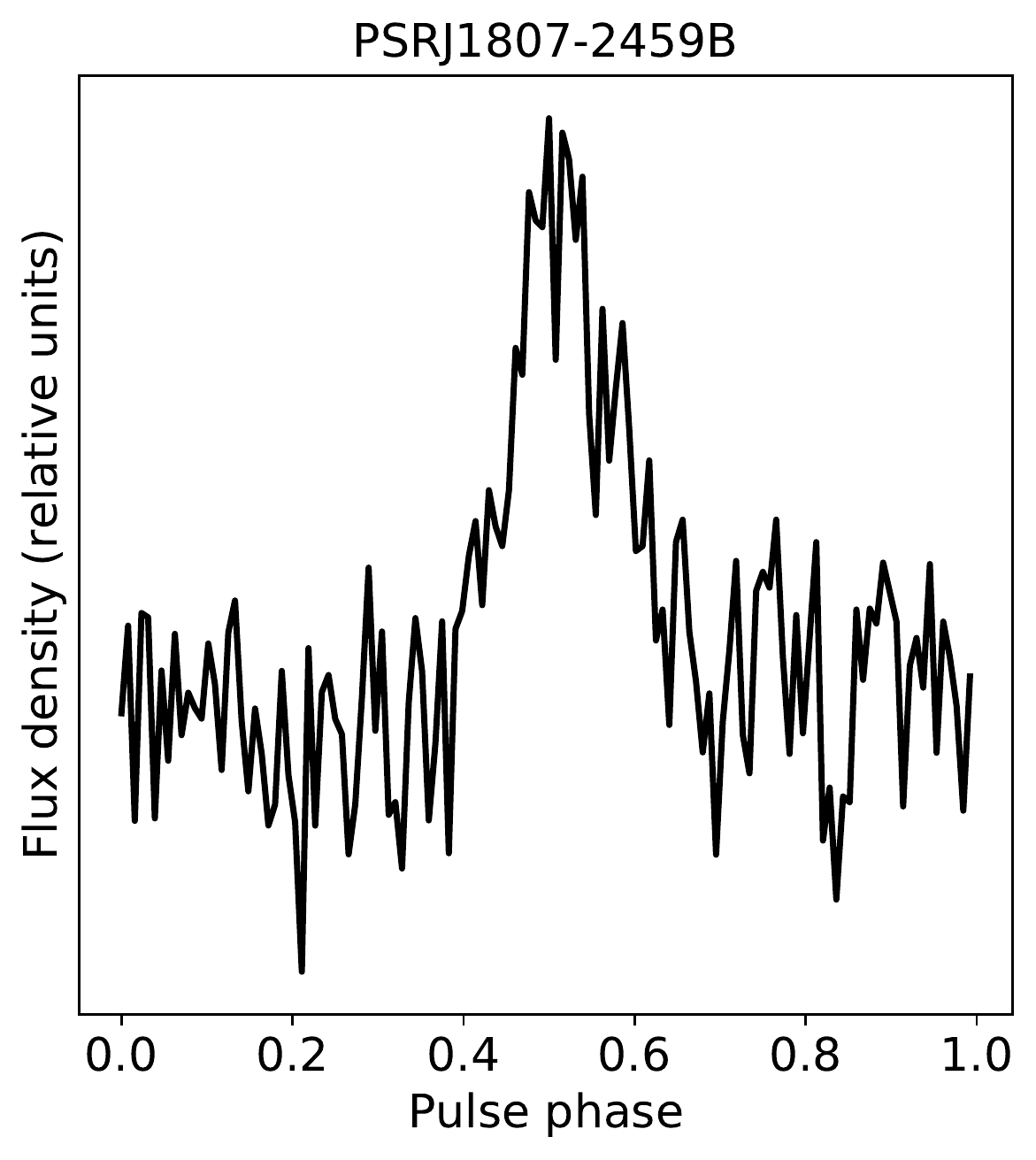}

    \label{fig:6544b}
\end{subfigure}
\\ \vskip 0.5 cm
\begin{subfigure}[b]{0.24\textwidth}

    \includegraphics[width=\textwidth]{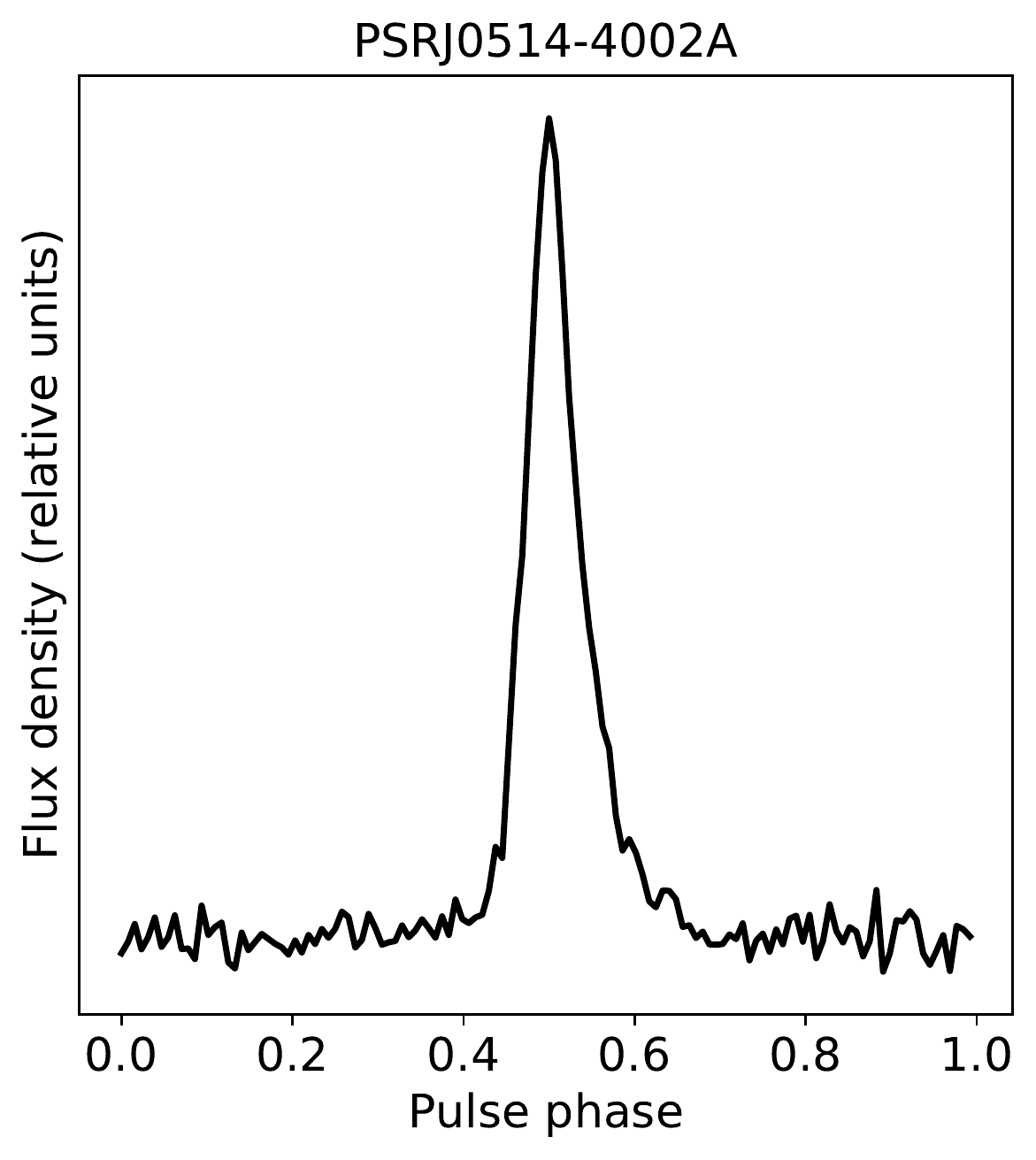}

    \label{fig:1851a}
\end{subfigure}
\begin{subfigure}[b]{0.24\textwidth}
    \includegraphics[width=\textwidth]{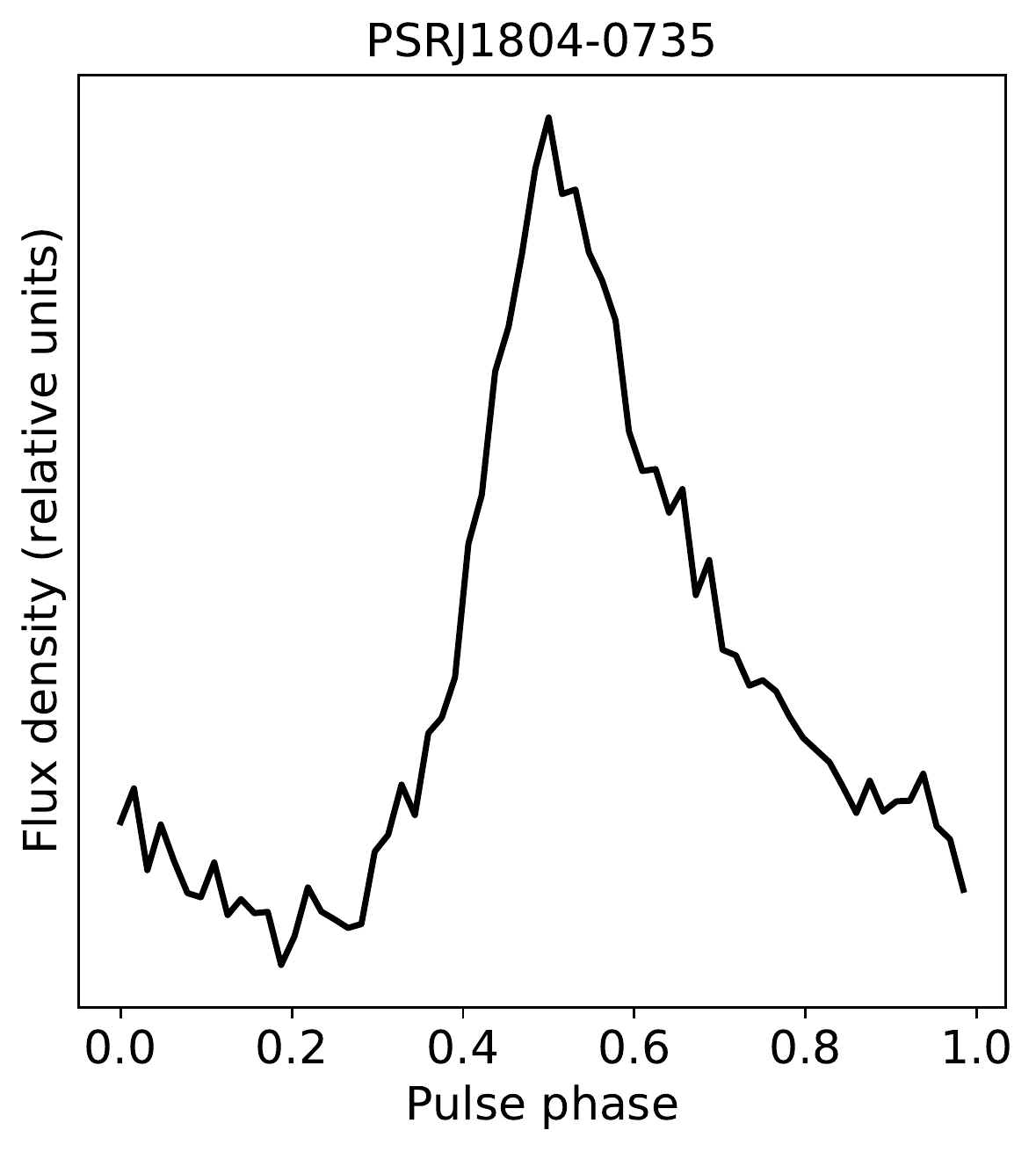}

    \label{fig:6539a}
\end{subfigure}
\caption {(Cont.) Intensity profile vs. rotational phase of previously known pulsars visible with uGMRT at sub-GHz frequencies.}
\label{fig:known_profiles}
\end{figure*}

\begin{table*}
\caption[]{Properties estimated for re-detected and newly discovered pulsars. $\rm F_c$ represents the central frequency of observation, $\rm S_{\rm min}$ is the minimum detection threshold for each cluster, $\rm S_{\rm obs}$ is the observed flux density, $\rm S_{\rm exp}$ is the expected flux density (assuming spectral index of -1.4 if unspecified by ATNF catalogue), $\tau_{\rm scat}$ is the measured scattering timescale at reference frequency $f_{\rm ref}$, and $\tau^{\rm th}_{\rm scat}$ is the theoretical estimate of scattering timescale at frequency $f_{\rm ref}$ (calculated from the relation by \citealt{Bhat+2004}).}
\label{tab:flux_estimates}
\footnotesize
\centering
\renewcommand{\arraystretch}{1.0}
\vskip 0.1cm
\begin{tabular}{lcccccccccc}
\hline
\hline
{Cluster} & {$\rm F_c$ } & {$\rm S_{\rm min}$ } & {Redetections}  & {S/N} & {$\rm S_{\rm obs}$} & {$\rm S_{\rm exp}$} & {Spectral} &{Scattering} & {$\tau_{\rm scat}$, $f_{\rm ref}$} &{$\tau^{\rm th}_{\rm scat}$ } \\ & (MHz) & (mJy) & & & (mJy) & (mJy) & Index & Index & (ms, MHz) & (ms)  
\\\midrule
{NGC~1851} & 400 & 0.206 & A & 126.6 & 2.717 & - & - & - & Not measurably scattered & 0.03
\\
\\
{Terzan~5} & 650 & 0.143 & A & 384.2 & 7.491 & 4.08 &$-$2.420 & $-$3.2(1) & 0.640(9), 640 & 5.48\\
& & & C & 33.4 & 0.915 & - & - & - & - & -  \\
& & & D & 12.2 & 0.309 & 0.19 & $-$1.797& - & - & - \\
& & & F & 10.5 & 0.223 &0.17 & $-$1.649& - &  - & -\\
& & & H & 12.2 & 0.310 & 0.07 & $-$2.694 & - & - & - \\
& & & I & 12.9 & 0.175 & - & - & - & 0.29(7), 700 & 3.57\\
& & & J & 18.9 & 0.199 &0.09& $-$2.088 & - & - & -\\
& & & L & 18.6 & 0.371 & 0.19 &$-$3.030 & - & - & -\\
& & & M & 15.0 & 0.503 & 0.26 & $-$1.407 & - & - & -\\
& & & N & 18.2 & 0.292 & 0.15 & $-$2.314 & - & 0.56(19), 625 & 5.49\\
& & & O & 14.2 & 0.344 & 0.57&$-$1.138 & - & - & -\\
& & & U & 10.3 & 0.345 & -&$-$2.732 & - & - & -\\
\\
{NGC~6440} & 650 & 0.099 & A & 178.8 & 4.234 & 1.72 & $-$2.197 & - & Not measurably scattered & 3.25\\
& & & B & 10.9&  0.332 & 0.22 & $-$1.747 & - & - & - \\
& & & D & 8.2 & 0.248 & 0.36 & $-$1.062 & - & - & - \\
\\
{NGC~6441} & 650 & 0.105 & A & 14.6 & 0.192 & 0.28 & $-$1.156 & - & - & - \\
& & & B & 12.9 & 0.432 & 0.17 & $-$2.188 & - & - & - \\
\\
{NGC~6539} & 400 & 0.099 & A & 94.1 & 2.263 & 3.08 & - & - & 3.80(83), 412
 & 7.82\\
 \\
 {NGC~6544} & 650 & 0.099 & A & 105.9 & 2.628 & 4.05 &$-$1.273 & $-$4.0(5) & 0.25(2), 662 & 0.25\\
& & & B & 21.2 & 0.414 & 0.42 & $-$1.378 & -  & 0.37(12), 600 & 0.41\\
\\
{NGC~6652} & 400 & 0.205 & B & 31.6 & 1.01 & - & - & - & 0.174(3), 330 & 0.142\\
\hline
\hline
\end{tabular}
\end{table*}

\section{Imaging the globular clusters}
\label{sec:imaging_gc}
Using the visibility data, we created radio maps for each of these clusters. 
The detection of steep-spectrum radio sources could  allow the discovery of bright radio pulsars, which may have been missed 
by previous time-domain pulsar surveys either due to the limited field of  view of large telescopes or due to the Doppler smearing
of highly accelerated pulsar systems.  
Imaging known pulsars in these clusters also provide arcsecond  localisations that can be used to refine the timing solutions for 
systems without precisely known positions.

The procedure used standard tasks incorporated in \texttt{CASA} \citep{McMullin+2007} to perform flagging, calibration, and imaging on the dataset. For the observations where a standard flux calibrator was not observed, we calculated a model for the flux density of the phase calibrator from other nearby epochs where a proper flux calibrator was observed and used it for the delay, bandpass, and gain calibrations. This was sufficient to provide reasonable estimates for the flux densities and spectral indices of radio sources. 

\begin{figure*}
\begin{subfigure}{.33\textwidth}
  \centering
  \includegraphics[width=1.0\linewidth,trim={1.5cm 1.5cm 1.5cm 1.5cm},clip]{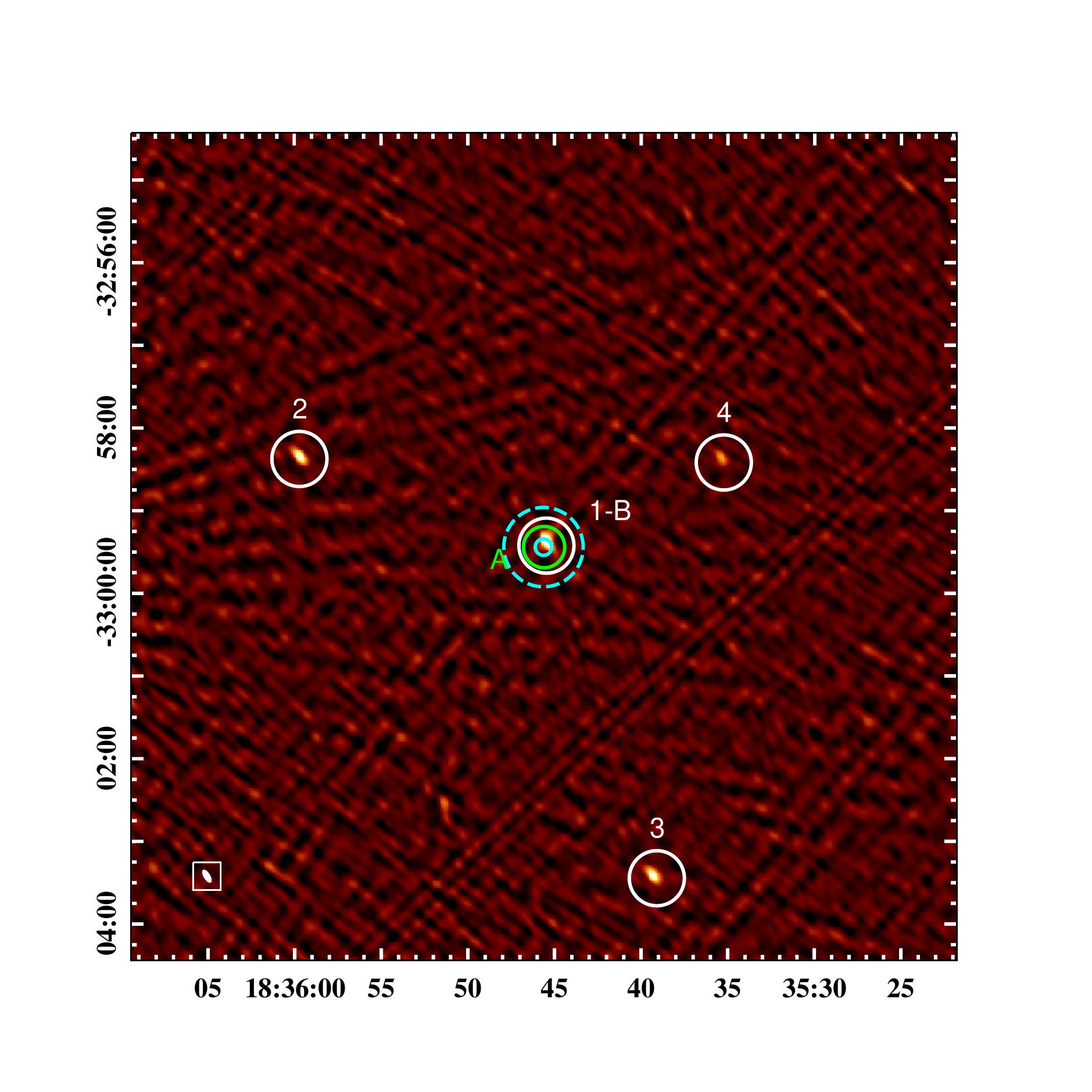}
  \caption{NGC~6652}
  \label{fig:sfig1}
\end{subfigure}%
\begin{subfigure}{.33\textwidth}
  \centering
  \includegraphics[width=1.0\linewidth,trim={1.5cm 1.5cm 1.5cm 1.5cm}]{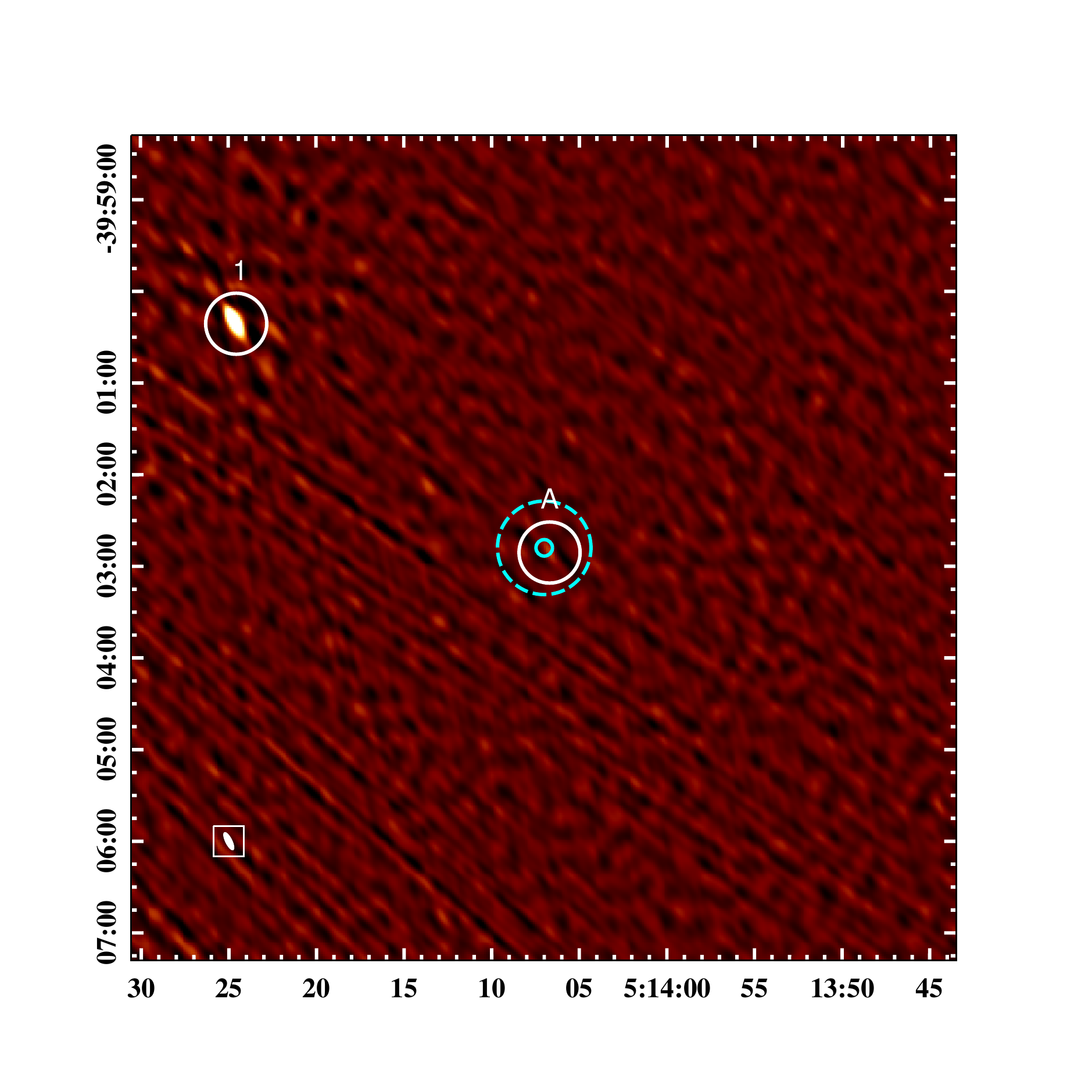}
  \caption{NGC~1851}
  \label{fig:sfig2}
\end{subfigure}
\begin{subfigure}{.33\textwidth}
  \centering
  \includegraphics[width=1.0\linewidth,trim={1.5cm 1.5cm 1.5cm 1.5cm}]{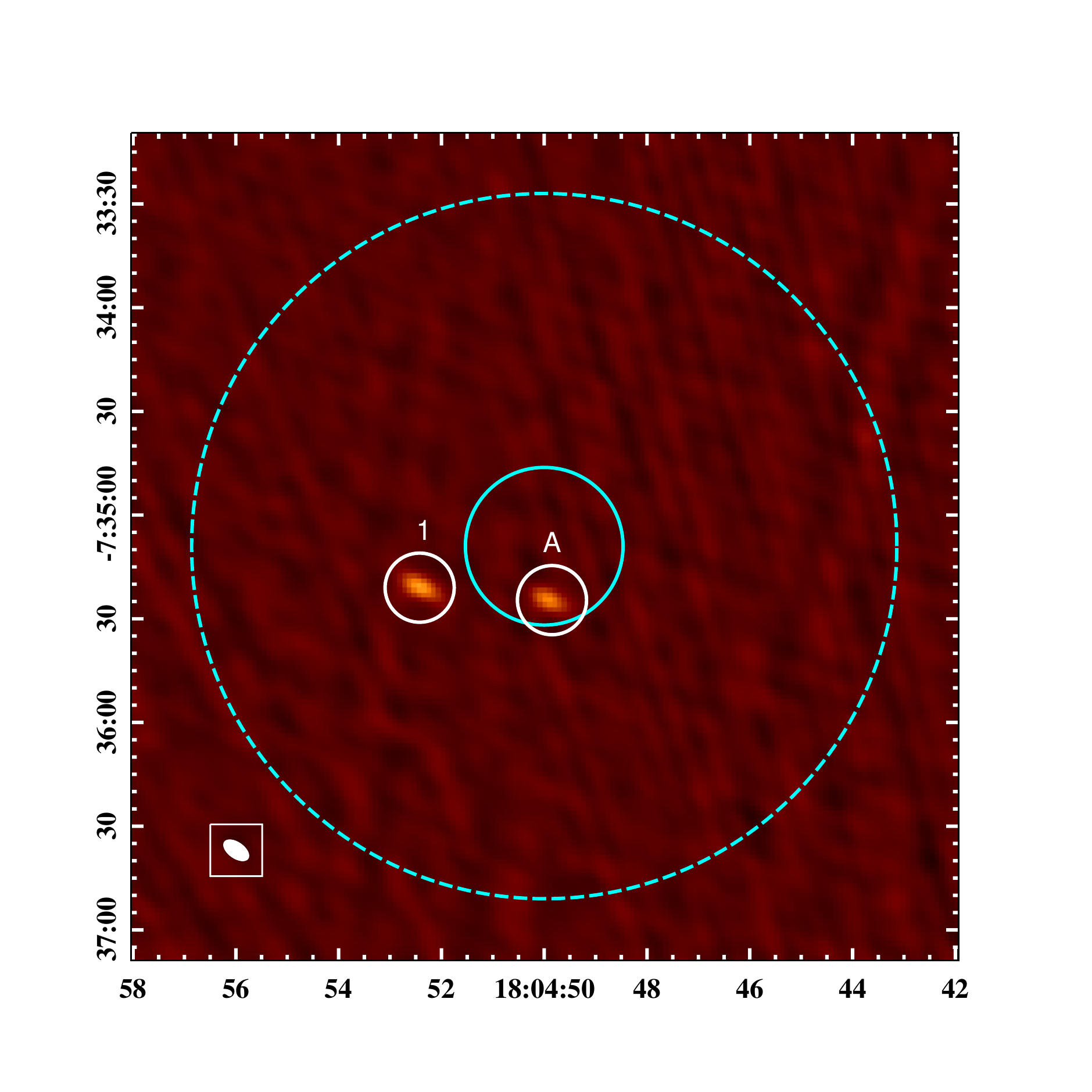}
  \caption{NGC~6539}
  \label{fig:sfig3}
\end{subfigure}
\\ \vskip 0.4 cm

\begin{subfigure}{.33\textwidth}
  \centering
  \includegraphics[width=1.0\linewidth,trim={1.5cm 1.5cm 1.5cm 1.5cm}]{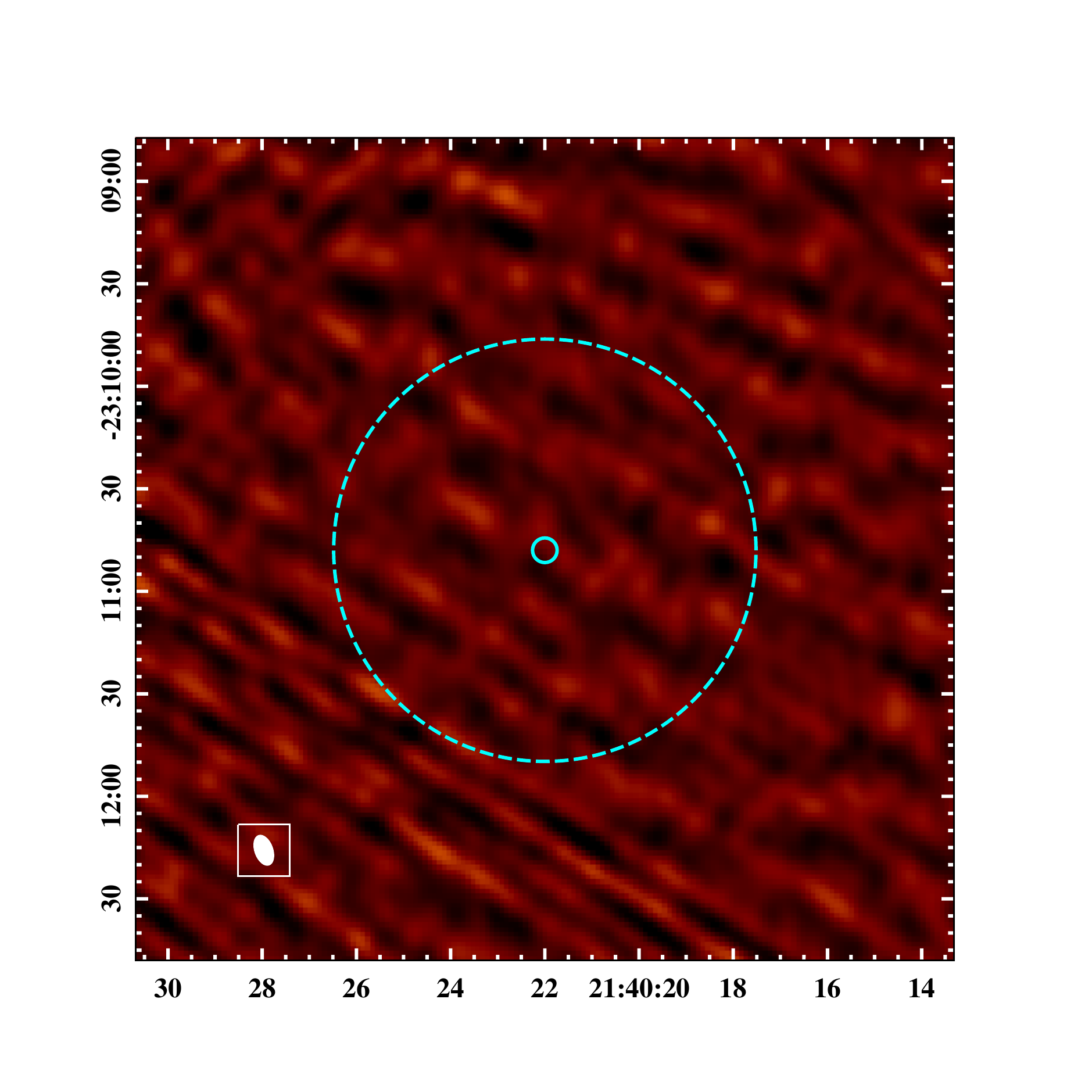}
  \caption{M~30}
  \label{fig:sfig4}
\end{subfigure}
\begin{subfigure}{.33\textwidth}
  \centering
  \includegraphics[width=1.0\linewidth,trim={1.5cm 1.5cm 1.5cm 1.5cm}]{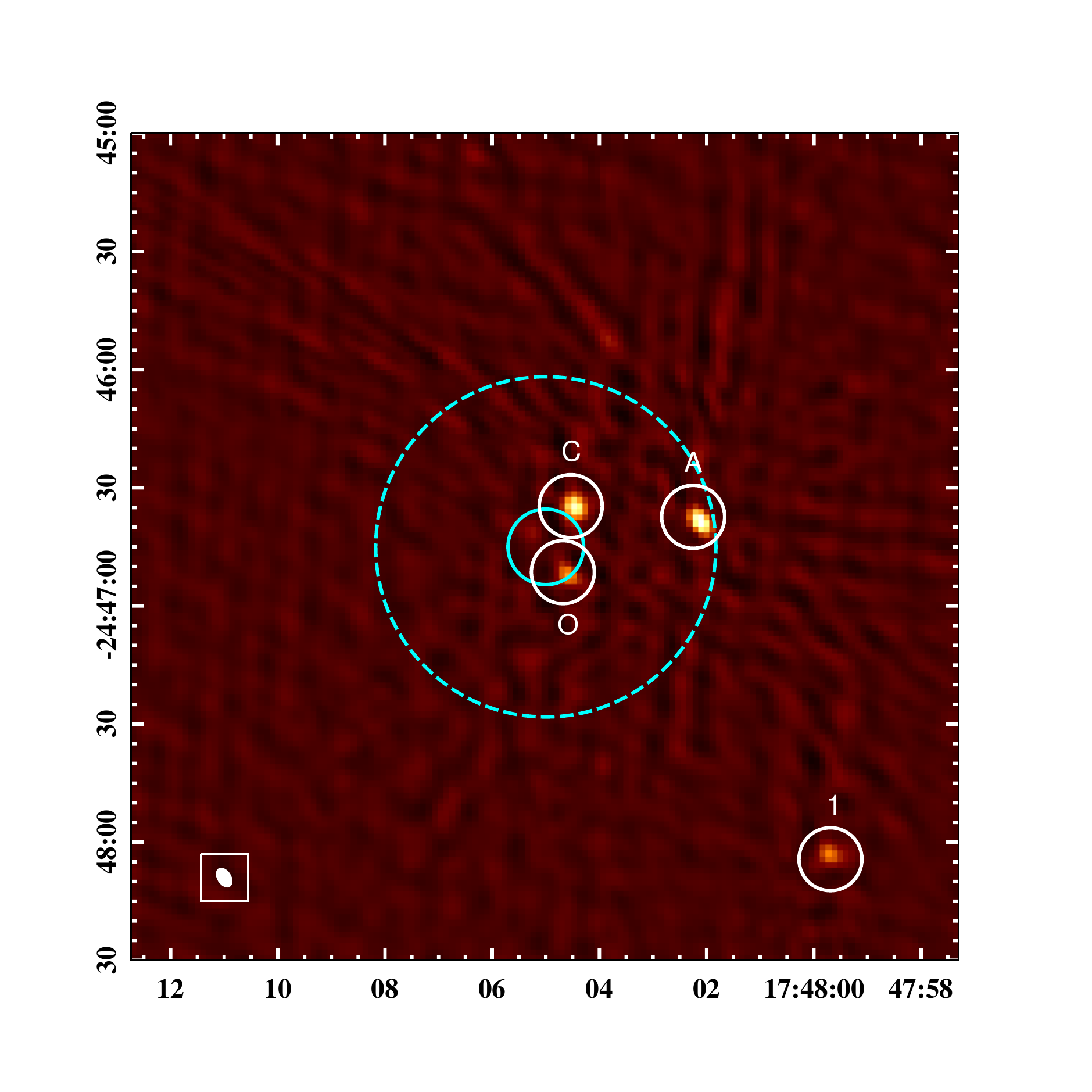}
  \caption{Terzan~5}
  \label{fig:sfig5}
\end{subfigure}
\begin{subfigure}{.33\textwidth}
  \centering
  \includegraphics[width=1.0\linewidth,trim={1.5cm 1.5cm 1.5cm 1.5cm}]{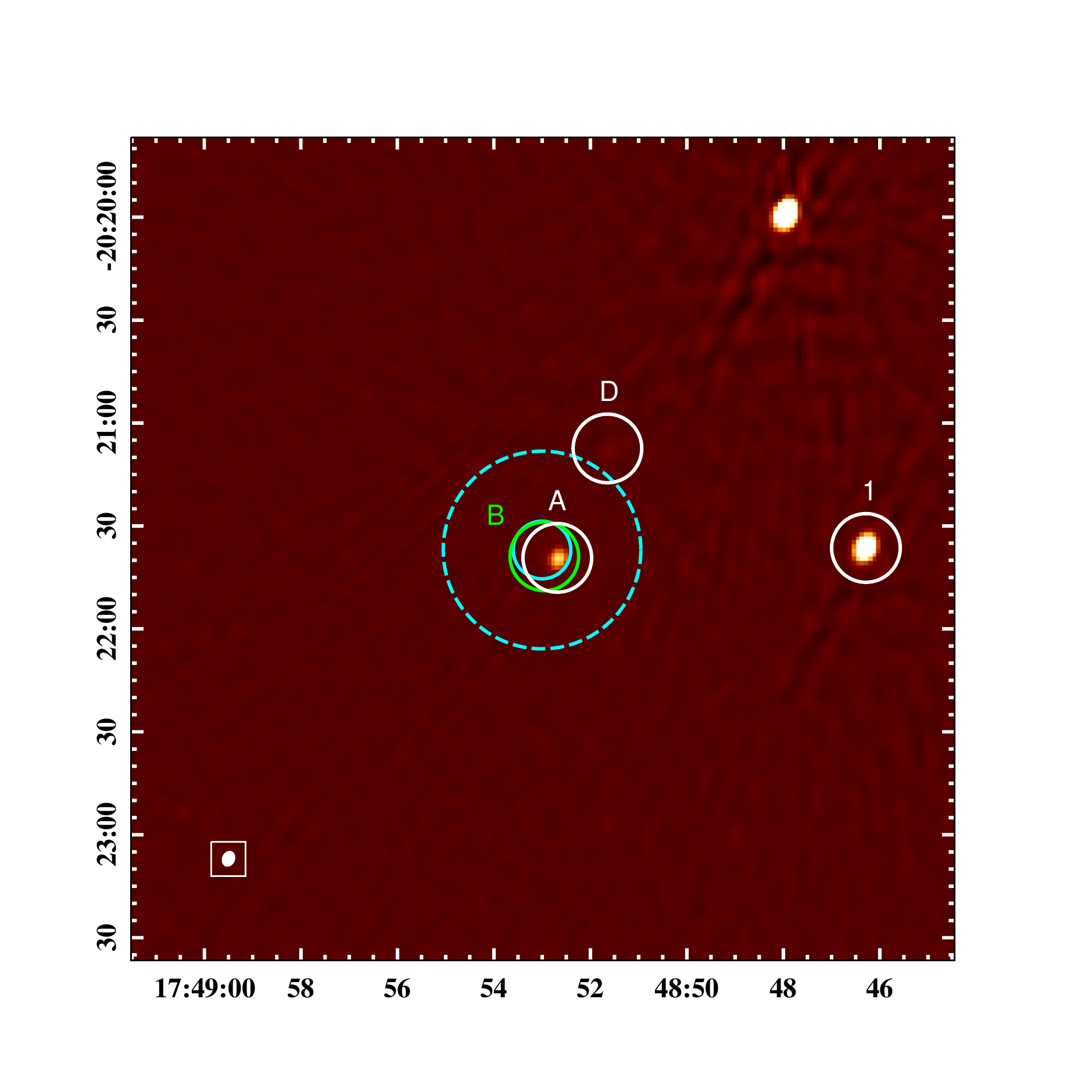}
  \caption{NGC~6440}
  \label{fig:sfig6}
\end{subfigure}
\\ \vskip 0.4 cm
\begin{subfigure}{.33\textwidth}
  \centering
  \includegraphics[width=1.0\linewidth,trim={1.5cm 1.5cm 1.5cm 1.5cm}]{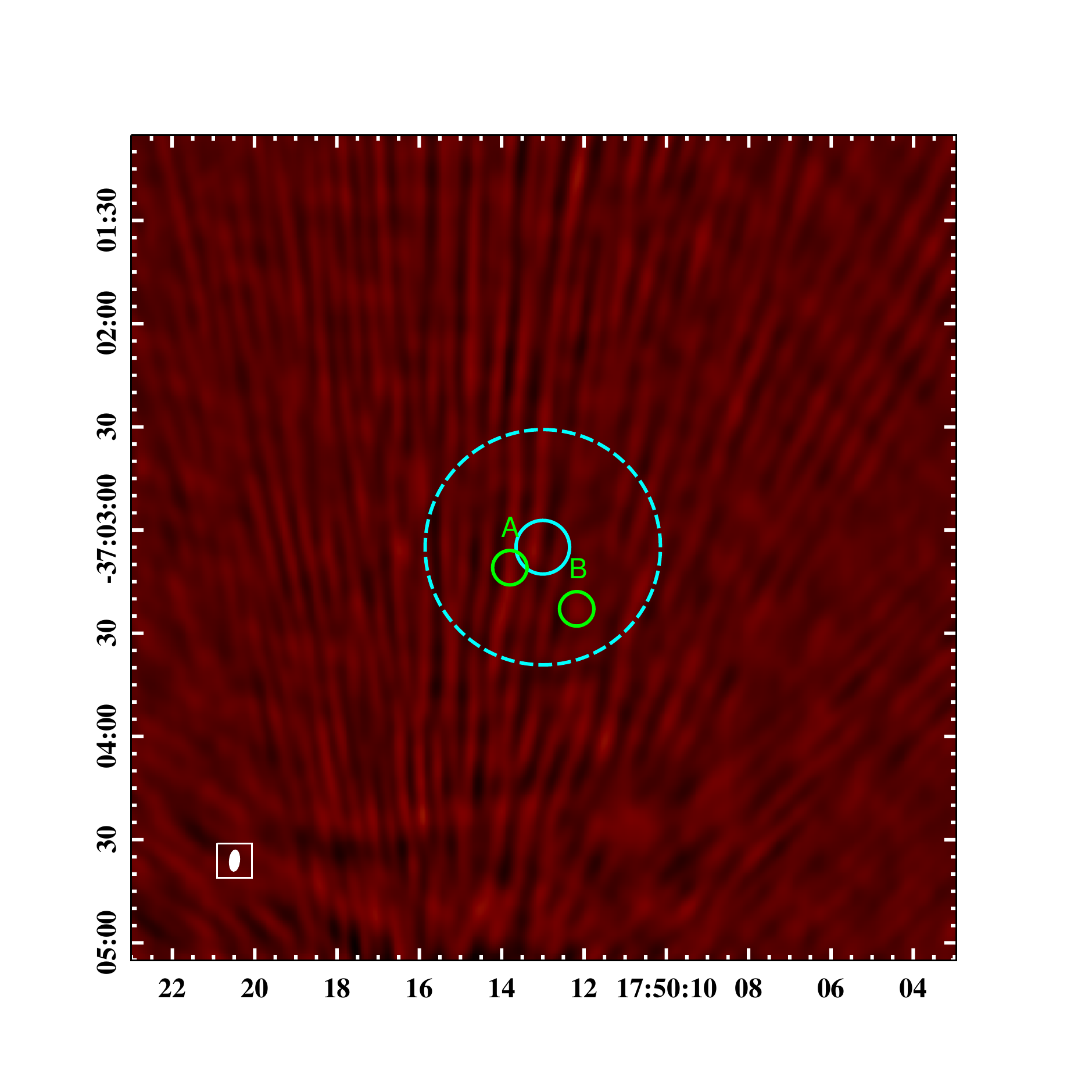}
  \caption{NGC~6441}
  \label{fig:sfig7}
\end{subfigure}
\begin{subfigure}{.33\textwidth}
  \centering
  \includegraphics[width=1.0\linewidth,trim={1.5cm 1.5cm 1.5cm 1.5cm}]{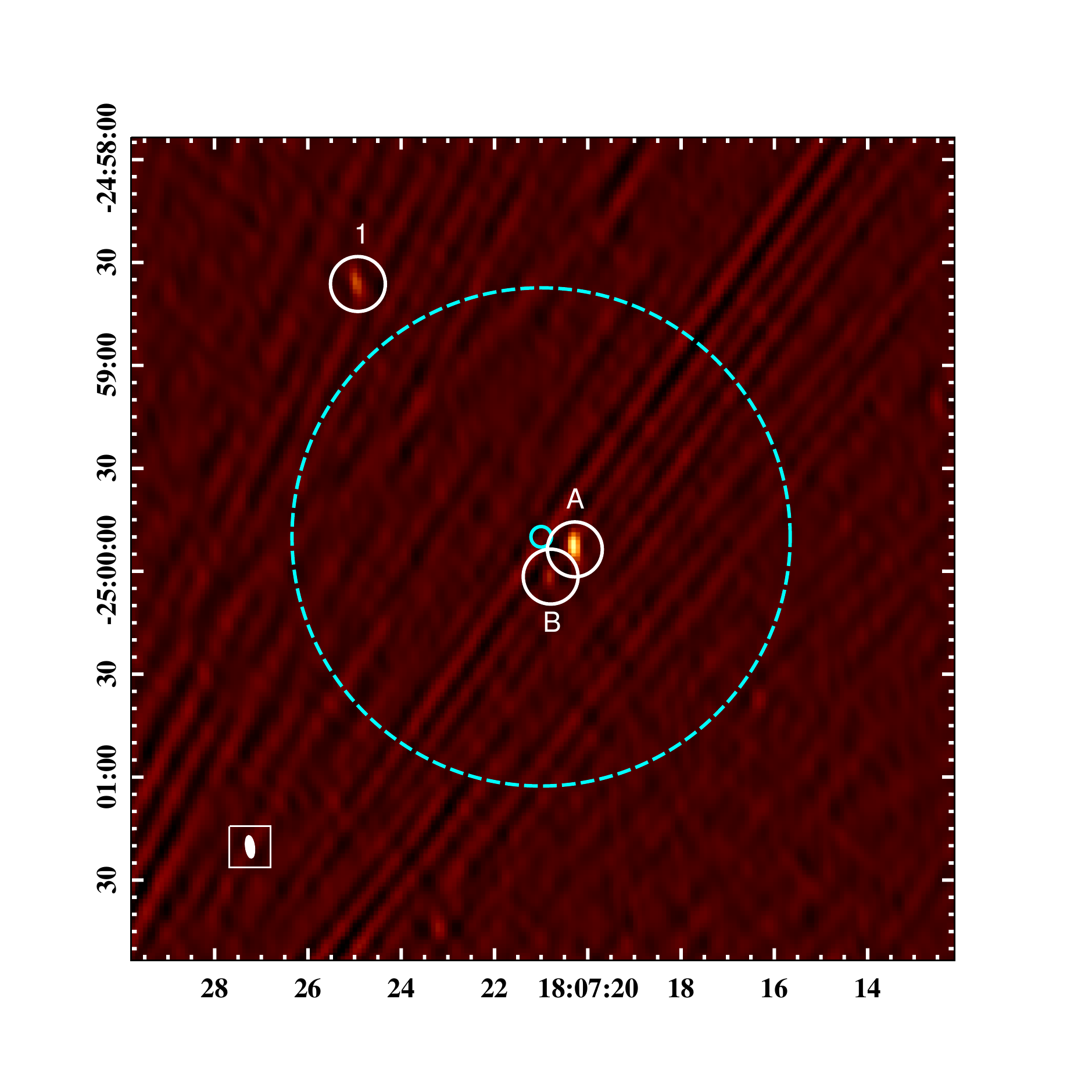}
  \caption{NGC~6544}
  \label{fig:sfig8}
\end{subfigure}
\caption{Radio Images of 8 GCs observed with uGMRT. X and Y axis represent RA (hh:mm:ss) and DEC ($\degr$:$\arcmin$:$\arcsec$), respectively. The cyan-coloured circle shows the core radius, and the dashed cyan circle shows the half-mass radius.} Sources circled in white are the detections in the image, while green-coloured circles show sources visible in the PA data but not in the radio image. In the lower left corner, we display the size and
orientation of the imaging beam.
\label{fig:pos_beams}
\end{figure*}

We used the \texttt{CASA} task \texttt{tclean} in the multi-term 
multi-frequency synthesis (MT-MFS) mode \citep{Rau+2011} to produce 
radio maps of each cluster with in-band spectral indices calculated 
for all sources. A few rounds of phase-only self-calibration were 
also performed for each of the clusters.
As the extent of the GCs ($<2\arcmin$) is small compared to the FWHM of the primary beam 
of the GMRT antennas ($69\arcmin$ at 400~MHz and $43\arcmin$ at 640~MHz), the flux densities 
of radio sources near clusters were negligibly affected by the beam response. 
Therefore, we did not correct for this effect in the images. The positions of all radio sources are determined with respect to the well-known positions of the phase calibrators in the International Celestial Reference Frame (ICRF). The pulsar timing uses the planetary ephemerides DE436, which is aligned with the ICRF to within 0.2 milliarcseconds. All the radio images are presented in Figure~\ref{fig:pos_beams}, and Table \ref{tab:image_props} summarises the image properties of each of these clusters.

In NGC~6652, we see four bright radio sources (labelled 1-4) near the 
core of this cluster. Source 1 is the brightest in the cluster 
and coincides with the timing position of the new pulsar, NGC6652B. 
This pulsar may be the steep-spectrum source mentioned by \citet{Tremou+18}, 
but without a reported position we cannot be sure.
Sources labelled 2, 3, and 4, are not associated with known pulsar 
positions in this cluster. Since all three sources are within the tidal radius and 
the two with measurable spectral indices (Table~\ref{tab:flux_image_sources}) that
are relatively steep ($\alpha < -1$), they may be pulsars. Observations to 
follow up on these sources are ongoing.   

In clusters NGC~1851, NGC~6539, NGC~6544, NGC~6440, and Terzan~5, 
we found five more radio sources not associated with any 
known pulsar position (one in each of these). 
The source in NGC~1851 was first noted by \citet{Freire+2004}, 
who found no pulsations at 327~MHz.  
For the other four clusters, there are 
higher frequency radio maps from the MAVERIC survey \citep{Tremou+18}. 
Using the 5 and 7~GHz radio point source catalogues of 
\citet{Shishkovsky+20}, we find no matches within 4~arcseconds of our 
source in NGC~6539.  Based on the 5.2~GHz RMS noise of 
$1.7~{\rm \mu Jy~beam}^{-1}$ for this cluster, we can set a 5$\sigma$ upper limit of $\alpha_{0.4-5} = -2.3$ for the spectral index of our source.  
For NGC~6544, we find that our source matches (within an arcsecond) 
with one in the \citet{Shishkovsky+20} catalogue.  Based on our 
650~MHz flux density measurement and the reported 5.0 and 7.4~GHz flux 
densities, we obtain a spectral index of $\alpha \approx -0.6$,
which could be a flat-spectrum pulsar or a background radio galaxy. 
For NGC~6440, we find a matching MAVERIC source within 0.4~arcseconds 
of our source.  Based on our measured 650~MHz flux density and the 
reported 5.0 and 7.1~GHz flux densities, we estimate a spectral index 
of $\alpha \approx -1$.  
For Terzan~5, we clearly see that our source is a background galaxy 
in the 3.4~GHz map by \citet{Urquhart+20}.

All the pulsar candidate point sources are well within the 
tidal radii of each of these clusters, indicating they could possibly 
be part of these clusters. Since these sources are at the outskirts 
of the clusters, they would have been missed by previous pulsar 
search surveys, which only observed their central regions. As these sources 
are now localised in the image, we can point all the GMRT antennas on their 
locations and achieve higher sensitivity and hopefully clarify their nature.

Estimates of flux density, spectral indices, and positions of all these visible radio sources are listed in 
Table~\ref{tab:flux_image_sources}. The statistical 1$\sigma$ uncertainties from the \texttt{CASA} task 
\texttt{imfit} are presented along with the measured values. To confirm the accuracy of these estimates, 
we compared the imaging positions with known timing positions for the pulsars with phase-connected solutions. 
We noticed a positional offset in these images of around 1-2 arcseconds in RA and DEC. To verify this, we compared the positions of bright sources in the field of view of each of these clusters with their 
positions in the NVSS catalogue of \citealt{Condon+1998}. Columns 5 and 6 show the mean systematic shift 
(along with the uncertainties) in RA and DEC for each of these clusters. 

\begin{figure}
\centering
        \includegraphics[width=\columnwidth]{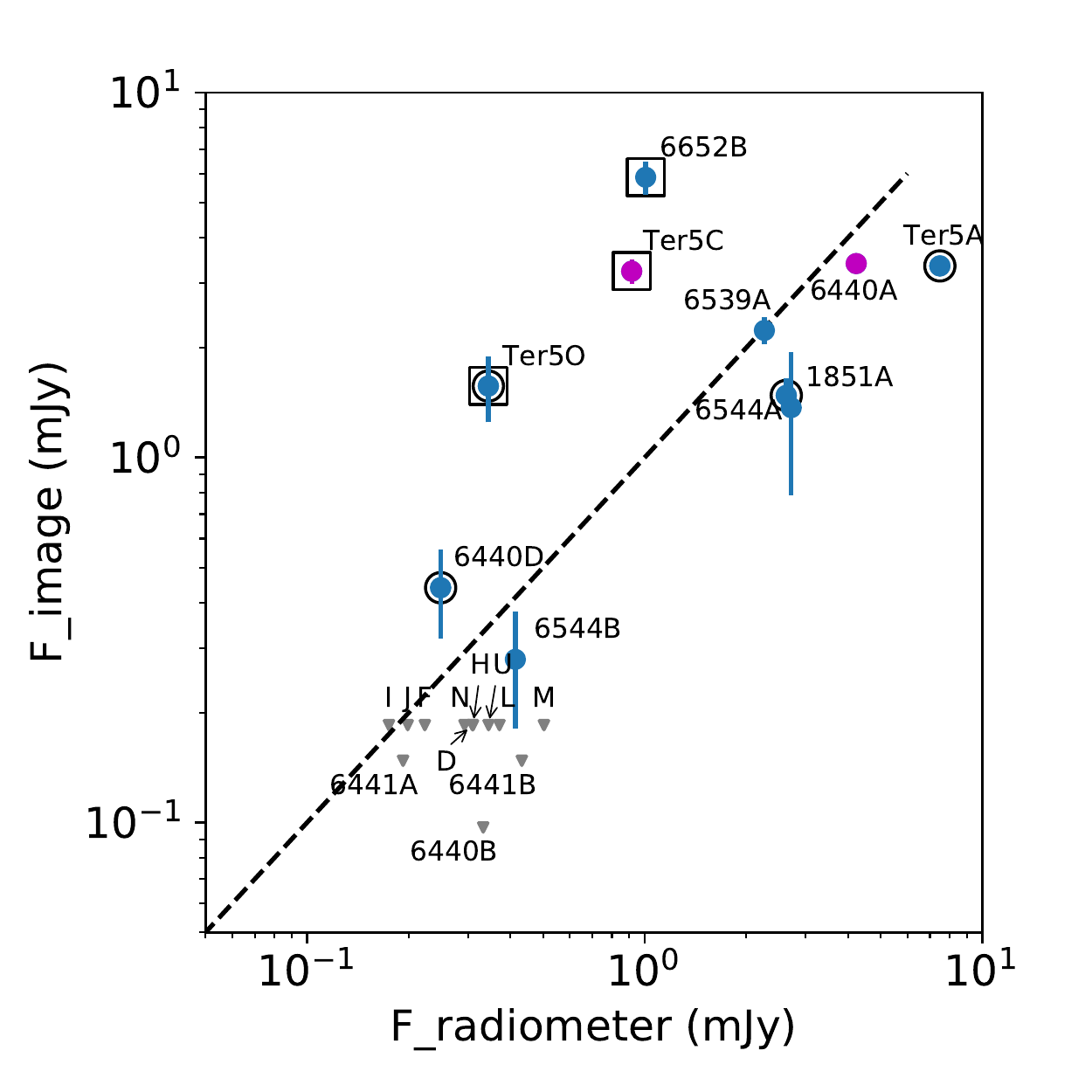}
        \caption{Comparison of flux density estimates from radio images and folded profiles at 400 MHz for pulsars NGC~1851A, NGC~6544A, NGC~6544B, NGC~6652B, and NGC~6539A, and at 650 MHz for the rest. Blue points represent binary systems, while magenta points represent isolated systems. Pulsars in eclipsing binary systems are marked with black circles, and pulsars with large duty cycles are marked with squares. Points in grey represent the upper limits for pulsars with no measured flux in imaging. The dashed line represents the linear expected trend. The only factor correlating with an imaging flux excess is the large duty cycle.}
        \squeezeup
        \label{fig:flux_comparison}
\end{figure}

A comparison between the flux density estimates from the radiometer equation (as discussed in Section \ref{subsec:characterization}) and from the images is shown in Figure~\ref{fig:flux_comparison}. Upper limits on the flux density of pulsars not visible in the image is set to be 3$\sigma$, that is, thrice the RMS noise of the respective cluster's image. We see that most of the pulsars follow a linear trend on the plot, except for PSR NGC~6652B, Ter~5C, and Ter~5O, where the flux obtained from imaging is significantly higher than that of folded profiles from the PA data.
There are several possible causes for this. First, these pulsars have broad pulse profiles, which could indicate that they have emission at all spin phases,
with the varying component of the flux density representing a relatively small amount of the total flux density. Indeed, a large
non-varying component should only be detectable in the radio images.
Such emission could be studied in detail in polarimetric studies of their pulse profiles by revealing, for instance, large amounts of polarised emission during the full rotation cycle. 
Second, in the case of an eclipsing binary system such as Ter~5O, the additional flux in the image can also include contributions from the continuum radio emission from the plasma in the interacting binary. In such systems, one could also see suppression of the pulsed emission, further skewing the ratio of continuum to pulsed emission.

In Figure~\ref{fig:flux_comparison}, we highlight which pulsars are in binaries or eclipsing binaries, and which ones have large duty cycles.
We can see that the only reliable predictor of an anomalous imaging flux is the large duty cycle, suggesting that the excess imaging flux is caused by emission at all spin phases, and not by interaction with the companion.

A surprising discovery is that one of these pulsars -our newly discovered pulsar, NGC~6652B-  is the brightest pulsar identified in all our images, despite its faint pulsations.
The position of the pulsar is close to the cluster centre, this means that an additional possible explanation for its large flux could be the overlap of several radio sources at that position. However, this is unlikely given that Source 1 is not at the centre of NGC~6652 (see Fig.~\ref{fig:cluster-position-6652}), but offset from it and centred instead in the timing position of NGC 6652B. If there were additional sources contributing to their flux, they should (statistically) be distributed closer to the centre and shift the position of Source 1 towards the centre of the cluster. Despite that, we cannot exclude the
possibility that additional radio sources are contributing to the large flux of NGC~6652B.  

\begin{table*}
\caption[]{Properties of the radio images.}
\label{tab:image_props}
\footnotesize
\centering
\renewcommand{\arraystretch}{1.0}
\begin{tabular}{crrrrr}
\hline
\hline
{Cluster} &{Epoch} &{Frequency} &{RMS noise} &{Synthesized beam} & {Position Angle} \\ & {(MJD)} & {(MHz)} & {(mJy/bm)} & ($\arcsec$) & ($\degr$) \\ \midrule
{NGC~1851} & 58051 &400 &0.220 &13.30 $\times$ 4.69 & 26.6 \\
{Terzan~5}  & 58332 &650 &0.062 &5.48 $\times$ 3.58 & 32.2 \\
{NGC~6440}  & 58363 &650 &0.032 &4.87 $\times$ 3.72 & $-$23.2  \\
{NGC~6441}  & 58332 &650 &0.049 &6.50 $\times$ 3.22 & $-$5.4 \\
{NGC~6539} & 57966 &400 &0.088 &8.58 $\times$ 4.80 & 54.8  \\
{NGC~6544}  & 58363 &650 &0.043 &6.99 $\times$ 2.94 & 7.5 \\
{NGC~6652}  & 58084 &400 &0.124 &10.87 $\times$ 4.84 &  31.6  \\
{M30} & 58102 &400 &0.220 &9.44 $\times$ 5.54 & 19.7 \\
\hline
\hline
\end{tabular}
\end{table*}

\begin{table*}
\caption[]{Properties of radio sources found in the images.}
\label{tab:flux_image_sources}
\footnotesize
\setlength{\tabcolsep}{1.5pt}
\centering
\renewcommand{\arraystretch}{1.2}
\vskip 0.1cm
\begin{tabular}{lcccccccccc}
\hline
\hline
{Cluster} & {Systematic} & {Systematic} & {Source ID / Name} &{RA (J2000)} &{DEC (J2000)}   & {Flux}&{Spectral Index} \\
& {shift in RA (s)} & {shift in DEC ($\arcsec$)} & &{(hh:mm:ss)} & {$\degr$:$\arcmin$:$\arcsec$} &{(mJy)}\\ \midrule
{NGC~1851} &  0.31(6)& $-$2.25(73)& J0514$-$4002A & 05:14:06.70 $\pm$ 0.11 & $-$40:02:50.87 $\pm$ 2.43 & 1.37 $\pm$ 0.58 & - \\ 
& & & 1 & 05:14:24.65 $\pm$ 0.013 & $-$40:00:19.99 $\pm$ 0.29 & 19.6 $\pm$ 1.3 & $-$1.27 $\pm$ 0.29\\ \\
{Terzan~5} & 0.17(10) & $-$0.82(1.07) & J1748$-$2446A & 17:48:02.11 $\pm$  0.003& $-$24:46:38.62 $\pm$ 0.06 & 3.35 $\pm$  0.13 & - \\
&  & & J1748$-$2446C & 17:48:04.46 $\pm$  0.008 & $-$24:46:35.06 $\pm$ 0.14  & 3.24 $\pm$ 0.32 & - \\
& & & J1748$-$2446O & 17:48:04.58 $\pm$ 0.02& $-$24:46:52.29 $\pm$ 0.21 &  1.57 $\pm$ 0.25 & - \\
& & & 1 & 17:47:59.68 $\pm$ 0.02& $-$24:48:03.23 $\pm$ 0.14 & 1.52 $\pm$ 0.21 & - \\ \\
{NGC~6440}  & 0.04(4) & $-$0.5(6) & J1748$-$2021A & 17:48:52.66 $\pm$  0.002 & $-$20:21:39.64 $\pm$  0.04  & 3.40 $\pm$ 0.11 & $-$2.02 $\pm$ 0.01 &  \\
& & & J1748$-$2021D & 17:48:51.59 $\pm$  0.03& $-$20:21:36.08 $\pm$  0.46  & 0.44 $\pm$ 0.12 & - \\
& & & 1 & 17:48:46.29 $\pm$  0.0008& $-$20:21:36.41 $\pm$  0.01 & 18.33 $\pm$ 0.20 & $-$0.94 $\pm$ 0.01 \\ \\
{NGC~6539} & - & - & B1802$-$07 & 18:04:49.91 $\pm$ 0.01 & $-$07:35:24.75 $\pm$ 0.12  & 2.23 $\pm$ 0.19 & -  \\
& & & 1 & 18:04:52.41 $\pm$ 0.015 & $-$07:35:29.92 $\pm$ 0.12 &  2.94 $\pm$ 0.21 & - \\ \\
{NGC~6544}  & 0.06(3) & 1.0(6) & J1807$-$2459A & 18:07:20.29 $\pm$ 0.004 & $-$24:59:52.74 $\pm$ 0.23  & 1.48 $\pm$ 0.17 & - \\
& & & J1807$-$2459B & 18:07:20.81 $\pm$ 0.007& $-$25:00:01.70 $\pm$ 0.48  & 0.28 $\pm$ 0.099 & - \\
& & & 1 & 18:07:24.94 $\pm$ 0.01& $-$24:58:36.14 $\pm$ 0.46 &  0.68 $\pm$  0.12 & - \\ \\

{NGC~6652}  & 0.11(13) & 1.88 (1.83) & 1 / {J1835$-$3259B} & 18:35:45.68 $\pm$ 0.01 & $-$32:59:23.46 $\pm$  0.31 & 5.86 $\pm$  0.60 & $-$3.57 $\pm$ 0.25  \\
& & & 2 & 18:35:59.67 $\pm$ 0.02 & $-$32:58:20.95 $\pm$ 0.34 &  3.65 $\pm$ 0.28 & $-$2.03 $\pm$ 0.38 \\
& & & 3 & 18:35:39.30 $\pm$ 0.01 & $-$33:03:25.01 $\pm$ 0.15 &  3.95 $\pm$ 0.22 & $-$1.08 $\pm$ 0.09 \\
& & & 4 & 18:35:35.39 $\pm$ 0.02 & $-$32:58:22.23 $\pm$  0.40 &  1.90 $\pm$ 0.19 & - \\ \\

\hline
\hline
\end{tabular}
\end{table*}

\section{Conclusions and discussion}
\label{sec:discussion}
\label{sec:conclusions}
 
In this paper, we present the results of a GC pulsar survey we performed with the new wide-bandwidth receivers of uGMRT to search for steep-spectrum pulsars at 400 MHz and 650 MHz. We observed eight GCs and searched each cluster for isolated and binary pulsar systems with segmented and full-length acceleration and jerk search techniques. We discovered a new MSP binary, J1835$-$3259B in NGC~6652. We presented the timing solution for this system with four years of GMRT observations and nine years of GBT observations, for a total time baseline of 10 years. The system is in a wide orbit binary of 1.19 days and has a small eccentricity of $3.5 \times 10^{-5}$. Assuming a pulsar mass of $1.4 \, M_{\odot}$, the median companion mass is 0.21 $M_\odot$. 

The observed orbital and spin period derivatives of the pulsar have contributions from accelerations due to the galactic field ($a_{\rm gal}$), GCs ($a_{\rm gc}$), and its composite proper motion ($\mu$):
\begin{equation}
    {\left(\frac{ {\dot{P}_{\rm b}}}{P_{\rm b}}\right)}_{\rm obs} = {\left(\frac{ {\dot{P}_{\rm b}}}{P_{\rm b}}\right)}_{\rm int} + \frac{a_{\rm gc}}{c}+ \frac{a_{\rm gal}}{c} + \frac{\mu^2 d}{c}
.\end{equation}

Assuming no contribution from the intrinsic orbital period derivative due to negligible gravitational wave damping in the system, the GC acceleration can hence be constrained to $\, 3.3 \pm 6.3 \times 10^{-9} \rm m \, \rm s^{-2}$; this does not
usefully constrain the mass model of the cluster.

Since both the observed spin and orbital period derivatives will be affected by similar accelerations on the system, the following relation can be used to estimate the intrinsic spin period derivative of the pulsar:
\begin{equation}
    \left( \frac{\dot{P}}{P} \right)_{\rm int} = \left(\frac{\dot{P}}{P}\right)_{\rm obs} - \left(\frac{ {\dot{P}_{\rm b}}}{P_{\rm b}}\right)_{\rm obs}\; .
\end{equation}

From the observed values of $\dot{P}$ and $\dot{P}_{\rm b}$, we estimate a 1$\sigma$ upper limit on the intrinsic spin period derivative, $ \dot{P}_{\rm int} \leq 6.7 \times 10^{-20} \, \rm s \rm s^{-1}$. 
This puts a lower limit of $0.43 \rm \, Gyr$ on the characteristic age and an upper limit of $3.5 \times 10^{8} \, \rm G$ on the surface magnetic field. These values are consistent with MSP systems in the Galactic disc that have helium WD companions, and also some MSP - helium WDs in GCs \citep{Freire+2017}.

We find that some of the previously known pulsars in these clusters have steep ($<-2$) spectra, despite the fact that
they were found in surveys at higher frequencies. Comparing the predicted scattering timescales to those measured, we also found the measured values to be considerably smaller than the predictions in most cases (see Table~\ref{tab:flux_estimates}). This means it was easier to detect these pulsars than might have been expected. However, it is important to be aware of the uncertainty around the predictions; \citealt{2015MNRAS.449.1570L} found that their estimate of the $\tau$-DM relationship predicted weaker scattering at large DMs than that of \citealt{Bhat+2004}, which we use here for the predicted scattering timescales. We note that the largest differences between predicted and measured scattering timescales in our sample correspond to those pulsars with larger DMs, in agreement with the result of \citealt{2015MNRAS.449.1570L}.

With the visibility dataset recorded simultaneously with uGMRT, we created radio images of these eight clusters.
From this, we estimated the flux density and spectral indices of all pulsars in these clusters.
Interestingly, the newly discovered pulsar NGC~6652B is the brightest of all pulsars among all of our images.
This and other pulsars with broad pulse profiles, such as Terzan 5 C and Terzan 5 O, are substantially brighter in
the images than they appear in pulsation studies, this shows that they have a large amount of DC radio emission, even
at the minimum of their pulse profiles. This raises the possibility that some of the sources in the images might be 
pulsars that are bright but also have small variability.

These radio images are also useful for looking for bright sources that may have 
been missed by limited beam size or biases and limitations of the pulsation searches. 
We identified three radio sources not associated with any known pulsars in 
NGC~6652 and one each in NGC~6539, Terzan~5, NGC~6440, and NGC~6544. 
Cross-checking previous surveys shows that the three sources in
NGC~6652 and one source in NGC~6539 have not been detected before. The source in Terzan~5 is 
a background galaxy \citep{Urquhart+20}, and the very bright source in NGC~1851 
was detected by \citep{Freire+2004} and has no pulsar counterpart; the other five sources are 
good pulsar candidates that we will follow up on in future pulsation 
searches.
The detection of unknown steep sources in the images of GCs allow us to identify 
undiscovered pulsars that may be present outside the core regions; these outer 
parts of the clusters are not usually covered when searching for pulsars in GCs. 
Thus, interferometers such as the GMRT will play a crucial role in a more complete 
characterisation of the pulsar population in GCs. 

\begin{acknowledgements}
We thank the referee for the detailed report and many constructive suggestions, which have contributed significantly to the quality of this work.
We thank the staff of the GMRT for help with the observations. The GMRT is operated
by the National Centre for Radio Astrophysics (NCRA) of the
Tata Institute of Fundamental Research (TIFR), India. The Green Bank Observatory is a facility of the National Science Foundation operated under cooperative agreement by Associated Universities, Inc. Data used in this analysis were taken under GBT projects AGBT09B 031, AGBT09C 072, AGBT10A 060, AGBT10A 082, AGBT10B 018, AGBT11B 070, and AGBT12A 388. Part of this research was carried out at the Jet Propulsion Laboratory, California Institute of Technology, under a contract with the National Aeronautics and Space Administration.
TG thanks Gregory Desvignes and Abhimanyu Susobhanan for useful discussions.
AR gratefully acknowledges continuing valuable support from the Max-Planck Society, as well as financial support by the research grant ``iPeska'' (P.I. Andrea Possenti) funded under the INAF national call Prin-SKA/CTA approved with the Presidential Decree 70/2016. MED acknowledges support from the National Science Foundation (NSF) Physics Frontier Center award 1430284, and from the Naval Research Laboratory by NASA under contract S-15633Y.
\end{acknowledgements}



\bibliographystyle{aa}
\bibliography{43062corr} 


\begin{appendix}
\section{}
\label{appendix:a}
Table \ref{tab:psr_cands} presents the properties of a few high-significance pulsar candidates found from the GC census presented in this work. Of these, two candidates in GCs NGC~6440 and NGC~6544 were found with exceptionally high acceleration.

In NGC~6440, the pulsar candidate 
was found in the segmented acceleration search with an extraordinary acceleration of $354 \rm \,  m\, s^{-2}$, rotating at a spin period of 3.35-ms. Its high acceleration could be due to its presence in a tight binary system, hence, if confirmed, it could be one of the few systems that can be used to test theories of gravity in the strong field regime. The candidate is only found in the first 20-min segment of the observation, which could be because of the failure of the assumption of constant acceleration for the remainder of the orbit. 
Similarly, in cluster NGC~6544, a 4.43-ms pulsar candidate was detected with an acceleration of $\sim$ $102\, \rm  m \, s^{-2}$. Considering the significance of all the candidates presented in Table \ref{tab:psr_cands} and their detections in multiple segments, we encourage future search surveys in these clusters -particularly at low frequencies- to confirm these candidates. 
\begin{table}[H]
\caption[]{Properties of significant pulsar candidates}
\label{tab:psr_cands}
\footnotesize
\begin{center}
\renewcommand{\arraystretch}{1.0}
\begin{tabular}{crrrrrrr}
\hline
\hline
{Cluster} & {Epoch} & {DM} &{Period} &{Acceleration} &{Segment length} & {Significance} & {Intensity profile} \\  & {(MJD)} & {(\dmunit)} & (ms) & ($\rm m\,s^{-2}$) & {(min)} & ($\sigma$) & \\ \midrule
{NGC~6441} & 58332 & 232.99 & 2.15 & 0 & full & 11.2 & \begin{minipage}{.1\textwidth}
      \includegraphics[width=\linewidth, height=15mm]{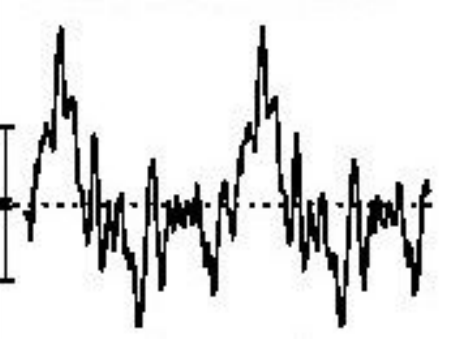}
    \end{minipage} \\
{NGC~6440} & 58363 & 227.92 & 3.35 & 354 & 20 & 10.6 & \begin{minipage}{.1\textwidth}
      \includegraphics[width=\linewidth, height=15mm]{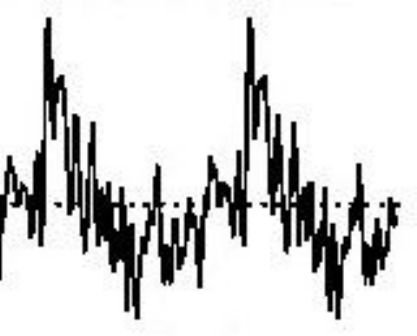}
    \end{minipage} \\
{NGC~6440} & 58363 & 215.65 & 2.61 & 0.1 & 20, full & 11.1 & \begin{minipage}{.1\textwidth}
      \includegraphics[width=\linewidth, height=15mm]{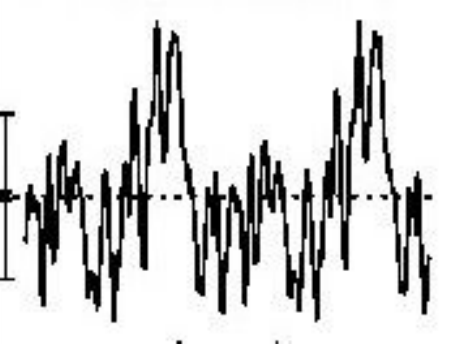}
    \end{minipage}\\
{NGC~6544} & 58363 & 130.68 & 4.43 & 102 & 20 & 9.7 & \begin{minipage}{.1\textwidth}
      \includegraphics[width=\linewidth, height=15mm]{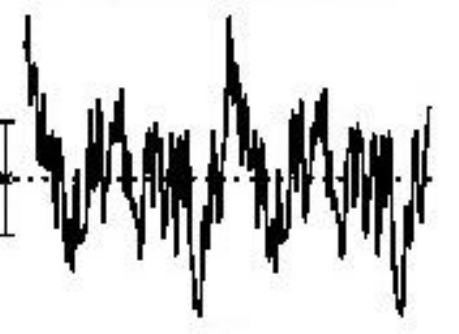}
    \end{minipage} \\
{NGC~6544} & 58363 & 132.35 & 5.17 & 0 & full & 7.9 & \begin{minipage}{.1\textwidth}
      \includegraphics[width=\linewidth, height=15mm]{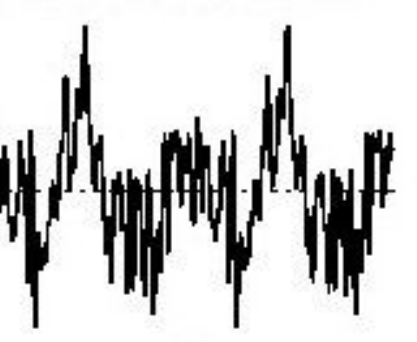}
    \end{minipage}\\
{Terzan~5} & 58332 & 219.90 & 2.93 & 0 & 20 & 9.8 & 
\begin{minipage}{.1\textwidth}
      \includegraphics[width=\linewidth, height=15mm]{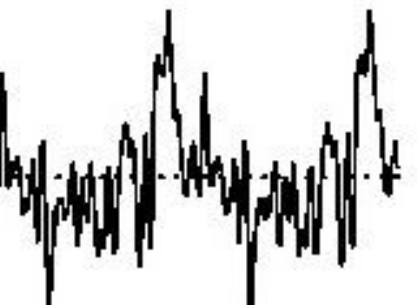}
    \end{minipage}\\
\hline
\hline
\end{tabular}
\end{center}
\end{table}
\FloatBarrier
\end{appendix}

\end{document}